\documentclass[twocolumn]{aastex6}

\def\adfs27{ADFS-27}

\received{20/01/2020}
\revised{17/04/2020}
\accepted{21/04/2020}

\usepackage{apjfonts}
\usepackage{aas_macros, xspace} 
\newcommand{\eq}{\,=\,}

\def\ts     {\thinspace}
\def\kms    {\ts km\ts s$^{-1}$}

\def\msol   {$M_{\odot}$}
\def\lsol   {$L_{\odot}$}
\def\lprime {K\,\kms\,pc$^2$}

\def\oh     {{\rm OH}($^2\Pi_{1/2}\,J$=3/2$\to$1/2)}
\def\bci    {[C{\scriptsize I}]($^3P_{2}$$\to$$^3P_{1}$)}
\def\cii    {[C{\scriptsize II}]($^2P_{3/2}$$\to$$^2P_{1/2}$)}
\def\nii    {[N{\scriptsize II}]($^3P_{1}$$\to$$^3P_{0}$)}
\def\aco    {{\rm CO}($J$=1$\to$0)}

\def\bco    {{\rm CO}($J$=2$\to$1)}

\def\dco    {{\rm CO}($J$=4$\to$3)}
\def\eco    {{\rm CO}($J$=5$\to$4)}
\def\fco    {{\rm CO}($J$=6$\to$5)}
\def\gco    {{\rm CO}($J$=7$\to$6)}

\def\pco    {{\rm CO}($J$=16$\to$15)}

\def\ahcn    {{\rm HCN}($J$=1$\to$0)}

\def\ahco    {{\rm HCO$^+$}($J$=1$\to$0)}

\def\ahnc    {{\rm HNC}($J$=1$\to$0)}
\def\acch    {{\rm CCH}($N$=1$\to$0)}

\def\alma     {Atacama Large (sub-)Millimeter Array (ALMA)}

\shorttitle{COLDz:\ Dusty Starbursts at $z$$>$5}
\shortauthors{Riechers et al.}

\begin{document}

\title{${}$\vspace{-6mm} \\ COLDz:\ A High Space Density of Massive Dusty Starburst Galaxies \\ $\sim$1 Billion Years after the Big Bang}


\author{Dominik A.\ Riechers\altaffilmark{1,2}}
\author{Jacqueline A.\ Hodge\altaffilmark{3}}
\author{Riccardo Pavesi\altaffilmark{1}}
\author{Emanuele Daddi\altaffilmark{4}}
\author{Roberto Decarli\altaffilmark{5}}
\author{\\ Rob J.\ Ivison\altaffilmark{6}}
\author{Chelsea E.\ Sharon\altaffilmark{7}}
\author{Ian Smail\altaffilmark{8}}
\author{Fabian Walter\altaffilmark{2,9}}
\author{Manuel Aravena\altaffilmark{10}}
\author{Peter L.\ Capak\altaffilmark{11}}
\author{\\ Christopher L.\ Carilli\altaffilmark{9,12}}
\author{Pierre Cox\altaffilmark{13}}
\author{Elisabete da Cunha\altaffilmark{14,15,16}}
\author{Helmut Dannerbauer\altaffilmark{17,18}}
\author{\\ Mark Dickinson\altaffilmark{19}}
\author{Roberto Neri\altaffilmark{20}}
\author{Jeff Wagg\altaffilmark{21}}

\altaffiltext{1}{Department of Astronomy, Cornell University, Space
  Sciences Building, Ithaca, NY 14853, USA}
\altaffiltext{2}{Max-Planck-Institut f\"ur Astronomie, K\"onigstuhl 17, D-69117 Heidelberg, Germany}
\altaffiltext{3}{Leiden Observatory, Leiden University, P.O. Box 9513,
  NL2300 RA Leiden, The Netherlands}
\altaffiltext{4}{Laboratoire AIM, CEA/DSM-CNRS-Univ.\ Paris Diderot, Irfu/Service d'Astrophysique, CEA Saclay, Orme des Merisiers, F-91191 Gif-sur-Yvette cedex, France}
\altaffiltext{5}{INAF - Osservatorio di Astrofisica e Scienza dello Spazio, via Gobetti 93/3, I-40129, Bologna, Italy}
\altaffiltext{6}{European Southern Observatory,
  Karl-Schwarzschild-Stra{\ss}e 2, D-85748 Garching, Germany}
\altaffiltext{7}{Yale-NUS College, \#01-220, 16 College Avenue West, Singapore 138527}
\altaffiltext{8}{Centre for Extragalactic Astronomy, Department of Physics, Durham University, South Road, Durham DH1 3LE, UK}
\altaffiltext{9}{National Radio Astronomy Observatory, Pete V. Domenici Array Science Center, P.O. Box O, Socorro, NM 87801, USA}
\altaffiltext{10}{N\'ucleo de Astronom\'ia, Facultad de Ingenier\'ia y Ciencias, Universidad Diego Portales, Av. Ej\'ercito 441, Santiago, Chile}
\altaffiltext{11}{Spitzer Science Center, California Institute of Technology, MC 220-6, 1200 East California Boulevard, Pasadena, CA 91125, USA}
\altaffiltext{12}{Cavendish Astrophysics Group, University of Cambridge, Cambridge, CB3 0HE, UK}
\altaffiltext{13}{Sorbonne Universit{\'e}, UPMC Universit{\'e} Paris 6 and CNRS, UMR 7095, Institut d'Astrophysique de Paris, 98bis boulevard Arago, F-75014 Paris, France}
\altaffiltext{14}{Research School of Astronomy and Astrophysics, Australian National University, Canberra, ACT 2611, Australia}
\altaffiltext{15}{International Centre for Radio Astronomy Research, University of Western Australia, 35 Stirling Hwy, Crawley, WA 6009, Australia}
\altaffiltext{16}{ARC Centre of Excellence for All Sky Astrophysics in 3 Dimensions (ASTRO 3D)}
\altaffiltext{17}{Instituto de Astrof\'isica de Canarias, E-38205 La Laguna, Tenerife, Spain}
\altaffiltext{18}{Universidad de La Laguna, Departamento de Astrof\'isica, E-38206 La Laguna, Tenerife, Spain}
\altaffiltext{19}{NSF's National Optical-Infrared Astronomy Research Laboratory, 950 North Cherry Avenue, Tucson, AZ 85719, USA}
\altaffiltext{20}{Institut de RadioAstronomie Millim\'etrique, 300 Rue
  de la Piscine, Domaine Universitaire, F-38406 Saint Martin d'H\'eres,
  France}
\altaffiltext{21}{SKA Organization, Lower Withington, Macclesfield, Cheshire SK11 9DL, UK}

 \email{riechers@cornell.edu}

\begin{abstract}

We report the detection of \bco\ emission from three massive dusty
starburst galaxies at $z$$>$5 through molecular line scans in the
NSF's Karl G.\ Jansky Very Large Array (VLA) CO Luminosity Density at
High Redshift (COLDz) survey. Redshifts for two of the sources,
HDF\,850.1 ($z$=5.183) and AzTEC-3 ($z$=5.298), were previously
known. We revise a previous redshift estimate for the third source
GN10 ($z$=5.303), which we have independently confirmed through
detections of CO $J$=1$\to$0, 5$\to$4, 6$\to$5, and [C{\sc ii}]
158\,$\mu$m emission with the VLA and the NOrthern Extended
Milllimeter Array (NOEMA). We find that two currently independently
confirmed CO sources in COLDz are ``optically dark'', and that three
of them are dust-obscured galaxies at $z$$>$5. Given our survey area
of $\sim$60\,arcmin$^2$, our results appear to imply a $\sim$6--55
times higher space density of such distant dusty systems within the
first billion years after the Big Bang than previously thought. At
least two of these $z$$>$5 galaxies show star-formation rate surface
densities consistent with so-called ``maximum'' starbursts, but we
find significant differences in CO excitation between them. This
result may suggest that different fractions of the massive gas
reservoirs are located in the dense, star-forming nuclear regions --
consistent with the more extended sizes of the [C{\sc ii}] emission
compared to the dust continuum and higher [C{\sc ii}]-to-far-infrared
luminosity ratios in those galaxies with lower gas excitation.  We
thus find substantial variations in the conditions for star formation
between $z$$>$5 dusty starbursts, which typically have dust
temperatures $\sim$57\%$\pm$25\% warmer than starbursts at $z$=2--3
due to their enhanced star formation activity.

\end{abstract}


\section{Introduction} \label{sec:intro}

Luminous dusty star-forming galaxies (DSFGs) represent the most
intense episodes of star formation throughout cosmic history (see,
e.g., Blain et al.\ \citeyear{blain02}; Casey et
al.\ \citeyear{casey14} for reviews). While the bulk of the population
likely existed at redshifts $z$$\sim$1 to 3.5 (e.g., Greve et
al.\ \citeyear{greve05}; Bothwell et al.\ \citeyear{bothwell13}), a
tail in their redshift distribution has been discovered over the past
decade (e.g., Capak et al.\ \citeyear{capak08}; Daddi et
al.\ \citeyear{daddi09a}; Coppin et al.\ \citeyear{coppin10}; Smol{\v
  c}i\'c et al.\ \citeyear{smolcic12b}), found to be reaching out to
$z$$>$5 (Riechers et al.\ \citeyear{riechers10a}; Capak et
al.\ \citeyear{capak11}), and subsequently, $z$$>$6 (Riechers et
al.\ \citeyear{riechers13b}). Dust emission in moderately luminous
galaxies\footnote{In this work, galaxies with infrared luminosities of
  10$^{11}$$<$$L_{\rm IR}$$<$10$^{12}$\,\lsol\ are considered to be
  moderately luminous, and those above as luminous.}  has now been
detected at $z$$>$8 (Tamura et al.\ \citeyear{tamura19}), but no
luminous DSFG is currently known at $z$$\geq$7 (e.g., Strandet et
al.\ \citeyear{strandet17}).

DSFGs in the $z$$>$5 tail are thought to be rare, but their level of
rarity is subject to debate (e.g., Asboth et al.\ \citeyear{asboth16};
Ivison et al.\ \citeyear{ivison16}; Bethermin et
al.\ \citeyear{bethermin15,bethermin17}; see also Simpson et
al.\ \citeyear{simpson14,simpson20}; Dudzeviciute et
al.\ \citeyear{dudzeviciute20}). A significant challenge in
determining the space density of such sources is the difficulty in
finding them in the first place. Given their distance, classical
techniques combining optical and radio identifications have been
largely unsuccessful due to the faintess or lack of detection at these
wavelengths, commonly leading to misidentifications given the
significant positional uncertainties of the classical sub/millimeter
single-dish surveys in which they are the most easily seen (e.g.,
Chapman et al.\ \citeyear{chapman05}; Cowie et
al.\ \citeyear{cowie09}, and references therein). Also, due to the
strong negative K correction at sub/millimeter wavelengths (e.g.,
Blain et al.\ \citeyear{blain02}), it remained challenging to pick out
the most distant DSFGs among the much more numerous specimen at
$z$$<$3.5. Over the past decade, many of these challenges were
overcome through new observational capabilities and selection
techniques, such as direct identifications based on interferometric
observations of the dust continuum emission (e.g., Younger et
al.\ \citeyear{younger07}; Smol{\v c}i\'c et
al.\ \citeyear{smolcic12b}; Simpson et
al.\ \citeyear{simpson14,simpson15}; Brisbin et
al.\ \citeyear{brisbin17}; Stach et al.\ \citeyear{stach18}; see also
earlier works by, e.g., Downes et al.\ \citeyear{downes99};
Dannerbauer et al.\ \citeyear{dannerbauer02}), redshift
identifications through targeted molecular line scans (e.g.,
Wei\ss\ et al.\ \citeyear{weiss09}; Riechers \citeyear{riechers11b}),
and target selection based on sub/millimeter colors or flux limits
(e.g., Cox et al.\ \citeyear{cox11}; Riechers et
al.\ \citeyear{riechers13b,riechers17}; Dowell et
al.\ \citeyear{dowell14}; Vieira et al.\ \citeyear{vieira10};
Wei\ss\ et al.\ \citeyear{weiss13}). Nevertheless, all of the current
studies only provide incomplete censuses of the $z$$>$5 DSFG
population due to biases in the selection, limited sensitivity in the
parent sub/millimeter surveys, and incomplete redshift confirmations
of existing samples.

Here we aim to follow a complementary approach to more traditional
studies that builds on the finding that all luminous DSFGs appear to
contain large molecular gas reservoirs that fuel their star formation,
and to be significantly metal-enriched, leading to bright CO line
emission. As such, these systems are preferentially picked up by
panoramic molecular line scan surveys, and they may even dominate
among detections at the highest redshifts, where most other galaxy
populations may exhibit only weak CO emission (e.g., Pavesi et
al.\ \citeyear{pavesi19}) due to a combination of lower characteristic
galaxy masses at earlier epochs, lower metallicity (which is thought
to lead to an increase in the $\alpha_{\rm CO}$ conversion factor,
i.e., a lower CO luminosity per unit molecular gas mass; see Bolatto
et al.\ \citeyear{bolatto13} for a review), and possibly lower
CO line excitation (e.g., Daddi et al.\ \citeyear{daddi15}). Support
for this idea was provided by the detection of the
``optically-dark''\footnote{See, e.g., Calabro et
  al.\ (\citeyear{calabro19}) and references therein for a more
  detailed discussion of the nature of such sources at lower
  redshift.} $z$=5.183 DSFG HDF\,850.1 in a molecular line scan in the
{\em Hubble} Deep Field North (Walter et al.\ \citeyear{walter12b}),
but the survey area of $\sim$0.5\,arcmin$^2$ was too small to make
more quantitative statements regarding the broader properties and
space density of such sources.\footnote{Also, while its redshift was
  not known at the time, the telescope pointing was chosen to include
  HDF\,850.1. As such, this measurement did not constitute an unbiased
  discovery of an ``optically-dark'' source.}

To build upon this encouraging finding, and to provide a better
understanding of the true space densities of the most distant DSFGs,
we here study the properties of dusty starbursts at $z$$>$5 found in
sensitive molecular line scans, based on the $\sim$60\,arcmin$^2$ VLA
COLDz survey data (Pavesi et al.\ \citeyear{pavesi18b}; Riechers et
al.\ \citeyear{riechers19}, hereafter P18, R19).\footnote{See {\tt
    http://coldz.astro.cornell.edu} for additional information.} We
report the detection of \bco\ emission from three systems initially
detected in 850\,$\mu$m and 1.1\,mm continuum surveys, HDF\,850.1,
AzTEC-3, and GN10 (Hughes et al.\ \citeyear{hughes98}; Pope et
al.\ \citeyear{pope05}; Scott et al.\ \citeyear{scott08}), two of
which had previous correct redshift identifications through CO
measurements (AzTEC-3 and HDF\,850.1; Riechers et
al.\ \citeyear{riechers10a}; Capak et al.\ \citeyear{capak11}; Walter
et al.\ \citeyear{walter12b}).  We also report higher-resolution
observations of AzTEC-3 and HDF\,850.1, and detailed follow-up
observations of the third system, the ``optically-dark'' galaxy GN10,
as well as CO line excitation modeling for the sample. Our analysis is
used to constrain the evolution of dust temperature with redshift, and
the space density of $z$$>$5 DSFGs. In Section 2, we describe all
observations, the results of which are given in Section 3. Section 4
provides a detailed analysis of our findings for the COLDz $z$$>$5
DSFG sample, which are discussed in the context of all currently known
$z$$>$5 DSFGs in Section 5. A summary and conclusions are provided in
Section 6. We provide additional line parameters and an alternative
spectral energy distribution fit for GN10 and additional observations
of two $z$$>$4 dusty starbursts, GN20.2a and b, in parts A, B, and C
of the Appendix, respectively. We use a concordance, flat $\Lambda$CDM
cosmology throughout, with $H_0$\eq69.6\,\kms\,Mpc$^{-1}$,
$\Omega_{\rm M}$\eq0.286, and $\Omega_{\Lambda}$\eq0.714 (Bennett et
al.\ \citeyear{bennett14}).

\section{Data} \label{sec:data}

\subsection{Very Large Array}

\subsubsection{COLDz survey data}

\bco\ line emission ($\nu_{\rm rest}$=230.5380\,GHz) toward
HDF\,850.1, AzTEC-3, and GN10 was detected in molecular line scans
with the VLA within the COLDz survey data (project IDs: 13A-398 and
14A-214; PI: Riechers). A detailed description of the data is given by
P18.  In brief, COLDz targeted two regions in the COSMOS and
GOODS-North survey fields at 35 and 34\,GHz (corresponding to
$\sim$8.7\,mm), covering areas of $\sim$9 and 51\,arcmin$^2$ in 7- and
57-point mosaics with the VLA, respectively. Observations were carried
out under good Ka band weather conditions for a total of 324\,hr
between 2013 January 26 and 2015 December 18 in the D and DnC array
configurations, as well as reconfigurations between C, DnC, and D
array. The correlator was set up with two intermediate frequency bands
(IFs) of 4\,GHz bandwidth (dual polarization) each in 3-bit mode,
centered at the frequencies indicated above. Gaps between individual
128\,MHz sub-bands were mitigated through frequency switching. The
radio quasars J1041+0610 and J1302+5748 were observed for complex gain
calibration and regular pointing corrections in the COSMOS and
GOODS-North fields, respectively. The absolute flux scale was derived
based on observations of 3C\,286.

Data reduction and imaging was performed with the \textsc{casa} 4.1
package,\footnote{\tt https://casa.nrao.edu} using the data pipeline
version 1.2.0. Imaging the data with natural baseline weighting yields
typical clean beam sizes of 2.5$''$, with variations between
individual pointings and across the large bandwidth.  Typical rms
noise levels are 60 and 100--200\,$\mu$Jy\,beam$^{-1}$ per 4\,MHz
($\sim$35\,\kms ) binned channel in the COSMOS and GOODS-North fields,
respectively. At the positions of GN10 and AzTEC-3 (for which maps
based on these data are shown in the following), the rms noise is 51
and 18\,$\mu$Jy\,beam$^{-1}$ per 76 and 60\,MHz ($\sim$623 and
491\,\kms ) binned channel at beam sizes of 1.95$''$$\times$1.67$''$
and 2.46$''$$\times$2.26$''$, respectively.

\subsubsection{GN10 \aco\ follow-up}

We observed \aco\ line emission toward GN10 using the VLA (project
ID:\ 16A-015; PI:\ Riechers). Observations were carried out under good
Ku band weather conditions for a total of 11\,hr during 5 tracks in C
array between 2016 February 02 and March 06. Two IFs with 1\,GHz
bandwidth (dual polarization) each in 8-bit mode were centered at
13.977 and 17.837\,GHz (corresponding to 2.1 and 1.7\,cm,
respectively) to cover the redshifted HCN, HCO$^+$, and
HNC($J$=1$\to$0) and \aco\ lines in GN10 ($\nu_{\rm rest}$=88.6318,
89.1885, 90.6636, and 115.2712\,GHz). Observations were carried out in
short cycles, spending between $\sim$330 and $\sim$470\,s on source,
bracketed by scans spending $\sim$75\,s on the gain calibrator
J1302+5748. Pointing was performed on the gain calibrator
approximately once per hour. The absolute flux scale was derived based
on observations of 3C\,286.

Data reduction and imaging was performed with the \textsc{casa} 5.4.2
package. Imaging the data with natural baseline weighting yields a
clean beam size of 1.75$''$$\times$1.28$''$ at an rms noise level of
21\,$\mu$Jy\,beam$^{-1}$ over 40\,MHz (656\,\kms ) at the
\aco\ line frequency. The rms noise near the HCN($J$=1$\to$0)
frequency is $\sim$27\,$\mu$Jy\,beam$^{-1}$ over 4\,MHz (85\,\kms ).
Imaging the data over the entire line-free bandwidth of 2.012\,GHz
yields an rms noise level of 1.49\,$\mu$Jy\,beam$^{-1}$ at a beam size
of 2.10$''$$\times$1.54$''$.

\subsubsection{GN10 1.3\,cm and 6.6\,mm continuum follow-up}

We observed continuum emission at 22.8649\,GHz (K band) and
45.6851\,GHz (Q band) toward GN10 (corresponding to 1.3\,cm and
6.6\,mm, respectively), using the VLA (project ID:\ AR693;
PI:\ Riechers).\footnote{These observations were tuned to the
  \aco\ and \bco\ emission lines at the previous, incorrect redshift
  estimate.} Observations were carried out under good K and Q band
observing conditions for a total of 44\,hr between 2009 July 19 and
2010 January 05. K-band observations were conducted for 4\,tracks in C
array, totaling 28\,hr, and Q-band observations were conducted for
2\,tracks in D array, totaling 16\,hr. All observations used the
previous generation correlator, covering two IFs of 50\,MHz bandwidth
(dual polarization) each at the tuning frequency and 300\,MHz (K band)
or 50\,MHz (Q band) above, respectively, in quasi-continuum
mode. Observations in C (D) array were carried out in short cycles,
spending 150\,s (200\,s) on source, bracketed by scans spending 60\,s
on the gain calibrator J13028+57486. Pointing was performed on the
gain calibrator approximately once per hour. The absolute flux scale
was derived based on observations of 3C\,286.

Data reduction and imaging was performed with the \textsc{aips}
package. Imaging the data with natural baseline weighting yields clean
beam sizes of 1.15$''$$\times$1.01$''$ and 1.82$''$$\times$1.68$''$ at
rms noise levels of 44 and 58\,$\mu$Jy\,beam$^{-1}$ over 100\,MHz in K
and Q band, respectively.

\subsubsection{HDF\,850.1 \bco\ high-resolution follow-up}

We observed \bco\ line emission toward HDF\,850.1 at higher spatial
resolution using the VLA (project ID:\ 16A-014;
PI:\ Riechers). Observations were carried out under good Ka band
weather conditions for a total of 8.8\,hr during 4 tracks in C array
between 2016 February 03 and 20. Two IFs with 4\,GHz bandwidth (dual
polarization) each in 3-bit mode were centered at 34\,GHz
(corresponding to 8.8\,mm) to cover the same frequency range as the D
array observations of the main survey. Gaps between sub-bands were
mitigated through frequency switching, using the same two setups with
a relative shift of 16\,MHz as in D array. Observations were carried
out in short cycles, spending $\sim$300\,s on source, bracketed by
scans spending $\sim$100\,s on the gain calibrator
J1302+5748. Pointing was performed on the gain calibrator
approximately once per hour. The absolute flux scale was derived based
on observations of 3C\,286.

Data reduction and imaging was performed with the \textsc{casa} 5.4.2
package. Imaging the data with natural baseline weighting yields a
clean beam size of 0.71$''$$\times$0.68$''$ at an rms noise level of
32.4\,$\mu$Jy\,beam$^{-1}$ over 66\,MHz (530\,\kms ) at the \bco\ line
frequency.

\subsection{NOEMA}

\subsubsection{GN10 \fco\ follow-up}

We observed \fco\ line emission ($\nu_{\rm rest}$=691.4731\,GHz)
toward GN10 using NOEMA (project ID:\ X--5;
PI:\ Riechers). Observations were carried out under good 3\,mm weather
conditions for 4 tracks in the A configuration between 2014 February
18 and 24, using 6 antennas (baseline range: 67--760\,m). This yielded
a total time of 13.8\,hr (16500\,visibilities) on source. Receivers
were tuned to 109.7037\,GHz (corresponding to 2.7\,mm). The correlator
was set up with a bandwidth of 3.6\,GHz (dual polarization).

Data reduction and imaging was performed with the \textsc{gildas}
package.\footnote{\tt https://www.iram.fr/IRAMFR/GILDAS/} Imaging the
data with natural or uniform baseline weighting yields clean beam
sizes of 0.82$''$$\times$0.71$''$ or 0.63$''$$\times$0.59$''$ at rms
noise levels of 73 or 90\,$\mu$Jy\,beam$^{-1}$ over 700\,MHz
(1912\,\kms ), respectively. Imaging the data with natural weighting
over the entire line-free bandwidth of 2.9\,GHz yields an rms noise
level of 31.6\,$\mu$Jy\,beam$^{-1}$.

\subsubsection{GN10 \cii\ follow-up}

We observed [C{\sc ii}] 158\,$\mu$m line emission ($\nu_{\rm
  rest}$=1900.5369\,GHz) toward GN10 using NOEMA (project ID:\ W14FH;
PI:\ Riechers). Observations were carried out under good 0.9\,mm
weather conditions for 1 track in the C configuration on 2015 April
15, using 6 antennas (baseline range: 21--172\,m). This yielded a
total time of 1.9\,hr (2249\,visibilities) on source. Receivers were
tuned to 301.524\,GHz (corresponding to 1.0\,mm). The correlator was
set up with a bandwidth of 3.6\,GHz (dual polarization).

Data reduction and imaging was performed with the \textsc{gildas}
package. Imaging the data with natural or uniform baseline weighting
yields clean beam sizes of 1.01$''$$\times$0.84$''$ or
0.81$''$$\times$0.76$''$ at rms noise levels of 0.62 or
0.71\,mJy\,beam$^{-1}$ over 800\,MHz (795\,\kms ),
respectively. Imaging the data with natural or uniform weighting over
the entire line-free bandwidth of 2.31\,GHz yields rms noise levels
of 324 or 365\,$\mu$Jy\,beam$^{-1}$.

\subsubsection{GN10 1.2 and 2\,mm continuum follow-up}

We observed continuum emission at 137.057\,GHz (2.2\,mm)\footnote{These
  observations were tuned to the \fco\ emission line at the previous,
  incorrect redshift estimate.} and 250.5\,GHz (1.2\,mm) toward GN10
using NOEMA (project IDs:\ T047 and T0B7; PI:\ Riechers). Observations
were carried out under good weather conditions for 3 tracks between
2009 June 4 and September 21 in the D configuration with 5\,antennas
(baseline range:\ 19--94\,m) at 2\,mm and for 2 tracks on 2011 January
23 and 24 in the A configuration with 6 antennas (baseline
range:\ 51--665\,m) at 1.2\,mm. This yielded a total of 7.1 and
3.7\,hr (17040 and 8940 visibilities) 6 antenna-equivalent on source
time at 2 and 1.2\,mm, respectively. Observations at 2\,mm were
carried out with the previous generation correlator with a bandwidth
of 1\,GHz (dual polarization). Observations at 1.2\,mm were carried
out with a bandwidth of 3.6\,GHz (dual polarization).

Data reduction and imaging was performed with the \textsc{gildas}
package. Imaging the 2\,mm data with natural baseline weighting yields
a clean beam size of 3.7$''$$\times$3.2$''$ at an rms noise level of
95\,$\mu$Jy\,beam$^{-1}$ over 1\,GHz bandwidth. Imaging the 1.2\,mm
data with natural or uniform baseline weighting yields clean beam
sizes of 0.45$''$$\times$0.38$''$ or 0.38$''$$\times$0.33$''$ at rms
noise levels of 0.35 or 0.43\,mJy\,beam$^{-1}$ over 3.6\,GHz,
respectively.

\begin{figure*}[th]
\epsscale{1.15}
\plotone{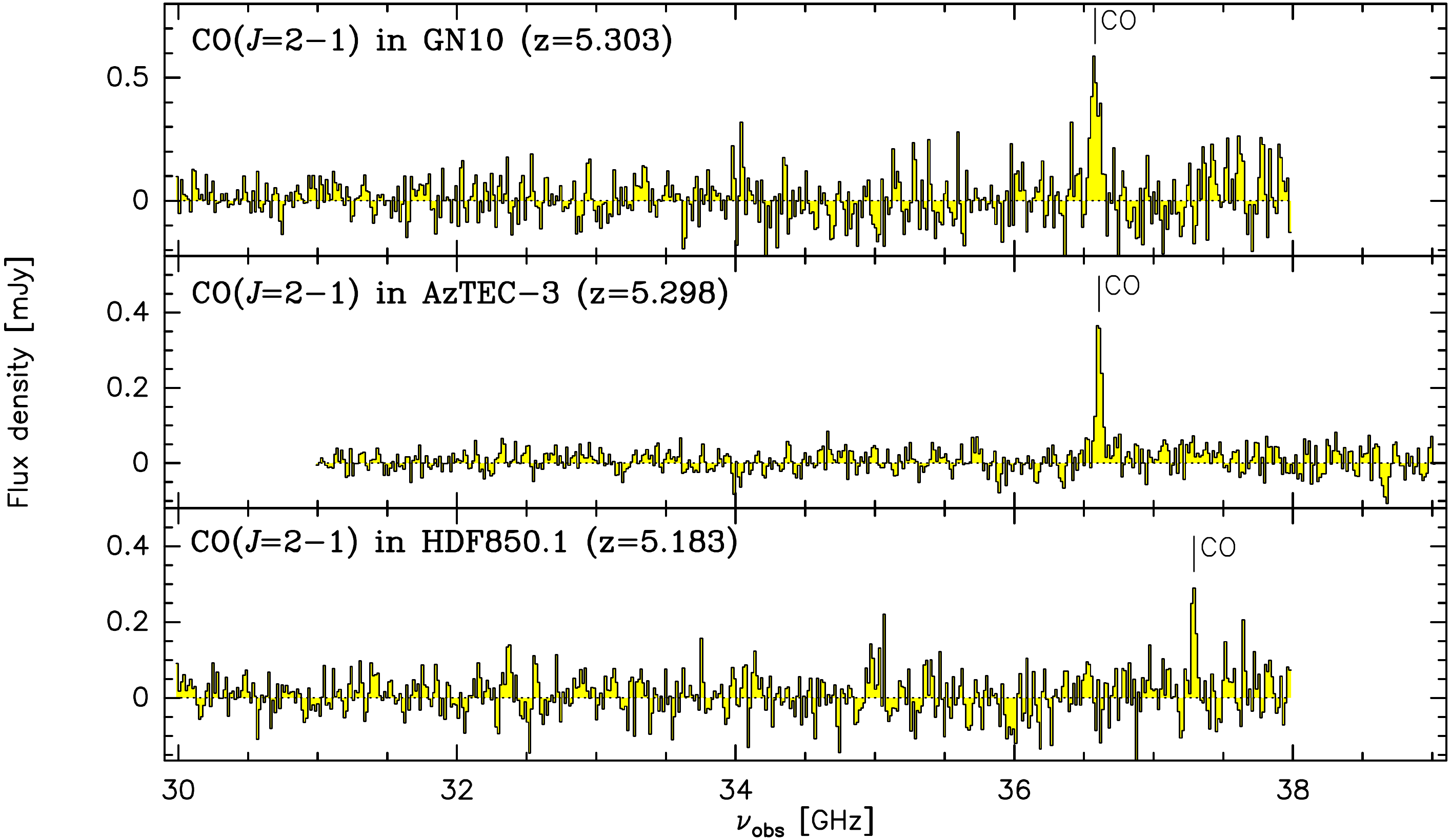}
\vspace{-2mm}

\caption{COLDz molecular line scan \bco\ spectra (histograms) of
  $z$$>$5 DSFGs, shown at 16\,MHz ($\sim$130\,\kms ) spectral
  resolution. Line emission in GN10, AzTEC-3 and HDF\,850.1 is
  detected at 8.6$\sigma$, 14.7$\sigma$, and 5.3$\sigma$ significance,
  respectively. \label{f1}}
%
\end{figure*}

\subsubsection{Archival:\ GN10 \eco }

Serendipitous \eco\ line emission ($\nu_{\rm rest}$=576.2679\,GHz) was
observed toward GN10 using NOEMA. These observations, taken from Daddi
et al.\ (\citeyear{daddi09b}), did not target GN10, which was offset
by 9.7$''$ or 19.7$''$ from the phase center for different tracks,
yielding primary beam attenuation factors of 1.08 or 1.30
respectively. Observations were carried out under good 3\,mm weather
conditions for 4 tracks in the B, C, and D configurations between 2008
May 04 and 2009 January 05, using 6 antennas (baseline range:
15--411\,m). This yielded a total time of 14.6\,hr
(25186\,visibilities) on source. Receivers were tuned to 91.375\,GHz
(corresponding to 3.3\,mm). The correlator was set up with a
bandwidth of 1\,GHz (dual polarization).

We adopted the data reduction performed by Daddi et
al.\ (\citeyear{daddi09b}), but re-imaged the data with the
\textsc{gildas} package. Imaging the data with natural baseline
weighting yields a clean beam size of 2.64$''$$\times$1.90$''$ at an
rms noise level of 66.6\,$\mu$Jy\,beam$^{-1}$ over 365.753\,MHz
(1200\,\kms ) at the phase center (96\,$\mu$Jy\,beam$^{-1}$ at the
position of GN10). Imaging the data over the line-free bandwidth
yields an rms noise level of 55\,$\mu$Jy\,beam$^{-1}$.

\subsubsection{AzTEC-3 \eco\ high-resolution follow-up}

We observed \eco\ line emission ($\nu_{\rm rest}$=576.2679\,GHz)
toward AzTEC-3 using NOEMA (project ID:\ U0D0;
PI:\ Riechers). Observations were carried out under good 3\,mm weather
conditions for 2 tracks in the A configuration between 2011 January 19
and February 04, using 6 antennas (baseline range:\ 100--760\,m). We
also used previous observations (project ID:\ T--F; PI:\ Riechers)
carried out for 1 track in the C configuration on 2010 April 1, using
6 antennas (baseline range:\ 15--176\,m; see Riechers et
al.\ \citeyear{riechers10a} for additional details). This yielded a
total time of 8.8\,hr (10560\,visibilities; 5340 in A configuration)
on source. Receivers were tuned to 91.558\,GHz (corresponding to
3.3\,mm). The correlator was set up with a bandwidth of 3.6\,GHz (dual
polarization).

Data reduction and imaging was performed with the \textsc{gildas}
package. Imaging the combined data with natural baseline weighting
yields a clean beam size of 2.21$''$$\times$1.43$''$ at an rms noise
level of 64\,$\mu$Jy\,beam$^{-1}$ over 280\,MHz (917\,\kms ). Imaging
the A configuration data only with uniform baseline weighting yields a
clean beam size of 1.39$''$$\times$0.85$''$ at an rms noise level of
96\,$\mu$Jy\,beam$^{-1}$ over 260\,MHz (852\,\kms ). Imaging the A
(A+C) configuration data with natural weighting over the entire
line-free bandwidth of 3.34\,GHz yields an rms noise level of 24.8
(18.8)\,$\mu$Jy\,beam$^{-1}$.


\begin{figure}[tbh]
\begin{deluxetable*}{ l c c c c c c c }
\tabletypesize{\scriptsize}
\tablecaption{Line fluxes and line luminosities for COLDz $z$$>$5 DSFGs. \label{t2x}}
\tablehead{
  Line & GN10 & & AzTEC-3 & & HDF\,850.1 & & References \\
       & {\em COLDz.GN.0} & & {\em COLDz.COS.0$^\star$} & & {\em COLDz.GN.31} & & \\
  \qquad\qquad\qquad (J2000.0)\tablenotemark{a} & \,(12:36:33.45, & +62:14:08.85)\qquad\,\, & \,(10:00:20.70, & +02:35:20.50)\qquad\,\, & \,(12:36:52.07, & +62:12:26.49)\qquad\,\, & \\
 & $I_{\rm line}$ & $L'_{\rm line}$ &  $I_{\rm line}$ & $L'_{\rm line}$ &  $I_{\rm line}$ & $L'_{\rm line}$ & \\
 & [Jy\,\kms ] & [10$^{10}$\,K\,\kms\,pc$^2$] & [Jy\,\kms ] & [10$^{10}$\,K\,\kms\,pc$^2$] & [Jy\,\kms ] & [10$^{10}$\,K\,\kms\,pc$^2$] & }
\startdata
${}$\aco\ & 0.054$\pm$0.017\tablenotemark{b} & 5.44$\pm$1.68 & & & $<$0.09 & $<$8.9 & 1, 2 \\
\bco\     & 0.295$\pm$0.035 & 7.47$\pm$ 0.90 & 0.199$\pm$0.018 & 5.02$\pm$0.44 & 0.148$\pm$0.057 & 3.62$\pm$1.39 & 1, 3 \\
          & & & {\em 0.23$\pm$0.04} & {\em 5.84$\pm$0.37} & {\em 0.17$\pm$0.04}   & {\em 4.15$\pm$0.98} & 4, 5 \\
\eco\     & 0.86$\pm$0.20 & 3.46$\pm$0.81 & 0.97$\pm$0.09 & 3.92$\pm$0.38 & 0.50$\pm$0.10 & 1.96$\pm$0.39 & 1, 6, 5 \\
          & & & {\em 0.92$\pm$0.09} & {\em 3.70$\pm$0.37} & & & 4 \\
\fco\     & 0.52$\pm$0.11 & 1.46$\pm$0.31 & 1.36$\pm$0.19 & 3.82$\pm$0.45 & 0.39$\pm$0.10 & 1.06$\pm$0.27 & 1, 4, 5 \\
\gco\     & & & & & 0.35$\pm$0.05 & 0.70$\pm$0.10 & 7 \\
\pco\     & & & $<$0.22       & $<$0.09 & & & 8 \\
\oh\      & & & 1.44$\pm$0.13 & 0.57$\pm$0.05 & & & 8 \\
\bci\     & & & & & 0.14$\pm$0.05 & 0.28$\pm$0.10 & 7 \\
\cii\     & 17.6$\pm$1.9 & 6.55$\pm$0.71 & 8.21$\pm$0.29 & 3.05$\pm$0.11 & 9.9$\pm$1.0 & 3.56$\pm$0.36 & 1, 8, 9 \\
          & {\em 16.2$\pm$1.4}\tablenotemark{c} & {\em 6.01$\pm$0.53} & {\em 7.8$\pm$0.4} & {\em 2.90$\pm$0.15} & {\em 14.6$\pm$0.3} & {\em 5.25$\pm$0.11} & 1, 10, 5 \\
\nii\     & & & 0.46$\pm$0.16 & 0.31$\pm$0.11 & & & 10 \\
\enddata
\tablenotetext{\rm a}{\bco\ centroid positions adopted from P18.}
\tablenotetext{\rm b}{A Gaussian fit to the line profile formally suggests 0.054$\pm$0.010\,Jy\,\kms, but we consider these uncertainties to be somewhat optimistic due to the increasing noise level towards the blue edge of the bandpass. We thus adopt more conservative error bars based on the signal-to-noise ratio of the detection in the moment-0 map.}
\tablenotetext{\rm c}{Main component only.}
{${}^\star$ Alternative ID:\ AS2COS0059.1 (Simpson et al.\ \citeyear{simpson20}).}
\tablerefs{${}$[1] this work; [2] Wagg et al.\ (\citeyear{wagg07}); [3] Pavesi et al.\ (\citeyear{pavesi18b}); [4] Riechers et al.\ (\citeyear{riechers10a}); [5] Walter et al.\ (\citeyear{walter12b}); [6] Daddi et al.\ (\citeyear{daddi09b}); [7] Decarli et al.\ (\citeyear{decarli14}); [8] Riechers et al.\ (\citeyear{riechers14b}); [9] Neri et al.\ (\citeyear{neri14}); [10] Pavesi et al.\ (\citeyear{pavesi16}).}
\end{deluxetable*}
\end{figure}


\subsubsection{AzTEC-3 1.2\,mm continuum follow-up}

We observed continuum emission at 250.0\,GHz (1.2\,mm) toward AzTEC-3
using NOEMA (project ID:\ U0D0; PI:\ Riechers). Observations were
carried out under good weather conditions for 3 tracks between 2011
January 25 and February 03 in the A configuration with 6 antennas
(baseline range:\ 100--760\,m) at 1.2\,mm. This yielded a total of
9.3\,hr (11160 visibilities) on source. Observations were carried out
with a bandwidth of 3.6\,GHz (dual polarization).

Data reduction and imaging was performed with the \textsc{gildas}
package. Imaging the data with natural or uniform baseline weighting
yields clean beam sizes of 0.62$''$$\times$0.25$''$ or
0.53$''$$\times$0.24$''$ at rms noise levels of 162 or
187\,$\mu$Jy\,beam$^{-1}$ over 3.6\,GHz, respectively.

\begin{figure}
\epsscale{1.15}
\plotone{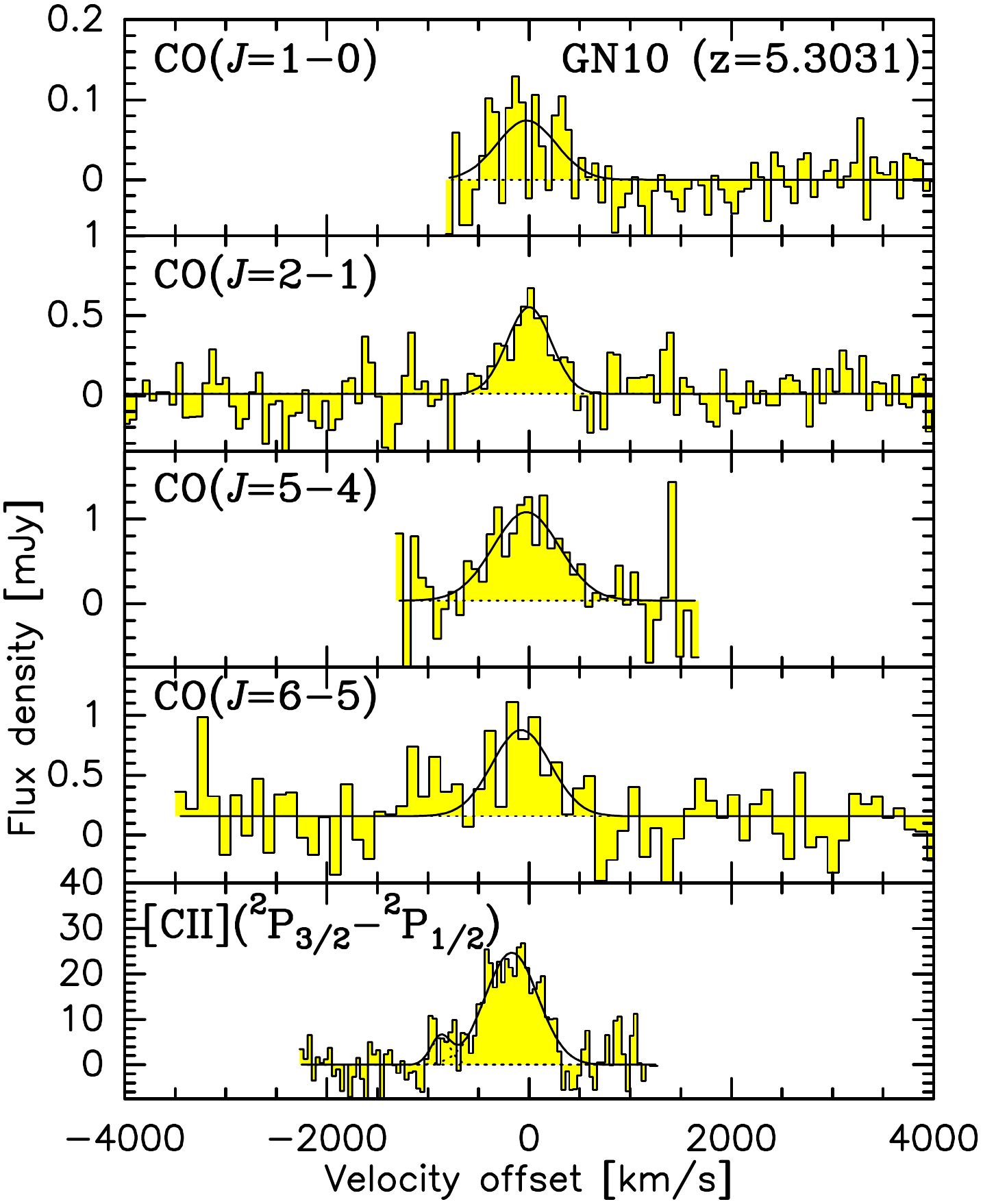}
\vspace{-2mm}

\caption{VLA and NOEMA line spectra of GN10 ($z$=5.3031; histograms)
  and Gaussian fits to the line profiles (black curves). Spectra are
  shown at resolutions of 66, 66, 75, 109, and 40\,\kms\ (4, 8, 23,
  40, and 40\,MHz; {\em top} to {\em bottom}), respectively. Continuum
  emission (9.55$\pm$0.73\,mJy) has been subtracted from the [C{\sc
      ii}] spectrum only. \label{f2}}
%
\end{figure}


\begin{figure}
\begin{deluxetable}{ l c l c }
\tabletypesize{\scriptsize}
\tablecaption{GN10 continuum photometry. \label{t1}}
\tablehead{
Wavelength & Flux density\tablenotemark{a} & Telescope & Reference \\
 ($\mu$m) & (mJy) & & }
\startdata
0.435 & $<$12$\times$10$^{-6}$           & {\em HST}/ACS & 1 \\
0.606 & $<$9$\times$10$^{-6}$            & {\em HST}/ACS & 1 \\ 
0.775 & $<$18$\times$10$^{-6}$           & {\em HST}/ACS & 1 \\ 
0.850 & $<$27$\times$10$^{-6}$           & {\em HST}/ACS & 1 \\ 
1.25  & $<$42$\times$10$^{-6}$           & Subaru/MOIRCS & 1 \\
1.60  & $<$15$\times$10$^{-6}$           & {\em HST}/NICMOS & 1 \\
2.15  & $<$42$\times$10$^{-6}$           & Subaru/MOIRCS & 1 \\
3.6   & (1.29$\pm$0.13)$\times$10$^{-3}$ & {\em Spitzer}/IRAC & 2\\
4.5   & (2.07$\pm$0.21)$\times$10$^{-3}$ & {\em Spitzer}/IRAC & 2 \\
5.8   & (2.96$\pm$0.37)$\times$10$^{-3}$ & {\em Spitzer}/IRAC & 2 \\
8.0   & (5.30$\pm$0.53)$\times$10$^{-3}$ & {\em Spitzer}/IRAC & 2 \\
16    & (17.5$\pm$6.3)$\times$10$^{-3}$  & {\em Spitzer}/IRS & 2 \\
24    & (33.4$\pm$7.9)$\times$10$^{-3}$  & {\em Spitzer}/MIPS & 2 \\
70    & $<$2.0                          & {\em Spitzer}/MIPS & 3 \\
110                    & $<$1.52        & {\em Herschel}/PACS & 2 \\
160                    & $<$5.3         & {\em Herschel}/PACS & 2 \\
160   & $<$30                           & {\em Spitzer}/MIPS & 3 \\
250\tablenotemark{b}   & 9.8$\pm$4.1    & {\em Herschel}/SPIRE & 2 \\
350\tablenotemark{b}   & 8.9$\pm$4.2    & {\em Herschel}/SPIRE & 2 \\
500\tablenotemark{b}   & 12.4$\pm$2.8   & {\em Herschel}/SPIRE & 2 \\
850   & 12.9$\pm$2.1                    & JCMT/SCUBA & 3 \\
      & 11.3$\pm$1.6                    & JCMT/SCUBA & 3 \\
870   & 12.0$\pm$1.4                    & SMA & 3 \\
995   & 9.55$\pm$0.73                   & NOEMA & 4 \\
1200  & 5.25$\pm$0.60                   & NOEMA & 4 \\
1250  & 5.0$\pm$1.0                     & NOEMA & 3 \\
2187  & 0.28$\pm$0.17                   & NOEMA & 4 \\
2733  & 0.148$\pm$0.032                 & NOEMA & 4 \\
3280  & $<$0.27                         & NOEMA & 5 \\
6560  & $<$0.174                        & VLA  & 4 \\
8820  & (8.1$\pm$4.2)$\times$10$^{-3}$  & VLA & 4 \\
13100 & $<$0.132                        & VLA & 4 \\
18850 & (4.3$\pm$1.5)$\times$10$^{-3}$   & VLA & 4 \\
210000 & (35.8$\pm$ 4.1)$\times$10$^{-3}$ & VLA & 2, 3 \\
\enddata
\tablenotetext{\rm a}{Limits are 3$\sigma$.}
\tablenotetext{\rm b}{De-blended fluxes. Uncertainties do not account for confusion noise, which
  formally is 5.9, 6.3, and 6.8\,mJy (1$\sigma$) at 250, 350, and 500\,$\mu$m,
  respectively (Nguyen et al.\ \citeyear{nguyen10}), but also is reduced though the de-blending process.}
\tablerefs{${}$[1] Wang et al.\ (\citeyear{wang09}) and references therein; [2] Liu et al.\ (\citeyear{liu18}); [3] Dannerbauer et al.\ (\citeyear{dannerbauer08}) and references therein; [4] this work; [5] Daddi et al.\ (\citeyear{daddi09b}).}
\end{deluxetable}
\end{figure}


\begin{figure*}
\epsscale{1.15}
\plotone{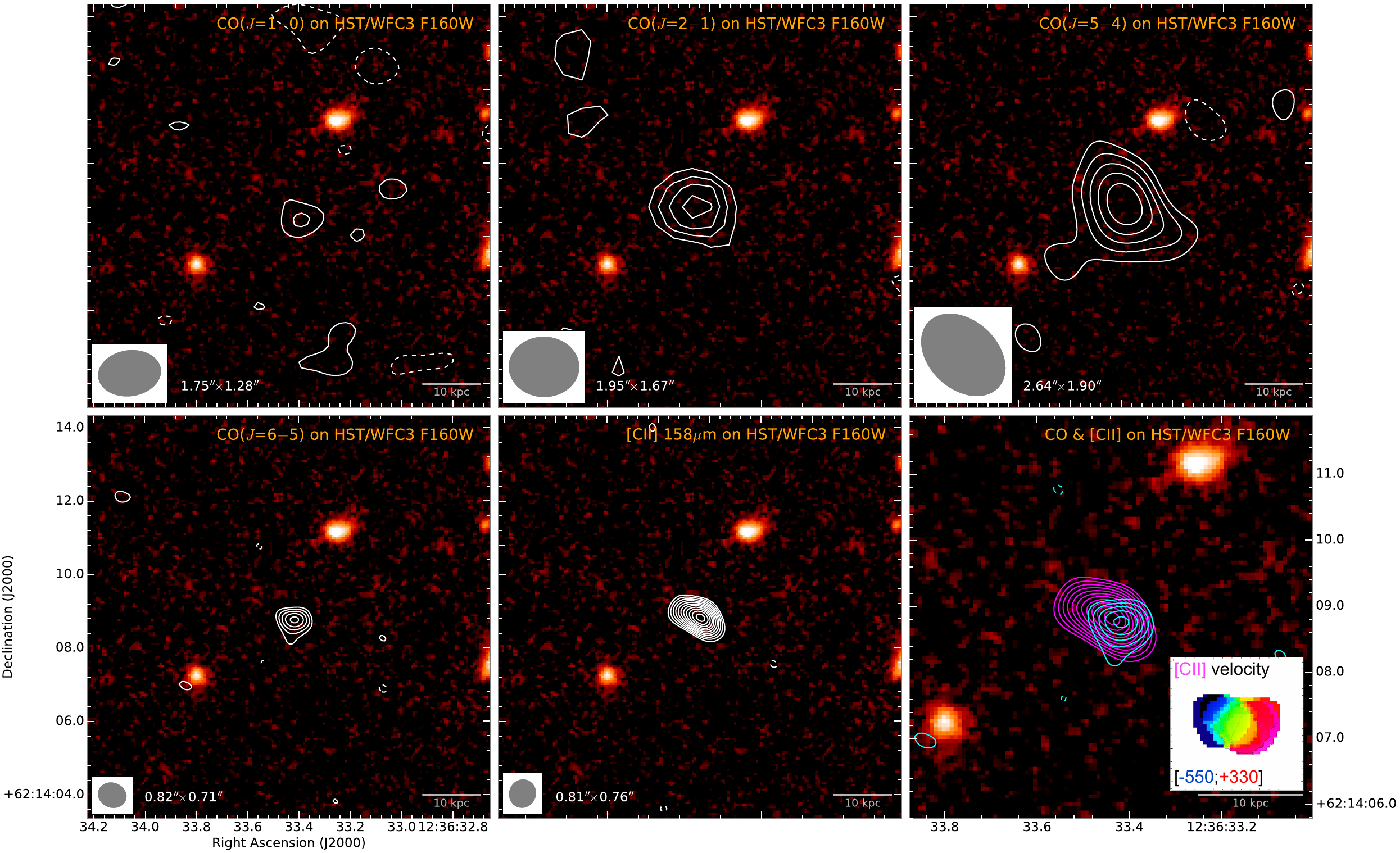}
\vspace{-2mm}

\caption{Velocity-integrated line contour maps overlaid on a {\em
    HST}/WFC3 F160W continuum image from the CANDELS survey toward
  GN10. Contour maps are averaged over 656, 623, 1199, 1913, and
  795\,\kms, respectively. Contours start at $\pm$2, 3, 2, 3, and
  4$\sigma$, and are shown in steps of 1$\sigma$=21, 51, 96, 73,
  and 710\,$\mu$Jy\,beam$^{-1}$ for the first five panels,
  respectively. The synthesized beam size is indicated in the bottom
  left corner of each panel where applicable. Last panel shows
  \fco\ (cyan) and [C{\sc ii}] (magenta) contours and is zoomed-in by
  a factor of 1.8 compared to all other panels. The inset in the last
  panel shows a velocity map over the central 880\,\kms\ of the [C{\sc
      ii}] emission (created from 80\,\kms\ velocity channels and
  adopting a detection threshold of 9\,mJy\,beam$^{-1}$). Velocity
  contours are shown in steps of 50\,\kms. \label{f3}}
%
\end{figure*}

\begin{figure*}
\epsscale{1.15}
\plotone{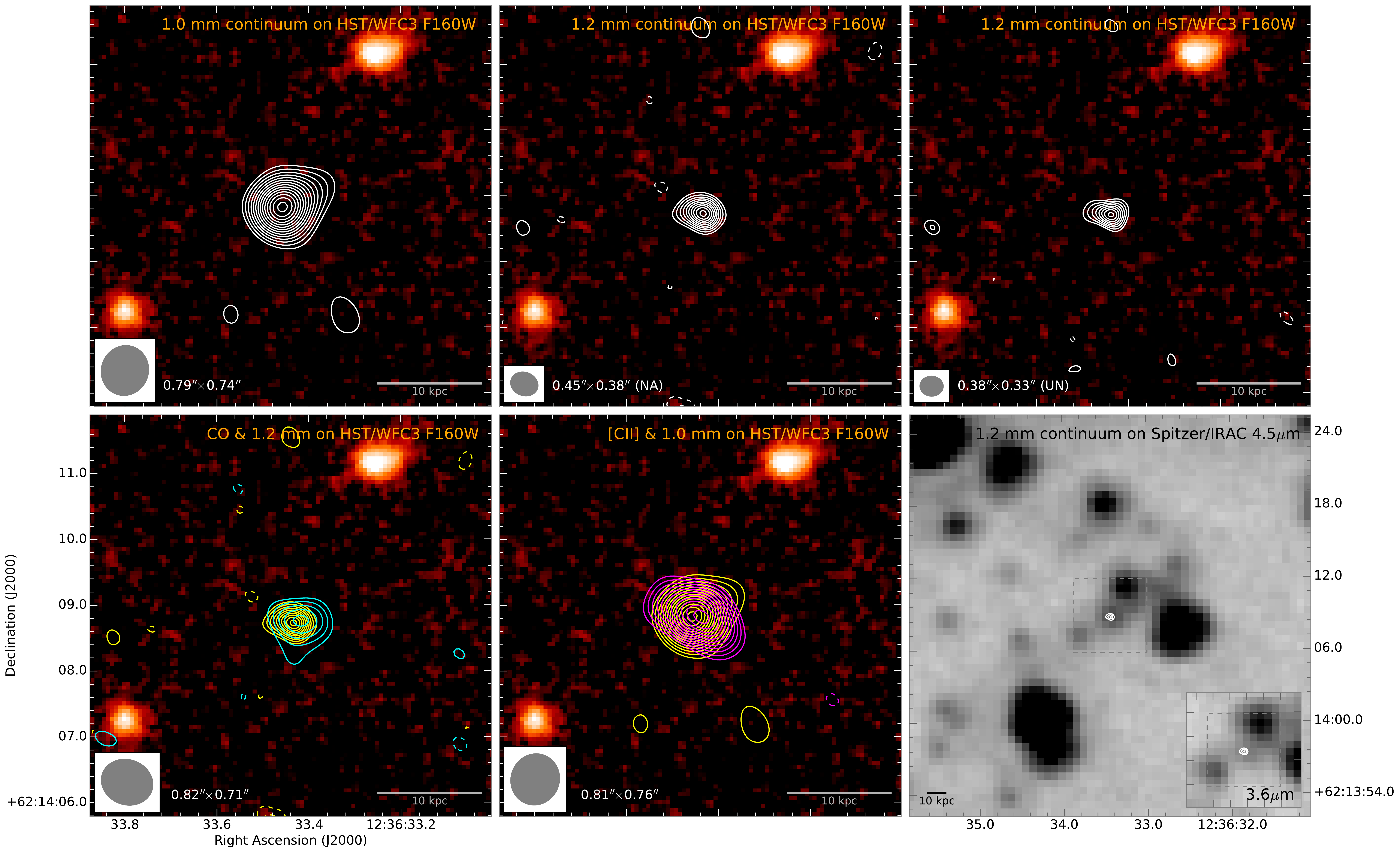}
\vspace{-2mm}

\caption{Rest-frame far-infrared continuum contour maps at
  observed-frame 1.0 ({\em top left}) and 1.2\,mm ({\em top middle}
  and {\em right}; imaged using natural and uniform baseline
  weighting, respectively), overlaid on a {\em HST}/WFC3 F160W
  continuum image toward GN10. Contours start at $\pm$4, 3, and
  3$\sigma$, and are shown in steps of 1$\sigma$=365, 352, and
  421\,$\mu$Jy\,beam$^{-1}$, respectively. The synthesized beam size
  is indicated in the bottom left corner of each panel where
  applicable. The 4$\sigma$ peaks south of GN10 in the {\em top left}
  panel are due to sidelobe residuals given imperfections in the
  calibration. {\em Bottom} panels show overlays of 1.2 and 1.0\,mm
  continuum (yellow) with \fco\ and [C{\sc ii}] emission ({\em bottom
    left} and {\em middle}; same contours as in Fig.~\ref{f3}), and
  1.2\,mm contours (natural weighting; shown in steps of
  $\pm$4$\sigma$) on {\em Spitzer}/IRAC 3.6 (inset) and 4.5\,$\mu$m
  images ({\em bottom right}). The dashed gray box in the last panel
  indicates the same area as shown in the other panels.
 \label{f4}}
%
\end{figure*}

\begin{figure*}
\epsscale{1.15}
\plotone{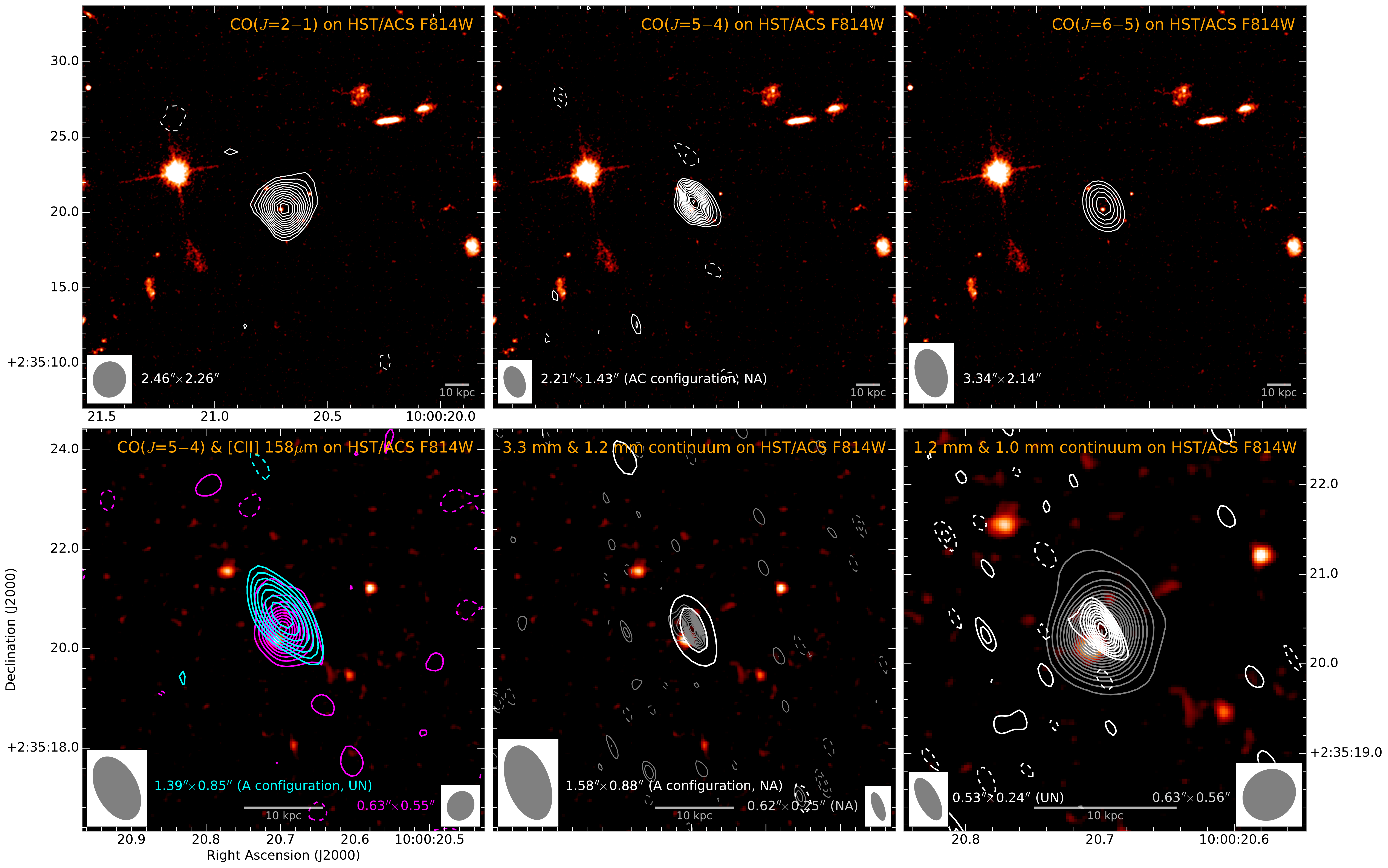}
\vspace{-2mm}

\caption{Velocity-integrated line and rest-frame far-infrared
  continuum contour maps overlaid on a {\em HST}/ACS F814W continuum
  image from the COSMOS survey toward AzTEC-3. \fco, [C{\sc ii}], and
  1.0\,mm continuum data are adopted from Riechers et
  al.\ (\citeyear{riechers10a,riechers14b}). Data are imaged with
  natural (NA) baseline weighting unless mentioned otherwise. Line
  contour maps in the {\em top row} include all available data and are
  averaged over 491, 917, 874\,\kms\ ({\em left} to {\em right}),
  respectively. Contours start at $\pm$3$\sigma$, and are shown in
  steps of 1$\sigma$=18, 64, and 224\,$\mu$Jy\,beam$^{-1}$,
  respectively. Line contour map in the {\em bottom left} panel
  includes only A configuration \eco\ data (cyan; imaged with uniform
  weighting). Contours are averaged over 852 (CO), and
  466\,\kms\ (magenta; [C{\sc ii}]), start at $\pm$3 and 4$\sigma$, and are
  shown in steps of 1 and 4$\sigma$, where 1$\sigma$=96 and
  200\,$\mu$Jy\,beam$^{-1}$, respectively. Continuum contour maps in
  the {\em bottom middle} panel include only A configuration 3.3\,mm
  continuum data (white). Contours start at $\pm$3$\sigma$, and are
  shown in steps of 1$\sigma$=24.8 (3.3\,mm) and
  162\,$\mu$Jy\,beam$^{-1}$ (1.2\,mm; gray), respectively. Contours in
  the {\em bottom right} panel start at $\pm$3 and 5$\sigma$, and are
  shown in steps of 1 (1.2\,mm; white, imaged with uniform weighting)
  and 5$\sigma$ (1.0\,mm; gray), where 1$\sigma$=187 and
  58\,$\mu$Jy\,beam$^{-1}$, respectively. The synthesized beam size is
  indicated in the bottom left (white or cyan contours) or right
  (magenta or gray) corner of each panel. Bottom panels are zoomed-in
  by a factor of 3.3, except last panel, which is zoomed-in by a
  factor of $\sim$6. \label{f12b}}
%
\end{figure*}

\begin{figure}
\epsscale{1.15}
\plotone{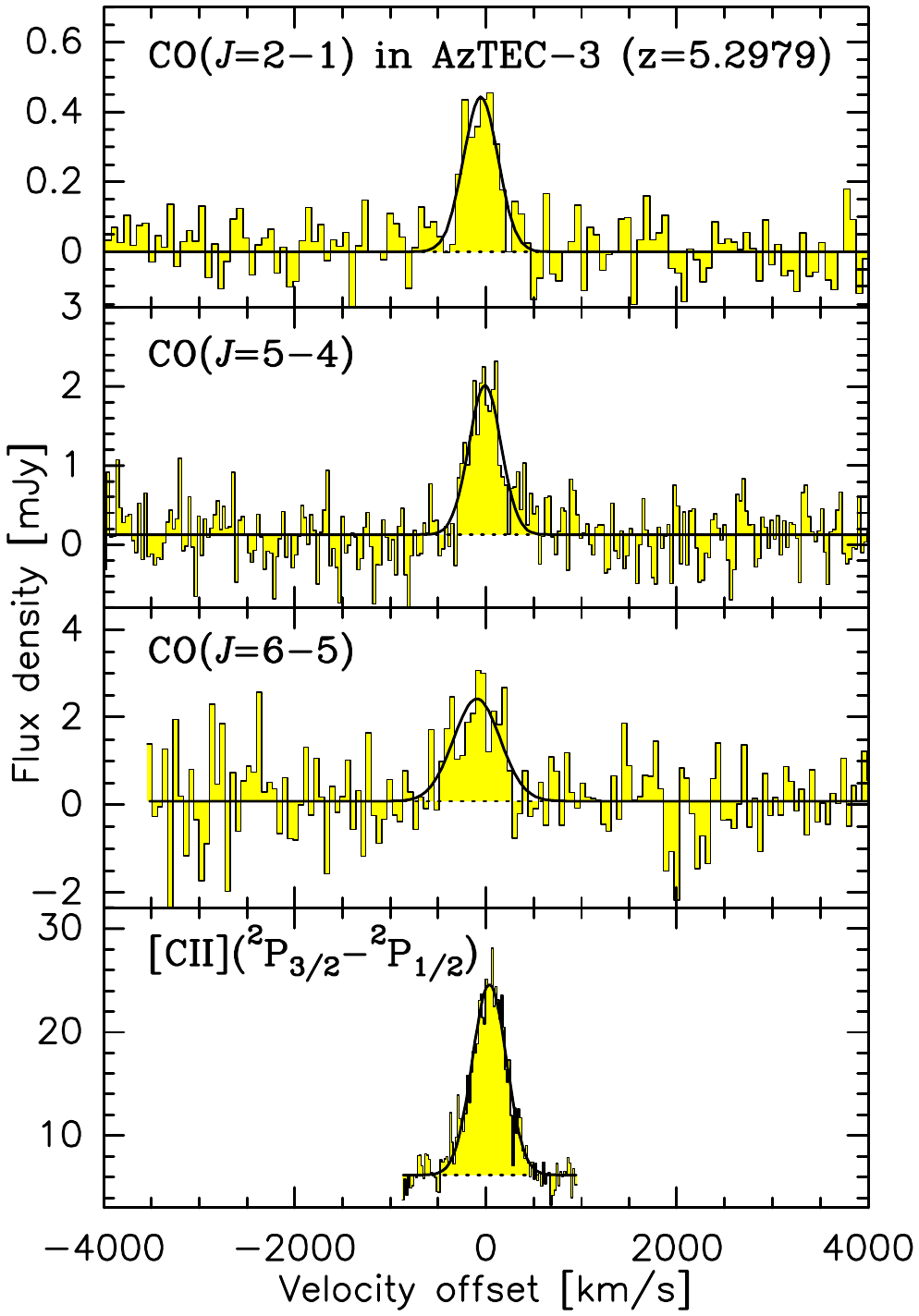}
\vspace{-2mm}

\caption{VLA, NOEMA, and ALMA line spectra of AzTEC-3 ($z$=5.2979;
  histograms) and Gaussian fits to the line profiles (black
  curves). \fco\ and [C{\sc ii}] data are adopted from Riechers et
  al.\ (\citeyear{riechers10a,riechers14b}). Spectra are shown at
  resolutions of 66, 33 55, and 20\,\kms\ (8, 10, 20, and 20\,MHz;
  {\em top} to {\em bottom}), respectively. \label{f12a}}
%
\end{figure}

\section{Results} \label{sec:results}

\subsection{COLDz Molecular Line Scan \bco\ Detections}

Our CO search in the COSMOS and GOODS-North fields carried out as part
of the COLDz molecular line scan survey (P18, R19) yielded four
matches\footnote{One match was found in the COSMOS field, and three
  matches were found in the $\sim$5.7 times larger GOODS-North field.}
with massive dusty star-forming galaxies initially selected in
single-dish bolometer surveys with the JCMT at 850\,$\mu$m or
1.1\,mm. One of the matches at 6.1$\sigma$ significance corresponds to
\aco\ emission associated with the $z$=2.488 DSFG GN19 (Pope et
al.\ \citeyear{pope05}; Riechers et al.\ \citeyear{riechers11c};
Ivison et al.\ \citeyear{ivison11}; see P18), and will not be
discussed further here.  Two of the matches correspond to
\bco\ emission in the $z$=5.183 and $z$=5.298 DSFGs HDF\,850.1 and
AzTEC-3 (Fig.~\ref{f1} and Table~\ref{t2x}), which are detected at
5.3$\sigma$ and 14.7$\sigma$ significance, respectively.  From
Gaussian fits to the line spectra and moment-0 maps, we find
\bco\ line FWHM of d$v_{\rm FWHM}$=(490$\pm$140) and
(424$\pm$44)\,\kms\ for HDF\,850.1 and AzTEC-3, yielding line fluxes
of $I_{\rm CO(2-1)}$=(0.148$\pm$0.057) and
(0.199$\pm$0.018)\,Jy\,\kms, respectively. These flux levels are
consistent with previous, lower-significance detections within the
relative uncertainties (Riechers et al.\ \citeyear{riechers10a};
Walter et al.\ \citeyear{walter12b}).

Unexpectedly at the time of observation, we also detect an emission
line at 8.6$\sigma$ significance toward the DSFG GN10 (Fig.~\ref{f1}),
previously thought to be at $z$=4.042 based on a single line detection
at 3\,mm and photometric redshift information (Daddi et
al.\ \citeyear{daddi09b}). We identify this line with \bco\ emission
at $z$=5.303, which implies that the line detected by Daddi et
al.\ (\citeyear{daddi09b}) corresponds to \eco\ emission, rather than
\dco\ emission. This identification was confirmed through the
successful detection of \aco, \fco, and [C{\sc ii}] emission at the same
redshift, as described in detail below (see Figs.~\ref{f2} and
\ref{f3}). This explains why our earlier attempts to detect \aco,
\bco, and \fco\ emission at $z$=4.042 (see Sect.~2) were unsuccessful.

\subsection{GN10 Follow-Up}

\subsubsection{Continuum Emission}

We detect strong continuum emission toward GN10 at 1.2 and 1.0\,mm,
and weak emission between 2.2\,mm and 1.9\,cm (see Fig.~\ref{f4}, and
Table~\ref{t1}). The flux keeps decreasing between 0.9 and 1.9\,cm,
and is $>$4--8 times lower than at 21\,cm. This suggests that the
emission detected up to 0.9\,cm (i.e., rest-frame 1.4\,mm) likely
still corresponds to thermal emission, but non-thermal emission may
start to significantly contribute at 1.9\,cm (i.e., rest-frame
3.0\,mm; see Fig.~\ref{f7}). The continuum emission is spatially
resolved along the major axis at 1.2\,mm by our observations with a
synthesized beam size of $\sim$0.35$''$. By fitting two-dimensional
Gaussian profiles to the emission in the visibility plane, we find a
size of (0.25$''$$\pm$0.07$''$)$\times$(0.10$''$$\pm$0.11$''$), which
corresponds to (1.6$\pm$0.4)$\times$(0.6$\pm$0.6)\,kpc$^2$.  A
circular Gaussian fit provides a full width at half power (FWHP)
diameter of 0.18$''$$\pm$0.05$''$, or (1.1$\pm$0.3)\,kpc. Due to the
agreement of the 1.2\,mm flux with a previous measurement at lower
spatial resolution at a close wavelength (Dannerbauer et
al.\ \citeyear{dannerbauer08}; see Table~\ref{t1}), and given the
baseline coverage down to $\sim$50\,m, it appears unlikely that the
1.2\,mm size measurement is biased towards low values due to missing
emission. Fits to the lower-resolution ($\sim$0.75$''$ beam size) data
at 1.0\,mm however suggest a size of
(0.58$''$$\pm$0.12$''$)$\times$(0.50$''$$\pm$0.10$''$), corresponding
to (3.6$\pm$0.7)$\times$(3.1$\pm$0.6)\,kpc$^2$, or a circular FWHP
diameter of 0.53$''$$\pm$0.08$''$ (3.3$\pm$0.5\,kpc$^2$). The
uncertainties for this measurement may be limited by interferometric
seeing due to phase noise (which is not factored into the fitting
errors), such that we treat the 1.0\,mm size measurement as an upper
limit only in the following. We however note that, in principle, the
dust emission at shorter wavelengths could appear more extended due to
an increasing dust optical depth as well, as discussed further in
Section 4. Also, the source shape could significantly deviate from a
Gaussian shape (such as a higher-index S\'ersic profile; see, e.g.,
discussion by Hodge et al.\ \citeyear{hodge16}), such that more
complex fitting procedures could yield different findings.

Despite its strong dust continuum emission at sub/millimeter
wavelengths, GN10 remains undetected up to observed-frame 2.2\,$\mu$m,
rendering it a ``K-band dropout'' (also see discussion by Wang et
al.\ \citeyear{wang09}; Daddi et al.\ \citeyear{daddi09b}, and
references therein). Even sensitive space-based imaging up to
1.6\,$\mu$m with the WFC3 camera on the {\em Hubble Space Telescope}
({\em HST}) from the CANDELS survey (Grogin et
al.\ \citeyear{grogin11}) yields no hint of emission due to dust
obscuration (Fig.~\ref{f4}), but mid-infrared continuum emission is
detected with {\em Spitzer}/IRAC longward of 3.6\,$\mu$m
(corresponding to $\sim$5700\,\AA\ in the rest frame; Dickinson et
al.\ \citeyear{dickinson04}; Giavalisco et
al.\ \citeyear{giavalisco04}) at the position of the millimeter-wave
dust continuum emission (Fig.~\ref{f4}). Thus, the stellar light in
the ``optically-dark'' galaxy GN10 is not entirely obscured by
dust.\footnote{Contributions to the rest-frame optical light by a
  dust-obscured active galactic nucleus in GN10 cannot be ruled out;
  see discussion in Sect.~4.}

\subsubsection{Line Emission}

We successfully detect \aco, \bco, \eco, \fco, and [C{\sc ii}]
emission toward GN10 at 3.3$\sigma$, 8.6$\sigma$, 6.9$\sigma$,
7.3$\sigma$, and 18.1$\sigma$ peak significance, respectively
(Figs.~\ref{f2} and \ref{f3} and Table~\ref{t2x}; see Appendix A for
further details). In combination, these lines provide an unambiguous
redshift identification. We extract the parameters of all emission
lines from Gaussian fitting to the line profiles. All CO lines are fit
with single Gaussian functions (see Appendix A for further
details). We find line peak fluxes of $S_{\rm line}$=(74$\pm$13),
(544$\pm$63), (1046$\pm$205), and (719$\pm$144)\,$\mu$Jy at FWHM line
widths of d$v_{\rm FWHM}$=(687$\pm$144), (512$\pm$72), (772$\pm$220),
and (681$\pm$173)\,\kms\ for the CO $J$=1$\to$0, 2$\to$1, 5$\to$4, and
6$\to$5 lines, respectively. The [C{\sc ii}] line is fit with two
Gaussian components, yielding $S_{\rm line}$=(24.7$\pm$1.6) and
(6.0$\pm$4.2)\,mJy and d$v_{\rm FWHM}$=(617$\pm$67) and
(227$\pm$243)\,\kms\ for the red- and blue-shifted components,
respectively. The parameters of the secondary component thus are only
poorly constrained, and we report [C{\sc ii}] fluxes including and
excluding this component in the following. It may correspond to a gas
outflow, or a close-by minor companion galaxy to the
DSFG.\footnote{The \fco\ spectrum shows excess positive signal at
  comparable velocities as the secondary [C{\sc ii}] component, but
  its significance is too low to permit further analysis.} Considering
only the main [C{\sc ii}] component, the widths of all lines are
consistent within the relative uncertainties.

Assuming the same widths as for the \bco\ line, we find 3$\sigma$
upper limits of $<$0.017\,Jy\,\kms\ for the HCN, HCO$^+$, and
HNC($J$=1$\to$0) lines.\footnote{The observations also cover the
  CCH($N$=1$\to$0) line, which is not detected down to a comparable
  limit (see Appendix).} This implies HCN, HCO$^+$, and HNC to CO line
luminosity ratio limits of order $<$40\%, which are only modestly
constraining given the expectation of few per cent to $\sim$20\%
ratios for distant starburst galaxies (see, e.g., Riechers et
al.\ \citeyear{riechers07b}; Oteo et al.\ \citeyear{oteo17b}, and
references therein).

The integrated line fluxes and line luminosities derived from these
measurements are summarized in Table~\ref{t2x}, together with those of
AzTEC-3 and HDF\,850.1, including values from the literature. The
\bco\ flux of GN10 reported here is somewhat lower than that found by
P18 using a different extraction method. Here we adopt our updated
value for consistency. Given the more complex velocity profile of the
[C{\sc ii}] line in GN10, we adopt the \bco-based redshift of
$z$=5.3031$\pm$0.0006 in the following. This measurement agrees within
1$\sigma$ with the \aco-, \eco-, and \fco-based measurements.\footnote{The
  redshift uncertainties for the \aco, \bco, \eco, \fco, and [C{\sc ii}] lines
  from fitting Gaussian functions to the line profiles are 60, 30, 71, 70,
  and 24\,\kms, respectively.} The systemic velocity centroid of the
[C{\sc ii}] line is slightly blueshifted, but emission is detected across
the same velocity range as in the CO $J$=1$\to$0 to 6$\to$5 lines. The
line peak offset thus is likely mainly due to internal variations in
the [C{\sc ii}]/CO ratio.

By fitting two-dimensional Gaussian profiles to the [C{\sc ii}]
emission, we find a size of
(1.04$''$$\pm$0.30$''$)$\times$(0.19$''$$\pm$0.10$''$).  This
corresponds to (6.5$\pm$1.9)$\times$(1.2$\pm$0.6)\,kpc$^2$.
Attempting to fit the \fco\ emission observed at comparable spatial
resolution yields a size of 0.42$''$$\times$$<$0.25$''$
(2.6$\times$$<$1.6\,kpc$^2$), but the fit does not converge well due
to the moderate signal-to-noise ratio of the detection. A circular fit
is consistent with a point source within the uncertainties. This
suggests that the [C{\sc ii}] emission is resolved at least along the
major axis, and that it appears to be more spatially extended than the
mid-$J$ CO and dust emission, which are consistent with having
comparable spatial extent within the current uncertainties. The [C{\sc
    ii}] emission shows a significant spatial velocity gradient across
the line emission in the main component alone (see Fig.~\ref{f3}).

\subsection{AzTEC-3 Follow-Up}

\subsubsection{Continuum Emission}

We detect strong continuum emission toward AzTEC-3 at 1.2\,mm, and
weak emission at 3.3\,mm at the same peak position
(Fig.~\ref{f12b}). We measure a continuum flux of
(118$\pm$25)\,$\mu$Jy at 3.3\,mm (i.e., rest-frame 520\,$\mu$m) from
the A configuration data, which is consistent with the 3$\sigma$ upper
limit of $<$0.12\,mJy obtained from the C configuration data (Riechers
et al.\ \citeyear{riechers10a}). The emission appears unresolved in
the A configuration data. Combining both data sets, we find a final
3.3\,mm flux of (126$\pm$19)\,$\mu$Jy. The continuum emission is
spatially resolved along the major axis at 1.2\,mm by our observations
with a synthesized beam size of $\sim$0.25$''$. By fitting
two-dimensional Gaussian profiles in the image plane, we find a size
of (0.45$''$$\pm$0.14$''$)$\times$(0.05$''$$^{+0.21''}_{-0.05''}$),
which corresponds to
(2.8$\pm$0.9)$\times$(0.3$^{+1.3}_{-0.3}$)\,kpc$^2$, and we measure a
1.2\,mm flux of (3.67$\pm$0.56)\,mJy. This is consistent with the size
of the 1.0\,mm continuum emission of
(0.40$''$$\pm$0.04$''$)$\times$(0.17$''$$^{+0.08''}_{-0.17''}$) and
the dust spectral energy distribution shape found by Riechers et
al.\ (\citeyear{riechers14b}). A circular Gaussian fit in the
visibility plane (which gives more weight to the longer baseline data)
yields a FWHP diameter of 0.14$''$$\pm$0.04$''$, or
(0.9$\pm$0.2)\,kpc, but a flux of only (2.75$\pm$0.29)\,mJy. Taken at
face value, this appears to suggest that $\sim$75\% of the emission
emerge from a compact region within the 2.5--3\,kpc diameter dust
reservoir.\footnote{We caution that the source shape could
  significantly deviate from a Gaussian shape, such that the structure
  and size of the dust emission could be more complex than indicated
  by these findings.}

\subsubsection{Line Emission}

We detect \eco\ emission toward AzTEC-3 at 15.0$\sigma$ peak
significance (Fig.~\ref{f12b}; CO $J$=2$\to$1 and 6$\to$5 are also
shown for reference). From a circular Gaussian fit in the visibility
plane to the moment-0 map, we find a line flux of $I_{\rm
  CO(5-4)}$=(0.97$\pm$0.09)\,Jy\,\kms, which is consistent with a
previous measurement by Riechers et
al.\ (\citeyear{riechers10a}). From Gaussian fitting to the line
profile (Fig.~\ref{f12a}; CO $J$=2$\to$1 and 6$\to$5 and [C{\sc ii}]
are also shown for reference), we obtain a peak flux of $S_{\rm
  CO(5-4)}$=(1.88$\pm$0.14)\,mJy and a FWHM of d$v_{\rm
  FWHM}$=(396$\pm$37)\,\kms, which is consistent with the previously
measured values, and those found for the \bco\ and [C{\sc ii}] lines
(Riechers et al.\ \citeyear{riechers14b}). The line may show an excess
in its red wing that is not captured well by the Gaussian fit, but the
significance of this feature is only moderate.  A circular Gaussian
fit in the visibility plane over the entire width of the emission of
917\,\kms\ (280\,MHz) suggests a FWHP source diameter of
0.44$''$$\pm$0.17$''$, which corresponds to (2.8$\pm$1.1)\,kpc. This
is comparable to the size of the 1.0 and 1.2\,mm continuum
emission. Fitting the source over 524\,\kms\ (160\,MHz) to capture
only the main component of the emission, we find
0.57$''$$\pm$0.15$''$, which corresponds to (3.5$\pm$0.9)\,kpc. This
is comparable to the size of the [C{\sc ii}] emission of
(0.63$''$$\pm$0.09$''$)$\times$(0.34$''$$^{+0.10''}_{-0.15''}$) found
by Riechers et al.\ (\citeyear{riechers14b}) over a similar velocity
range (466\,\kms ). There appears to be a small spatial offset between
the peaks of the \eco\ and [C{\sc ii}] emission (which is consistent
with the dust continuum peak; Fig.~\ref{f12b}). However, this offset
becomes insignificant ($\lesssim$1$\sigma$) when excluding the red CO
line wing, i.e., when considering comparable velocity ranges, which
results in a shift of the peak position by $\sim$0.1$''$
($\lesssim$2$\sigma$ significance shift) relative to the broader
velocity range. If confirmed, the emission in the red line wing thus
may correspond to a spatially offset kinematic component, but
additional data are required to investigate this finding in more
detail.

\begin{figure}[tbh]

\epsscale{1.15}
\plotone{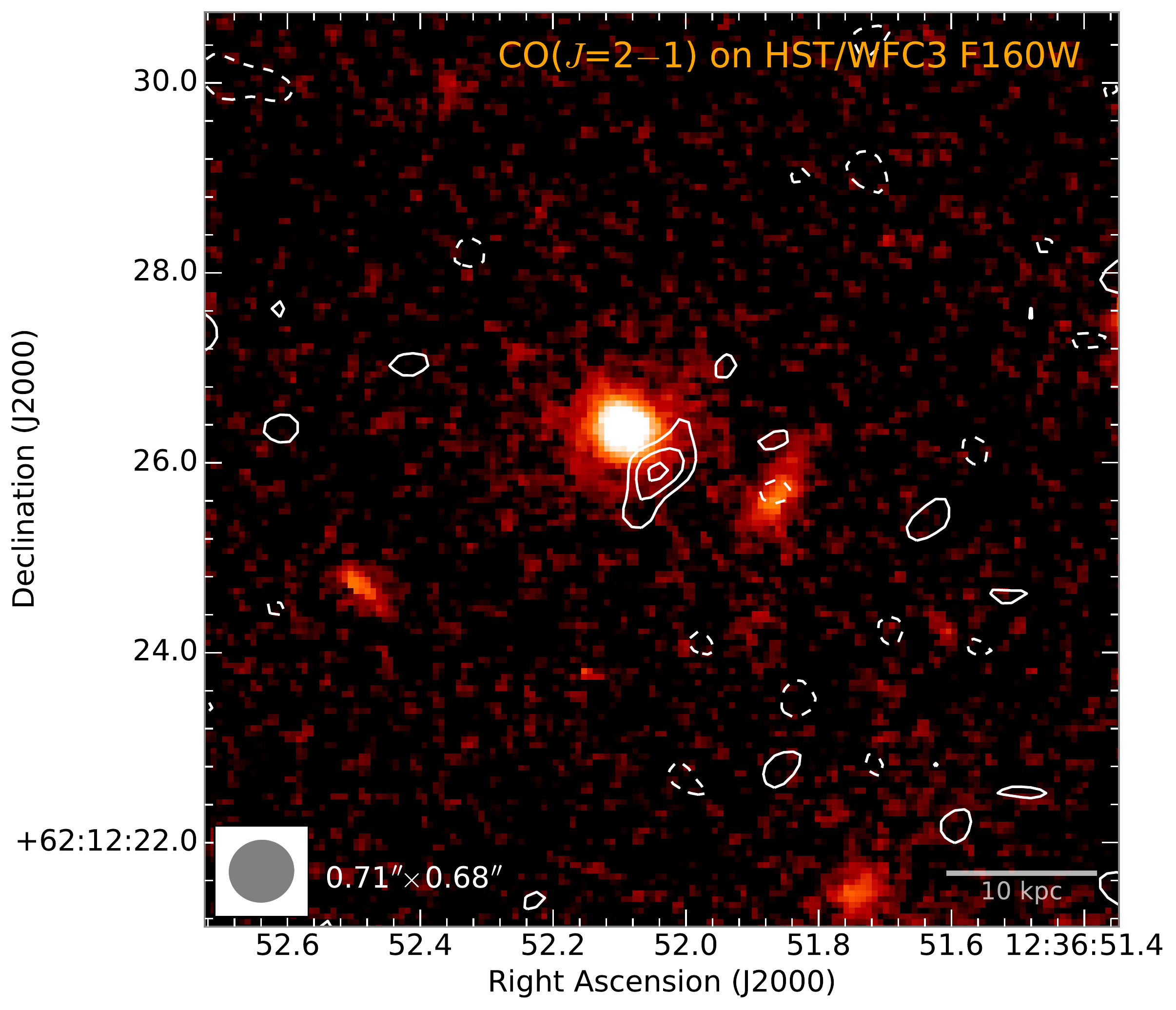}
\vspace{-2mm}

\caption{Velocity-integrated \bco\ line contour map (VLA C array data
  only) overlaid on a {\em HST}/WFC3 F160W continuum image from the
  CANDELS survey toward HDF\,850.1. Contour map is averaged over
  530\,\kms. Contours start at $\pm$2$\sigma$, and are shown in steps
  of 1$\sigma$=32.4\,$\mu$Jy\,beam$^{-1}$. The synthesized beam size
  is indicated in the bottom left corner.\label{f11}}
%
\end{figure}

\begin{figure*}[tbh]
\epsscale{1.05}
\plotone{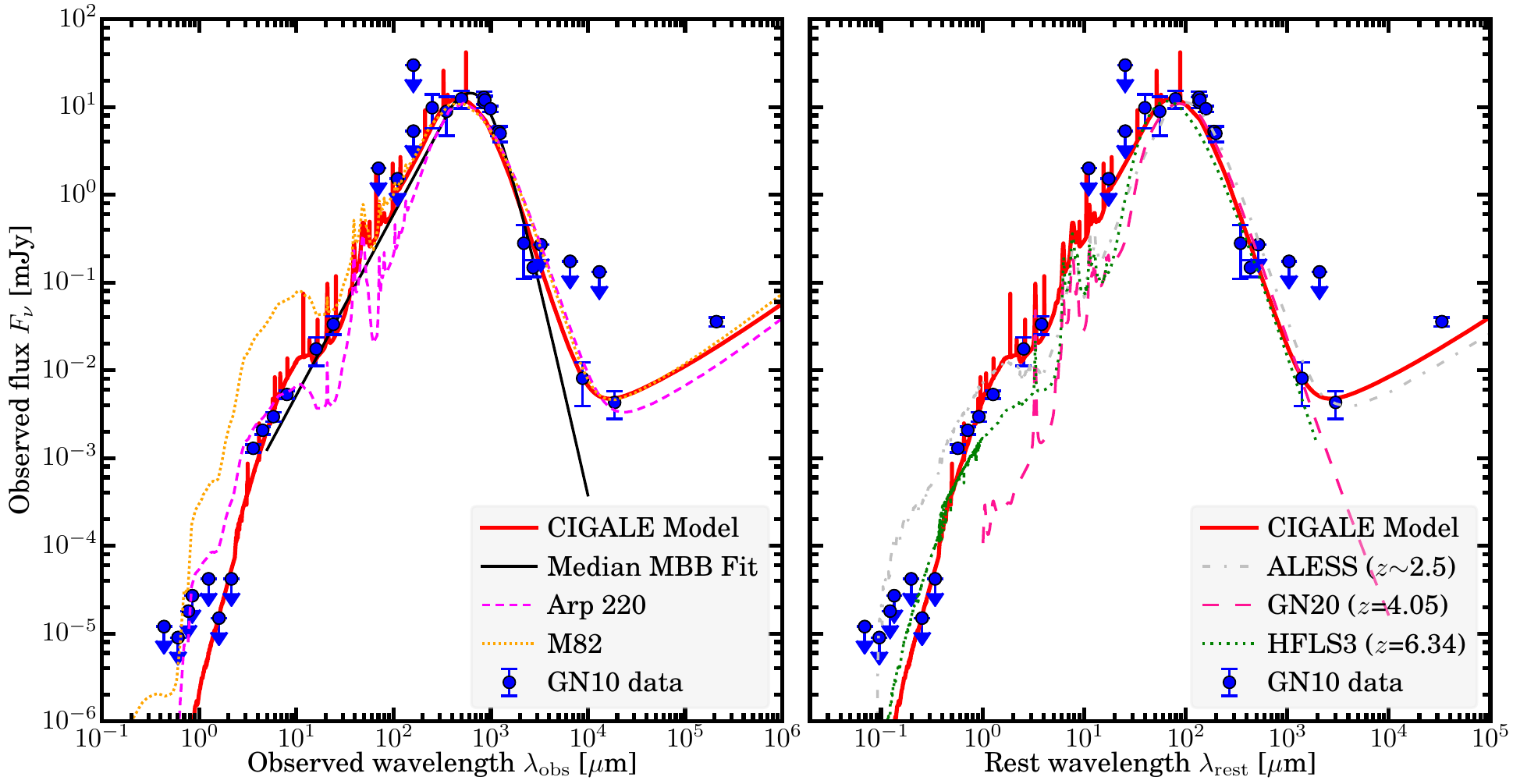}
\vspace{-2mm}

\caption{Spectral energy distribution of GN10 compared to well-known
  starbursts and a sample composite from the literature, showing its
  unusual rest-frame optical/infrared colors even when compared to
  other dust-obscured galaxies. {\em Left:} Continuum photometry
  (points), overlaid with median modified black body fit (MBB; black
  line) and \textsc{cigale} (red line) models. SED fits for the nearby
  starbursts M82 (dotted orange line) and Arp\,220 (dashed magenta
  line) are shown for comparison (Silva et al.\ \citeyear{silva98}),
  normalized to the 500\,$\mu$m flux of GN10. {\em Right:} Same points
  and red line, but also showing SED fits for the $z$=4.06 and
  $z$=6.34 DSFGs GN20 (dashed pink line) and HFLS3 (dotted green
  line), as well as a composite fit to DSFGs in the ALESS survey
  (dash-dotted gray line) for comparison (Magdis et
  al.\ \citeyear{magdis11}; Riechers et al.\ \citeyear{riechers13b};
  Swinbank et al.\ \citeyear{swinbank14}), normalized in the same
  way.\label{f7}}
%
\end{figure*}

\subsection{HDF\,850.1 Follow-Up}

We have imaged the \bco\ emission in HDF\,850.1 at $\sim$3 times
higher resolution than in the COLDz main survey data and previous
observations by Walter et al.\ (\citeyear{walter12b}). At a linear
resolution of $\sim$4.3\,kpc, the emission is detected at 4.5$\sigma$
peak significance, and also spatially resolved (Fig.~\ref{f11}). A
two-dimensional Gaussian fit to the emission in the image plane
suggests a deconvolved size of
(1.06$''$$\pm$0.23$''$)$\times$(0.49$''$$\pm$0.05$''$), which
corresponds to (6.7$\pm$1.5)$\times$(3.1$\pm$0.3)\,kpc$^2$. Given the
moderate signal-to-noise ratio of the detection, the real uncertainty
on the minor axis extent of the emission is likely of order 50\%. The
orientation and size of the emission are consistent with that seen in
the rest-frame 158\,$\mu$m dust continuum emission (Neri et
al.\ \citeyear{neri14}), and the peak positions agree to within one
synthesized beam size. No stellar light is detected at the position of
the CO emission even in deep {\em HST}/WFC3 imaging at observed-frame
1.6\,$\mu$m due to dust obscuration, which independently confirms
HDF\,850.1 as an ``optically-dark'' galaxy. It is strongly blended
with a bright foreground elliptical galaxy in {\em Spitzer}/IRAC
imaging longward of 3.6\,$\mu$m, such that only moderately
constraining upper limits can be obtained (see also discussion by Pope
et al.\ \citeyear{pope06}).

\section{Analysis of Individual Sources} \label{sec:analysis}

\subsection{Properties of GN10}

Given the new redshift identification of GN10, we here describe in
detail the determination of its key physical properties.

\subsubsection{Spectral Energy Distribution}

To extract physical parameters from the spectral energy distribution
(SED) of GN10, we have followed two approaches, as summarized in
Fig.~\ref{f7} and Table~\ref{t4}. First, we have fit the far-infrared
peak of the SED with a modified black body (MBB) routine, where the
MBB is joined to a smooth power law with slope $\alpha$ toward
observed-frame mid-infrared wavelengths (e.g., Blain et
al.\ \citeyear{blain03}, and references therein). The dust temperature
$T_{\rm dust}$, dust emissivity parameter $\beta_{\rm IR}$, and the
wavelength $\lambda_0$ where the optical depth becomes unity are used
as fitting parameters. In addition, the flux $f_{158\mu{\rm m}}^{\rm
  rest}$ at rest-frame 158\,$\mu$m is used as a normalization
parameter. The Markov Chain Monte Carlo (MCMC) and Nested Sampling
package \textsc{emcee} (Foreman-Mackey et al.\ \citeyear{foreman13})
is used to explore the posterior parameter distribution. By
integrating the fitted functions, we obtain far-infrared ($L_{\rm
  FIR}$) and total infrared ($L_{\rm IR}$) luminosities, which we then
use to determine dust-obscured star-formation rates SFR$_{\rm IR}$
under the assumption of a Kennicutt (\citeyear{kennicutt98})
conversion for a Chabrier (\citeyear{chabrier03}) stellar initial mass
function.  By adopting a mass absorption coefficient of
$\kappa_{\nu}$=2.64\,m$^2$\,kg$^{-1}$ at 125\,$\mu$m (Dunne et
al.\ \citeyear{dunne03}), we also estimate a dust mass $M_{\rm dust}$
from the fits, finding a high value in excess of 10$^9$\,\msol\ (see
Table~\ref{t4}).

We also fit the full optical to radio wavelength photometry of GN10
using \textsc{cigale,} the Code Investigating GALaxy Emission (Noll et
al.\ \citeyear{noll09}; Serra et al.\ \citeyear{serra11}), using a
broad range in star-formation histories and metallicities and a
standard dust attenuation law (Calzetti \citeyear{calzetti01}). We
find parameters that are broadly consistent with the MBB fitting where
applicable. Due to its high level of dust obscuration, GN10 remains
undetected shortward of observed-frame 3.6\,$\mu$m (rest-frame
$\sim$5700\,\AA ), which leaves some parameters only poorly
constrained. Thus, we only adopt the stellar mass $M_\star$ from the
\textsc{cigale} fit in the following. We find a high value in excess of
10$^{11}$\,\msol, i.e., $\sim$100 times the dust mass, which we
consider to be reliable to within a factor of a few (see
Table~\ref{t4}).\footnote{See Appendix B for an alternative $M_\star$
  value obtained with the \textsc{magphys} code, which we consider to be
  consistent within the uncertainties. See also Liu et
  al.\ (\citeyear{liu18}) for a discussion of the potential
  uncertainties associated with the determination of $M_\star$ for the
  most distant DSFGs, and Simpson et
  al.\ (\citeyear{simpson14,simpson17a}) for a more detailed
  discussion of the uncertainties of $M_\star$ estimates for DSFGs in
  general.}

From our MBB fits, we find that the dust turns optically thick at
$\sim$170\,$\mu$m (i.e., between observed-frame 1.0 and 1.2\,mm; see
Table~\ref{t4}), similar to the values found for other $z$$>$4 DSFGs
(see, e.g., Riechers et al.\ \citeyear{riechers13b,riechers14b};
Simpson et al.\ \citeyear{simpson17a}), such that dust extinction may
impact the observed [C{\sc ii}] line luminosity at 158\,$\mu$m. This would
be in agreement with a larger apparent dust emission size at 1.0\,mm
than at 1.2\,mm as found above due to dust optical depth effects, but
higher resolution 1.0\,mm imaging is required to further investigate
this finding. We also find a moderately high dust temperature of
(50$\pm$9)\,K.\footnote{See Appendix for an alternative,
  luminosity-averaged $T_{\rm dust}$ value obtained by fitting
  multiple dust components with the \textsc{magphys} code.}


\begin{figure}
\begin{deluxetable}{ l l l }
\tabletypesize{\scriptsize}
\tablecaption{GN10 MBB and \textsc{cigale} SED modeling parameters. \label{t4}}
\tablehead{
Fit Parameter\hspace{1.5cm} & unit\hspace{1.5cm} & value\tablenotemark{a}\hspace{1.5cm} }
\startdata
$T_{\rm dust}$               & K     & 50.1$^{+9.1}_{-9.1}$ \\
$\beta_{\rm IR}$             &       & 2.98$^{+0.18}_{-0.17}$ \\
$\lambda_{0}^{\rm rest}$      & $\mu$m & 168$^{+22}_{-25}$ \\
$\alpha$                   &        & 2.06$^{+0.15}_{-0.11}$ \\
$f_{158\mu{\rm m}}^{\rm rest}$\tablenotemark{b} & mJy   & 8.3$^{+0.5}_{-0.5}$ \\
$M_{\rm dust}$              & 10$^9$\,\msol\ & 1.11$^{+0.44}_{-0.27}$ \\
$L_{\rm FIR}$\tablenotemark{c} & 10$^{13}$\,\lsol\ & 0.58$^{+0.11}_{-0.09}$ \\
$L_{\rm IR}$\tablenotemark{d} & 10$^{13}$\,\lsol\ & 1.03$^{+0.19}_{-0.15}$ \\
SFR$_{\rm IR}$              & \msol\,yr$^{-1}$ & 1030$^{+190}_{-150}$ \\
$M_{\star}$\tablenotemark{e}     & 10$^{11}$\,\msol\ & 1.19($\pm$0.06) \\
\enddata 
\tablenotetext{\rm a}{Values as stated are 50$^{\rm th}$ percentiles. Lower and upper error bars are stated as 16$^{\rm th}$ and 84$^{\rm th}$ percentiles, respectively.}
\tablenotetext{\rm b}{Fit normalization parameter.}
\tablenotetext{\rm c}{Integrated between rest-frame 42.5 and 122.5\,$\mu$m.}
\tablenotetext{\rm d}{Integrated between rest-frame 8 and 1000\,$\mu$m.}
\tablenotetext{\rm e}{Parameter adopted from \textsc{cigale}. Quoted fitting uncertainties underestimate systematic effects, and thus, are not adopted to make any conclusions in the following.}
\end{deluxetable}
\end{figure}


\begin{figure}
\begin{deluxetable}{ l c c }
\tabletypesize{\scriptsize}
\tablecaption{GN10 [C{\sc ii}] dynamical modeling parameters. \label{t2}}
\tablehead{
Fit Parameter & unit & value\tablenotemark{a} }
\startdata
$$[C{\sc ii}] FWHM diameter & arcsec  & 1.03$^{+0.11}_{-0.10}$ \\
                      & kpc     & 6.4$^{+0.7}_{-0.7}$ \\
Velocity scale length & arcsec  & 0.25$^{+0.29}_{-0.18}$ \\
                      & kpc     & 1.6$^{+1.8}_{-1.1}$ \\
Disk inclination      & degrees & 80$^{+7}_{-8}$ \\
Position angle        & degrees & 206$^{+5}_{-5}$ \\
Maximum Velocity      & \kms    & 320$^{+100}_{-80}$ \\
Gas dispersion        & \kms    & 220$^{+25}_{-25}$ \\
Dust FWHM diameter    & arcsec  & 0.56$^{+0.05}_{-0.05}$ \\
                      & kpc     & 3.5$^{+0.3}_{-0.3}$ \\
$M_{\rm dyn}$ ($r$=3.66\,kpc) & 10$^{10}$\,\msol & 4.5$^{+2.1}_{-1.5}$ \\
$M_{\rm dyn}$ ($r$=5.49\,kpc) & 10$^{10}$\,\msol & 8.6$^{+3.6}_{-2.8}$ \\
\enddata 
\tablenotetext{\rm a}{Values as stated are 50$^{\rm th}$ percentiles. Lower and upper error bars are stated as 16$^{\rm th}$ and 84$^{\rm th}$ percentiles, respectively.}
\end{deluxetable}
\vspace{-12mm}
\end{figure}



\begin{figure*}[tbh]
\epsscale{1.15}
\plotone{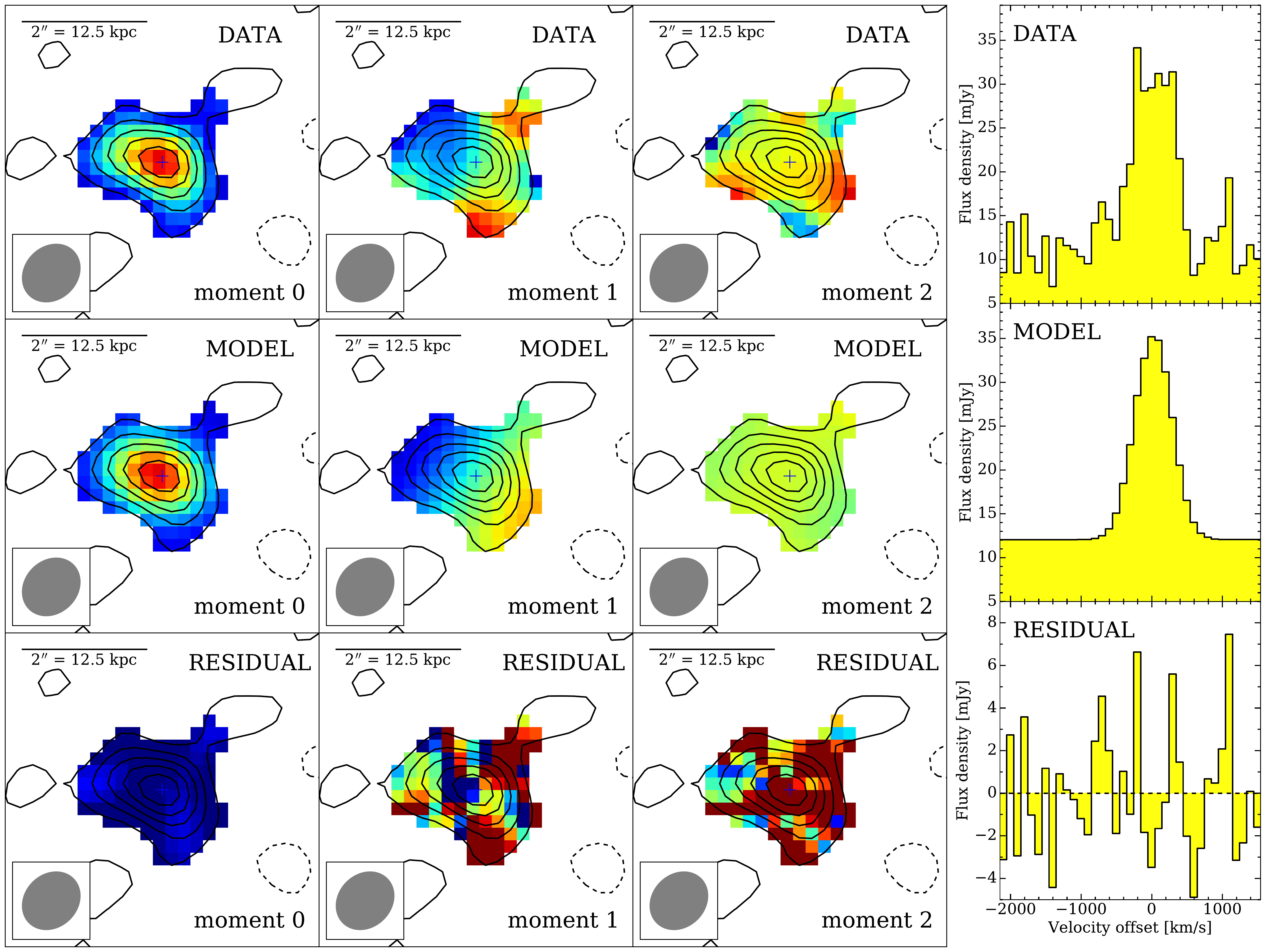}
\vspace{-2mm}

\caption{Visibility space dynamical modeling results for the [C{\sc
      ii}] emission toward GN10. Intensity (moment-0), velocity
  (moment-1), and velocity dispersion (moment-2) maps ({\em left}) for
  the data ({\em top}), model ({\em middle}), and ``data--model''
  residuals ({\em bottom}), and spectra (histograms; {\em
    right}). Median model fits and residuals are shown. Moment-0
  contours are overlaid on all panels, and are shown in steps of
  $\pm$2$\sigma$. Data are imaged with ``natural'' baseline
  weighting. Beam size is shown in the bottom left corner of each map
  panel. Continuum emission was included as part of the additional
  fitting parameters for the rotating disk model. \label{f5}}
%
\end{figure*}

\subsubsection{Star-Formation Rate Surface Density}

Based on the 1.2\,mm size of GN10, we find a source surface area of
(0.79$\pm$0.44)\,kpc$^2$ (0.99$\pm$0.30\,kpc$^2$; second quoted values
indicate results from circular Gaussian fits). Its flux thus
corresponds to a source-averaged rest-frame brightness temperature of
$T_{\rm b}$=(24.9$\pm$8.1)\,K (20.0$\pm$2.9\,K) at rest-frame
190\,$\mu$m, or 50\%$\pm$18\% (40\%$\pm$9\%) of the dust temperature
obtained from SED fitting. This suggests that the dust emission is
likely at least moderately optically thick, and that it fills the bulk
of the source surface area within its inferred size. From our SED fit,
we determine a dust optical depth of $\tau_{190\,\mu{\rm
    m}}$=0.69$\pm$0.29 (i.e. 1$-$exp($-\tau_{190\,\mu{\rm
    m}}$)=50\%$^{+13\%}_{-17\%}$), which is consistent with this
finding.  Using the $L_{\rm IR}$ derived in the previous subsection,
this size corresponds to an IR luminosity surface density of
$\Sigma_{\rm IR}$=(7.5$\pm$4.4)$\times$10$^{12}$\,\lsol\,kpc$^{-2}$
(6.0$\pm$2.0$\times$10$^{12}$\,\lsol\,kpc$^{-2}$), or a SFR surface
density of $\Sigma_{\rm
  SFR}$=(750$\pm$440)\,\msol\,yr$^{-1}$\,kpc$^{-2}$
(600$\pm$210\,\msol\,yr$^{-1}$\,kpc$^{-2}$).

\subsubsection{Dynamical Mass Estimate}

To obtain a dynamical mass estimate from our resolved line
  emission map, we have fitted visibility-plane dynamical models of a
rotating disk to the [C{\sc ii}] emission from GN10
(Fig.~\ref{f5}; see Pavesi et al.\ \citeyear{pavesi18a} for further
details on the modeling approach). Rotating circular disk models of
the [C{\sc ii}] emission are generated through the {\em KinMSpy} code
(Davis et al.\ \citeyear{davis13}), super-imposed on continuum
emission which is fitted by a circular two-dimensional Gaussian
function.\footnote{We include the continuum emission in the
    fitting to properly account for uncertainties associated with
    continuum modeling and subtraction.} The fitting parameters are
the disk center position, systemic velocity, gas dispersion, FWHM size
of the spatial light profile of the Gaussian disk, maximum velocity,
velocity scale length, inclination, position angle, and line flux. The
continuum flux and FWHM size are determined based on the emission in
the line-free channels. The fitting method employs
MCMC and Nested Sampling techniques as implemented in \textsc{emcee}
(Foreman-Mackey et al.\ \citeyear{foreman13}) and \textsc{MultiNest}
for {\tt python} (Feroz et al.\ \citeyear{feroz09}; Buchner et
al.\ \citeyear{buchner14}) to sample the posterior distribution of the
model parameters and to calculate the model evidence. To optimize the
parameter sampling, non-constraining, uniform priors are chosen for
additive parameters, logarithmic priors for scale parameters, and a
sine prior for the inclination angle. The data are fitted well by the
disk model, leaving no significant residuals in the moment-0 map or
spectrum. The results for all parameters are given in
Table~\ref{t2}. We find a median dynamical mass of $M_{\rm
  dyn}$=8.6$^{+3.6}_{-2.8}$(4.5$^{+2.1}_{-1.5}$)$\times$10$^{10}$\,\msol\ out
to a radial distance of 5.5 (3.7)\,kpc from the center. Given the FWHM
diameter of the [C{\sc ii}] emission of 6.4$^{+0.7}_{-0.7}$\,kpc, the
derived values are barely sufficient to host the estimated stellar
mass of $\sim$1.2$\times$10$^{11}$\,\msol\ if the stellar
component has a similar extent to the [C{\sc ii}] emission. This could
either indicate that the stellar mass in GN10 is overestimated, e.g.,
due to the model fitting a solution which has too old a
stellar population or the contribution of a dust-obscured active
galactic nucleus (AGN) to the mid-infrared emission (see, e.g.,
Riechers et al.\ \citeyear{riechers14a}), or that the kinematic
structure of GN10 is not dominated by rotation (such that the
dynamical mass is underestimated). Observations of the stellar and gas
components at higher spatial resolution are required to distinguish
between these scenarios.

\subsubsection{``Underluminous'' \aco\ Emission?}

Assuming optically thick emission, the integrated \bco\ to
\aco\ brightness temperature ratio $r_{21}$=1.37$\pm$0.45 (compared to
1.84$\pm$0.38 in line profile peak brightness temperature) toward GN10
suggests that the \aco\ line could be underluminous compared to
expectations for thermalized or sub-thermal gas excitation.  Given the
comparable spatial resolution of the \bco\ and \aco\ observations,
this is unlikely to be due to resolution effects.  If significant,
this effect may be due to the high cosmic microwave background (CMB)
temperature at $z$=5.3 ($T_{\rm CMB}$$\simeq$17.2\,K), compared to the
excitation potential above ground of the \aco\ transition (which
corresponds to an excitation temperature of $T_{\rm ex}$=5.5\,K). The
CMB acts as both a source of excitation and as a background for the CO
emission.

Studies at $z$$\sim$2--3 have found evidence for enhanced \aco\ line
widths and line strengths in some DSFGs due to the presence of low
density, low surface brightness gas, for which a low-$T_{\rm ex}$ line
like \aco\ is an ideal tracer (e.g., Riechers et
al.\ \citeyear{riechers11e}; Ivison et al.\ \citeyear{ivison11}).  The
CO brightness temperature itself is measured as a contrast to the
CMB. Thus, low surface brightness emission as traced by \aco\ may be
disproportionately affected by CMB effects toward higher redshifts
(e.g., by heating colder gas to $T_{\rm CMB}$, thereby reducing the
observable brightness temperature contrast), such that a preferential
impact toward weakened \aco\ line emission may be expected (e.g., da
Cunha et al.\ \citeyear{dacunha13b}; Zhang et
al.\ \citeyear{zhang16}).\footnote{In this scenario, a disproportional
  impact on emission from cold dust due to a reduced contrast toward
  and increased heating by the CMB would also be expected, resulting
  in a higher apparent dust temperature due to changes in the SED
  shape. This would be consistent with the relatively high measured
  $T_{\rm dust}$ of GN10.} Thus, a reduced \aco\ line flux compared to
higher-$J$ lines due to the CMB appears plausible to explain the
observed line spectra toward GN10. However, while not affected as
strongly, some impact on the \bco\ line (having an excitation
potential above ground corresponding to $T_{\rm ex}$=16.6\,K, i.e.,
$\sim$$T_{\rm CMB}$($z$=5.3)) would also be expected in this scenario,
yielding a reduced impact on $r_{21}$.

Apart from effects due to the CMB, another possible scenario is that
the low-$J$ CO line emission may not be optically thick in some
regions.  Given the higher expected optical depth in the \bco\ line
compared to \aco, $r_{21}$$>$1 would be possible in this scenario. In
principle, self absorption due to cold gas in the foreground of the
warmer gas located in the star-forming regions could also
disproportionately affect the strength of low-$J$ CO lines in some
source geometries.  While this finding is intriguing, higher
signal-to-noise measurements in the future are desirable to further
investigate this effect and its origin based on a detailed comparison
of the line profiles.

\subsection{Properties of AzTEC-3}

We here update some of the key properties on AzTEC-3, based on the new
information available in this work.

\subsubsection{Spectral Energy Distribution}

Using the new and updated 1--3\,mm photometry from this work, Pavesi
et al.\ (\citeyear{pavesi16}), and Magnelli et
al.\ (\citeyear{magnelli19}), we have re-fit the SED of AzTEC-3 with
the same routine as used by Riechers et al.\ (\citeyear{riechers14b}),
which is similar to that used for GN10 above. We find $T_{\rm
  dust}$=92$^{+15}_{-16}$\,K, $\beta_{\rm IR}$=2.09$\pm$0.21,
$\lambda_0^{\rm rest}$=181$^{+33}_{-34}$,
$\alpha$=10.65$^{+6.69}_{-6.42}$, and $L_{\rm
  FIR}$=(1.12$\pm$0.16$)\times$10$^{13}$\,\lsol. These values are
consistent with those found by Riechers et
al.\ (\citeyear{riechers14b}).  We also find a total infrared
luminosity of $L_{\rm
  IR}$=(2.55$^{+0.73}_{-0.74}$)$\times$10$^{13}$\,\lsol. The
relatively large uncertainties compared to $L_{\rm FIR}$ are due to
the limited reliability of the {\em Herschel}/SPIRE 250--500\,$\mu$m
photometry near the peak of the SED. Taken at face value, this
suggests SFR$_{\rm
  IR}$=(2500$\pm$700)\,\msol\,yr$^{-1}$.\footnote{Given the
  uncertainties on $L_{\rm IR}$, previous works adopted $L_{\rm FIR}$
  to determine the SFR of AzTEC-3. We adopt this updated value here
  for internal consistency of the analysis presented in this work.}

\subsubsection{Star-Formation Rate Surface Density}

Based on the 1.2\,mm size of AzTEC-3, we find a source surface area of
$<$(2.95$\pm$0.45)\,kpc$^2$ (0.62$\pm$0.17\,kpc$^2$; second
  quoted values indicate results from circular Gaussian fits, which
account for the compact component only). Its flux thus corresponds to
a source-averaged rest-frame brightness temperature of $T_{\rm
  b}$$>$(4.7$\pm$0.7)\,K (16.8$\pm$2.2\,K) at rest-frame 190\,$\mu$m,
or $>$5\%$\pm$1\% (18\%$\pm$4\%) of the dust temperature obtained from
SED fitting. This suggests that AzTEC-3 has significant structure on
scales significantly smaller than the $\sim$0.25$''$ beam of our
1.2\,mm observations. Using the $L_{\rm IR}$ derived in the previous
section, this size corresponds to $\Sigma_{\rm
  IR}$$>$(5.0$\pm$1.6)$\times$10$^{12}$\,\lsol\,kpc$^{-2}$
(18.0$\pm$7.1$\times$10$^{12}$\,\lsol\,kpc$^{-2}$), or $\Sigma_{\rm
  SFR}$$>$(500$\pm$160)\,\msol\,yr$^{-1}$\,kpc$^{-2}$
(1800$\pm$700\,\msol\,yr$^{-1}$\,kpc$^{-2}$).\footnote{We here assume
  that the fraction of the flux contained by the compact component at
  1.2\,mm is constant with wavelength, which may yield a lower limit
  on $\Sigma_{\rm IR}$ if this component is warmer than the rest of
  the source, or an upper limit if it has a higher optical depth.}

\subsection{Properties of HDF\,850.1}

The new high-resolution \bco\ data and a combination of constraints
from the literature allow us to determine some additional key
properties of HDF\,850.1, as detailed in the following.

\subsubsection{CO Luminosity Surface Density}

For HDF\,850.1, the size of the gas reservoir measured from the
high-resolution \bco\ observations at its observed $L'_{\rm CO(2-1)}$
implies a CO luminosity surface density of $\Sigma_{\rm
  CO(2-1)}$=(4.8$\pm$1.8)$\times$10$^{5}$\,\lsol\,kpc$^{-2}$. We also
find a rest-frame \bco\ peak brightness temperature of $T_{\rm b}^{\rm
  CO}$=1.6$\pm$0.4\,K at the $\sim$0.7$''$ resolution of our
observations, which agrees to within $\sim$7\% with the
source-averaged value. This modest value is consistent with the fact
that the source is resolved over less than two beams in our current
data.

\subsubsection{Star-Formation Rate Surface Density}

Adopting the apparent $L_{\rm
  IR}$=(8.7$\pm$1.0)$\times$10$^{12}$\,\lsol\ reported by Walter et
al.\ (\citeyear{walter12b}) and the rest-frame 158\,$\mu$m dust
continuum size of (0.9$''$$\pm$0.1$''$)$\times$(0.3$''$$\pm$0.1$''$)
reported by Neri et al.\ (\citeyear{neri14}), we find an apparent
physical source size of (5.7$\pm$0.6)$\times$(1.9$\pm$0.6)\,kpc$^2$
and a source-averaged $\Sigma_{\rm
  IR}$=6.0$\pm$0.9$\times$10$^{11}$\,\lsol\,kpc$^{-2}$ for
HDF\,850.1. This corresponds to $\Sigma_{\rm
  SFR}$$\sim$(60$\pm$10)\,\msol\,yr$^{-1}$\,kpc$^{-2}$.\footnote{The
  lensing magnification factor cancels out of the surface density
  calculations, such that the resulting properties are conserved under
  lensing, barring potential differential lensing effects.} We also
find a source-averaged rest-frame dust brightness temperature of
$T_{\rm b}$=1.4$\pm$0.3\,K at rest-frame 158\,$\mu$m, or
$\sim$4\%$\pm$1\% of its dust temperature. This suggests that the dust
has significant substructure on scales much smaller than the
$\sim$0.35$''$ beam of the observations presented by Neri et
al.\ (\citeyear{neri14}). It also is comparable to the CO brightness
temperature on comparable scales.\\[10mm]

\section{Discussion of the COLDz $z$$>$5 DSFG sample} \label{sec:discussion}

We here describe the COLDz $z$$>$5 DSFG sample in more detail, and
place it into context with other known $z$$>$5 DSFGs and the space
densities and clustering properties of DSFGs at the highest
redshifts. Key properties are summarized in Tables \ref{t2z} and
  \ref{t5}.

\subsection{Star-Formation Rate Surface Densities}

The SFR surface densities of $\Sigma_{\rm SFR}$=(750$\pm$440) and
(1800$\pm$700)\,\msol\,yr$^{-1}$\,kpc$^{-2}$ found above for GN10 and
the central region of AzTEC-3 are even higher than the values found
for some other compact starbursts like ADFS-27 ($z$=5.66;
430$\pm$90\,\msol\,yr$^{-1}$\,kpc$^{-2}$) and HFLS3 ($z$=6.34;
480$\pm$30\,\msol\,yr$^{-1}$\,kpc$^{-2}$; Riechers et
al.\ \citeyear{riechers13b,riechers17}). The source-averaged value of
$>$(500$\pm$160)\,\msol\,yr$^{-1}$\,kpc$^{-2}$ for AzTEC-3 is
comparable to these sources. Like these systems, GN10 and AzTEC-3 thus
show the properties expected for so-called ``maximum starbursts''. At
face value, the peak $\Sigma_{\rm SFR}$ in AzTEC-3 may slightly exceed
the expected Eddington limit for starburst disks that are supported by
radiation pressure (e.g., Thompson et al.\ \citeyear{thompson05};
Andrews \& Thompson \citeyear{at11}), but is potentially consistent
under the assumption of a more complex source geometry. On the other
hand, the high $\Sigma_{\rm SFR}$ value of GN10 could be boosted by an
obscured AGN contribution to the dust heating. GN10 exhibits strong
0.5--8\,keV X-ray emission.\footnote{AzTEC-3 is not detected at X-ray
  wavelengths.}  Using Equation 15 of Ranalli et
al.\ (\citeyear{ranalli03}), its observed
(absorption-corrected)\footnote{From fitting a Galactic absorption
  plus power-law model, an effective photon index
  $\Gamma$=1.40$^{+1.46}_{-1.36}$ was found for GN10, but it was not
  possible to simultaneously fit the absorbing column $N_{\rm H}$ and
  $\Gamma$ due to the limited photon counts. The data are consistent
  with no intrinsic absorption within the uncertainties, such that the
  absorption-corrected $L_{\rm X}$ could be considered an upper limit
  (Laird et al.\ \citeyear{laird10}) -- or, alternatively, a lower
  limit in case it is heavily absorbed.} rest-frame 2--10\,keV X-ray
luminosity of $L_{\rm X}$=5.6(12.5)$\times$10$^{42}$\,erg\,s$^{-1}$
(Laird et al.\ \citeyear{laird10}) corresponds to a SFR$_{\rm X}$ of
1100(2500)\,\msol\,yr$^{-1}$. Given its SFR$_{\rm
  IR}$=1030$^{+190}_{-150}$\,\msol\,yr$^{-1}$, its $L_{\rm X}$ remains
consistent with intense star formation, but a contribution from a
modestly luminous obscured AGN cannot be ruled out, depending on the
(relatively uncertain) absorption correction required. This would also
be consistent with a possible excess mid-infrared emission due to an
obscured AGN, which may be favored by some of the SED fits.

In contrast, the source-averaged $\Sigma_{\rm SFR}$ in HDF\,850.1 is
only $\sim$(60$\pm$10)\,\msol\,yr$^{-1}$\,kpc$^{-2}$, i.e., $\sim$12
times lower than in GN10, and $\sim$30 times lower than the peak value
in AzTEC-3. On the other hand, the values for GN10 and HDF\,850.1
would become comparable when assuming the larger 1.0\,mm continuum
size limit for GN10 instead of the smaller 1.2\,mm continuum size
adopted above. As such, the source-averaged $\Sigma_{\rm SFR}$ may be
more similar when accounting for potentially extended dust emission if
present, and thus, significantly lower than in AzTEC-3 on
average. Such more modest, ``sub-Eddington'' $\Sigma_{\rm SFR}$ on few
kpc scales would be consistent with what is found from high-resolution
studies of lower-$z$ DSFG samples on average (e.g., Simpson et
al.\ \citeyear{simpson15}; Hodge et al.\ \citeyear{hodge16,hodge19}).
For reference, using the sizes and SFRs for $z$$>$4 DSFGs in the
AS2UDS sample (Gullberg et al.\ \citeyear{gullberg19}; Dudzeviciute et
al.\ \citeyear{dudzeviciute20}), we find a median $\Sigma_{\rm
  SFR}$=(200$\pm$100)\,\msol\,yr$^{-1}$\,kpc$^{-2}$, where the error
corresponds to the bootstrap uncertainty on the median. Extending the
sample down to $z$$>$3.5 yields $\Sigma_{\rm
  SFR}$=(170$\pm$20)\,\msol\,yr$^{-1}$\,kpc$^{-2}$. However, a more
precise measurement of the 1.0\,mm continuum morphology of GN10 is
required to make firm statements about extended dust emission. The
intriguingly high peak $\Sigma_{\rm SFR}$ of AzTEC-3 will be explored
further in future work.

\subsection{CO Large Velocity Gradient Modeling}

\begin{figure}
\epsscale{1.15}
\plotone{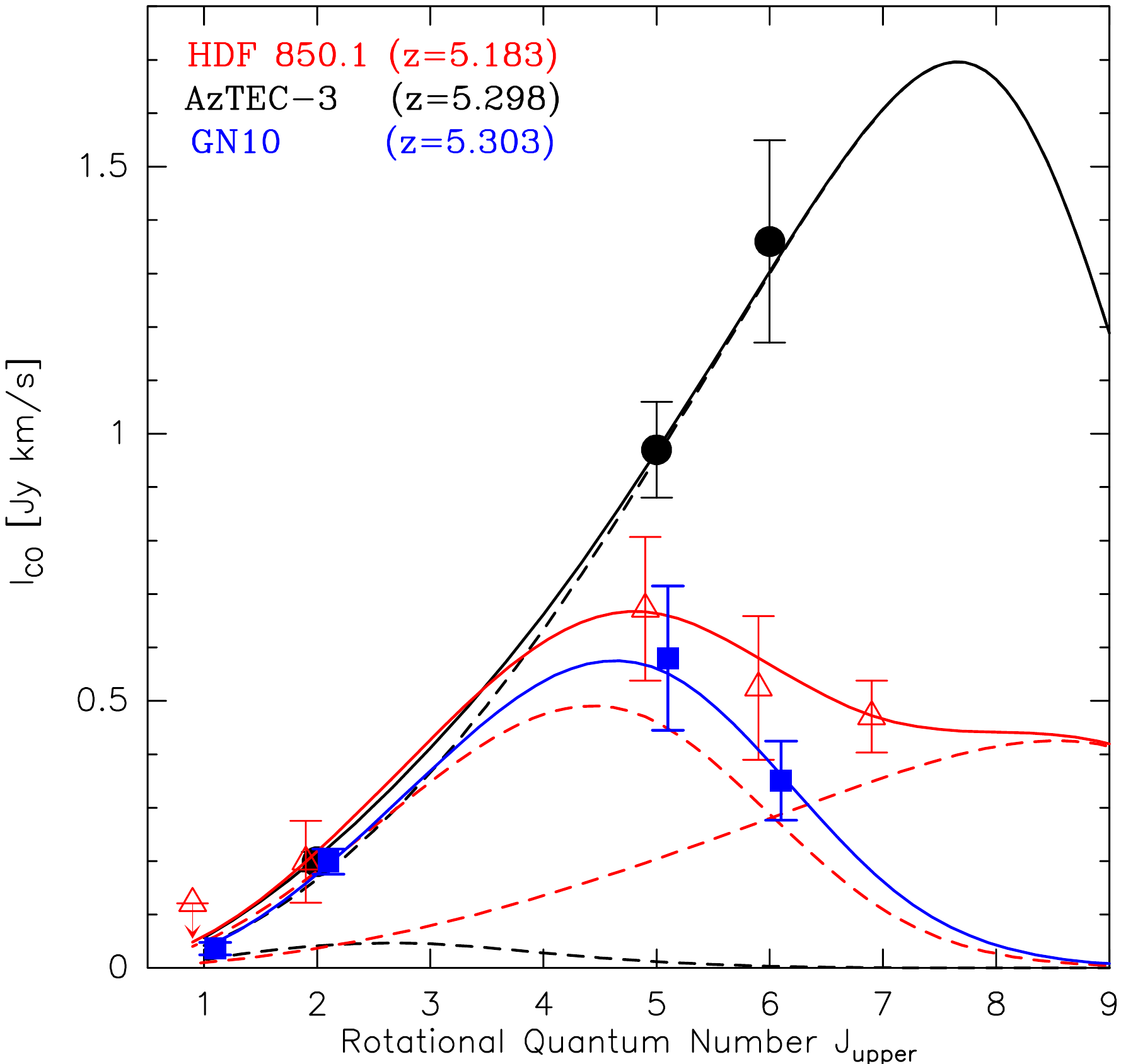}
\vspace{-2mm}

\caption{CO excitation ladders (points) and LVG modeling (lines) of
  the three COLDz $z$$>$5 DSFGs analyzed in this work. Line fluxes are
  normalized to the strength of the \bco\ line in AzTEC-3. Sources are
  slightly offset from each other in $J_{\rm upper}$ to improve
  clarity. The models (solid lines) of AzTEC-3 and HDF\,850.1 consist
  of two gas components each (dashed lines). \label{f6}}
%
\end{figure}

To constrain the line radiative transfer properties of the three COLDz
$z$$>$5 DSFGs based on the observed CO line excitation, we calculated
a series of large velocity gradient (LVG) models, which treat the gas
kinetic temperature $T_{\rm kin}$ and the gas density $\rho_{\rm gas}$
as free parameters (Fig.~\ref{f6}). For all calculations, the H$_2$
ortho-to-para ratio was fixed to 3:1, the CMB temperature was set to
the values appropriate for our targets (16.85--17.18\,K at $z$=5.183
to 5.303), and the Flower et al.\ (\citeyear{flower01}) CO collision
rates were adopted. We also used a ratio between the CO abundance and
velocity gradient of 10$^{-5}$\,pc (\kms )$^{-1}$ (see, e.g.,
Wei\ss\ et al.\ \citeyear{weiss05a,weiss07}; Riechers et
al.\ \citeyear{riechers06c}).

Our model of GN10 consists of a single LVG component with $T_{\rm
  kin}$=40\,K and $\rho_{\rm gas}$=10$^{3.6}$\,cm$^{-3}$, suggesting
the presence of gas at moderate excitation.\footnote{The model is
  consistent with high optical depth in all CO lines.} This model
would imply that the CO emission should fill the bulk of the
[C{\sc ii}]-emitting region, which perhaps suggests that the warm, compact
nucleus traced by the 190\,$\mu$m dust emission is embedded in a more
extended cold gas reservoir.\footnote{Both radial variations in
  temperature and optical depth may contribute to the observed effect;
  see also discussion by Calistro Rivera et
  al.\ (\citeyear{calistro18}).}

For AzTEC-3, we adopt the LVG model discussed by Riechers et
al.\ (\citeyear{riechers10a}). This model consists of a low-excitation
component with $T_{\rm kin}$=30\,K and $\rho_{\rm
  gas}$=10$^{2.5}$\,cm$^{-3}$ and a high-excitation component with
$T_{\rm kin}$=45\,K and $\rho_{\rm gas}$=10$^{4.5}$\,cm$^{-3}$. Based
on the updated \bco\ flux, we reduce the area filling factor of the
lower-temperature component.  This component now contributes only
$\sim$20\% to the \bco\ line flux. This model suggests an expected
\aco\ flux of 0.057\,Jy\,\kms, or a \bco\ to \aco\ line ratio of
$r_{21}$=0.91. If the high-excitation gas were distributed over the
same 3.9$\times$2.1\,kpc$^2$ area as the [C{\sc ii}] emission imaged
by Riechers et al.\ (\citeyear{riechers14b}), the LVG model would
imply an area filling factor of close to 50\%. Assuming the extent of
the main \eco\ component determined above instead would imply a
filling factor of just above 30\%. The low-excitation gas, on the
other hand, would need to be distributed over an area at least twice
as large in linear extent as the [C{\sc ii}] emission, corresponding
to an area approximately half as large as the current observed limit
on the \bco\ diameter of $\sim$8\,kpc. In light of these findings, we
will re-visit these models in future work, once more observational
constraints are available.

The bulk emission in HDF\,850.1 can be modeled with an LVG component
with the same parameters as for GN10 (suggesting an expected CO
$J$=1$\to$0 flux of 0.036\,Jy\,\kms ), but the model does not
reproduce the ratio between the \fco\ and \gco\ lines well. Thus, to
improve the model fit, it is necessary to add a second,
high-excitation gas component with $T_{\rm kin}$=70\,K and $\rho_{\rm
  gas}$=10$^{4.5}$\,cm$^{-3}$.\footnote{Since no detections above
  $J_{\rm upper}$=7 exist, we caution that the parameters of this
  second component are only poorly constrained by the data.} The
moderate-and high-excitation components in this model fill $\sim$20\%
and $\sim$1.4\% of the [C{\sc ii}]-emitting area imaged by Neri et
al.\ (\citeyear{neri14}), respectively. Assuming the \bco\ size
determined above instead yields $\sim$7.5\% lower values, which are
indistinguishable within the uncertainties.

With current observational constraints, these LVG model solutions are
not unique, but they illustrate that the gas excitation in all sources
can be modeled with parameters that span a similar range as the
parameters found in nearby spiral galaxies and starbursts (e.g.,
Wei\ss\ et al.\ \citeyear{weiss05a}; G\"usten et
al.\ \citeyear{gusten06}). Moreover, we find clear differences in the
gas excitation between $z$$>$5 DSFGs, showing that AzTEC-3 is likely
in a more extreme phase of its evolution as reflected in its high
molecular gas excitation, perhaps due to a high gas density in its
nuclear region or due to contributions from a buried AGN to the gas
excitation. All of the sources in our sample follow the \eco\ -- FIR
luminosity relation for star-forming galaxies (Fig.~\ref{f8}; e.g.,
Daddi et al.\ \citeyear{daddi15}; Liu et al.\ \citeyear{liu15}), which
shows that their level of star-formation activity (as traced by
$L_{\rm FIR}$) is consistent with what is expected based on the
properties of the warm, dense molecular gas (as traced by CO
$J$=5$\to$4). This is in agreement with the idea that a lower inferred
CO excitation among galaxies in our sample is likely related to a
lower fraction of dense molecular gas in the star-forming regions.


\begin{figure}[t]
\begin{deluxetable*}{ l c c c c }
\tabletypesize{\scriptsize}
\tablecaption{Derived properties for COLDz $z$$>$5 DSFGs. \label{t2z}}
\tablehead{
  Quantity & unit & GN10 & AzTEC-3 & HDF\,850.1}
\startdata
$D_{\rm dust}$ (major$\times$minor axis) & kpc$^2$ & (1.6$\pm$0.4)$\times$(0.6$\pm$0.6) & (2.8$\pm$0.9)$\times$(0.3$^{+1.3}_{-0.3}$) & (5.7$\pm$0.6)$\times$(1.9$\pm$0.6) \\
SFR$_{\rm IR}$ & \msol\,yr$^{-1}$ & 1030$^{+190}_{-150}$ & 2500$\pm$700 & 870$\pm$100 \\
$\Sigma_{\rm SFR}$ & \msol\,yr$^{-1}$\,kpc$^{-2}$ & 600$\pm$210 & $>$500$\pm$160 & 60$\pm$10 \\
$\Sigma_{\rm SFR}^{\rm peak}$ & \msol\,yr$^{-1}$\,kpc$^{-2}$ & 750$\pm$440 & 1800$\pm$700 & \\
$M_{\rm gas}$ & 10$^{10}$\msol & 7.1$\pm$0.9 & 5.7$\pm$0.5 & 2.2$\pm$0.8 \\
$M_{\rm dust}$ & 10$^{8}$\msol & 11.1$^{+4.4}_{-2.7}$ & 2.66$^{+0.74}_{-0.80}$ & 1.72$\pm$0.31 \\
GDR=$M_{\rm gas}$/$M_{\rm dust}$ & & 65$^{+20}_{-25}$ & 215$^{+65}_{-65}$ & 130$\pm$50 \\
$M_{\rm dyn}$ & 10$^{10}$\msol & 8.6$^{+3.6}_{-2.8}$ & 9.7$\pm$1.6 & 7.5$\pm$3.7 \\
$f_{\rm gas}$=$M_{\rm gas}$/$M_{\rm dyn}$ & & 83\%$^{+29\%}_{-36\%}$ & 59\%$\pm$11\% & 29\%$\pm$18\% \\
$\alpha_{\rm CO}^{\rm dyn}$ & \msol\,(K km\,s$^{-1}$\,pc$^2$)$^{-1}$ & $<$1.2 & $<$1.7 & $<$3.4 \\
$\tau_{\rm dep}$=$M_{\rm gas}$/SFR$_{\rm IR}$ & Myr & 70$\pm$15 & 22$\pm$7 & 40$\pm$15 \\
$L_{\rm CII}$/$L_{\rm FIR}$ & 10$^{-3}$ & 2.5$\pm$0.5 & 0.6$\pm$0.1 & 1.3$\pm$0.3 \\ 
$L_{\rm CII}$/$L_{\rm CO(1-0)}$ & & 4150$\pm$650 & 2400$\pm$300 & 4500$\pm$1800 \\
\enddata
\end{deluxetable*}
\end{figure}


\subsection{Gas Masses, Gas Surface Densities, Gas Fractions, Gas-to-Dust Ratios, Gas Depletion Times, and Conversion Factor}

Adopting the \aco\ fluxes from the LVG modeling and $\alpha_{\rm
  CO}$=1\,\msol\,(\lprime )$^{-1}$ to convert the resulting
\aco\ luminosities to molecular gas masses $M_{\rm gas}$, we find
$M_{\rm gas}$=(7.1$\pm$0.9), (5.7$\pm$0.5), and
(2.2$\pm$0.8)$\times$10$^{10}$\,\msol\ for GN10, AzTEC-3, and
HDF\,850.1,\footnote{Gas and dust masses and sizes for HDF\,850.1 are
  corrected by a factor of $\mu_{\rm L}$=1.6 to account for
  gravitational lensing magnification (Neri et
  al.\ \citeyear{neri14}).} respectively, where the overall
uncertainties are dominated by systematic uncertainties in
$\alpha_{\rm CO}$.\footnote{Statistical uncertainties in $r_{21}$ from
  the LVG modeling likely contribute only at the 10\%--15\% level.}
For GN10, this corresponds to 83\%$^{+29\%}_{-36\%}$ of the dynamical
mass estimate, which is representative under the assumption that the
\aco\ emission is as extended as the [C{\sc ii}] emission on which the
dynamical modeling was carried out. Adopting the dynamical mass
estimate of $M_{\rm dyn}$=(9.7$\pm$1.6)$\times$10$^{10}$\,\msol\ found
for AzTEC-3 by Riechers et al.\ (\citeyear{riechers14b}), its updated
gas mass corresponds to 59\%$\pm$11\% of the dynamical mass under the
same assumptions. Using an isotropic virial estimator (e.g., Engel et
al.\ \citeyear{engel10}) and the [C{\sc ii}] sizes and line widths of
its two components derived by Neri et al.\ (\citeyear{neri14}), we
find a combined dynamical mass of $M_{\rm
  dyn}$=7.5$\times$10$^{10}$\,\msol\ for HDF850.1.\footnote{The
  uncertainties of this estimate are dominated by systematic effects
  due to the choice of fitting method, which corresponds to at least
  50\%.}  As such, its gas mass corresponds to 29\%$\pm$18\% of the
dynamical mass under the same assumptions as for GN10.  Conversely,
based on their observed $L^\prime_{\rm CO}$, the dynamical masses of
GN10, AzTEC-3, and HDF\,850.1 at face value imply upper limits of
$\alpha_{\rm CO}$$<$1.2, $<$1.7, and $<$3.4\,\msol\,(\lprime )$^{-1}$,
respectively. For HDF\,850.1, the size of the gas reservoir measured
from the high-resolution \bco\ observations implies a molecular gas
mass surface density of $\Sigma_{\rm
  gas}$=(1.3$\pm$0.5)$\times$10$^{3}$\,\msol\,pc$^{-2}$. This suggests
that HDF\,850.1 follows the spatially-resolved star formation law,
which has been determined at lower redshift (e.g., Hodge et
al.\ \citeyear{hodge15}).

Given the dust masses of (11.1$^{+4.4}_{-2.7}$),
(2.66$^{+0.74}_{-0.80}$), and
(1.72$\pm$0.31)$\times$10$^{8}$\,\msol\ for GN10, AzTEC-3, and
HDF\,850.1 (this work; Riechers et al.\ \citeyear{riechers14b}; Walter
et al.\ \citeyear{walter12b}), the gas masses correspond to
gas-to-dust ratios of $\sim$65$^{+20}_{-25}$, 215$^{+65}_{-65}$, and
130$\pm$50, respectively, which are in the range of values found for
nearby infrared-luminous galaxies (e.g., Wilson et
al.\ \citeyear{wilson08}). Given the SFRs of (1030$^{+190}_{-150}$),
(2500$\pm$700), and (870$\pm$80)\,\msol\,yr$^{-1}$, (this work;
Riechers et al.\ \citeyear{riechers14b}; Walter et
al.\ \citeyear{walter12b}; Neri et al.\ \citeyear{neri14}), they also
yield gas depletion times of $\tau_{\rm dep}$=(70$\pm$15), (22$\pm$7),
and (40$\pm$15)\,Myr, respectively. This is consistent with short,
$\lesssim$100\,Myr starburst phases, as also typically found for
lower-redshift DSFGs (e.g., Simpson et al.\ \citeyear{simpson14};
Dudzeviciute et al.\ \citeyear{dudzeviciute20}), with the lowest value
found for the source showing the highest gas excitation.

\subsection{[C{\sc ii}]/FIR and [C{\sc ii}]/CO Luminosity Ratios}

To further investigate the difference in conditions for star formation
among the three COLDz $z$$>$5 DSFGs, we here consider their \cii/FIR
and \cii/\aco\ luminosity ratios (e.g., Stacey et
al.\ \citeyear{stacey91,stacey10}). We adopt the \aco\ fluxes from the
LVG modeling. We find [C{\sc ii}]/FIR ratios of $L_{\rm CII}$/$L_{\rm
  FIR}$=(2.5$\pm$0.5), (0.6$\pm$0.1), and
(1.3$\pm$0.3)$\times$10$^{-3}$ for GN10, AzTEC-3, and HDF\,850.1,
respectively, i.e., a factor of $\sim$4 variation. These values are
consistent with what is expected for luminous starburst galaxies
(e.g., Gracia-Carpio et al.\ \citeyear{gracia11}). The differences
between sources mirror those seen in the CO line ladders, indicating
lower [C{\sc ii}]/FIR ratios with increasing CO excitation. In GN10, the
\fco\ emission interestingly appears to show a comparable extent as
the 1.2\,mm continuum emission (see Fig.~\ref{f4}), and thus, is more
compact than the [C{\sc ii}] emission. Thus, the source-averaged CO
excitation may be comparatively low, but it may be significantly
higher in the nuclear region with the highest $\Sigma_{\rm
  SFR}$. Given the larger spatial extent of the [C{\sc ii}] emission, the
[C{\sc ii}]/FIR ratio likely also is substantially lower in this region than
the source average, consistent with the apparent global trend between
[C{\sc ii}]/FIR and CO excitation.

We also find [C{\sc ii}]/CO ratios of $L_{\rm CII}$/$L_{\rm
  CO(1-0)}$$\simeq$4150$\pm$650,\footnote{Adopting the measured
  \aco\ luminosity instead of that inferred from the LVG modeling
  would yield a ratio of $\simeq$5400 for GN10.} 2400$\pm$300, and
4500$\pm$1800 for GN10, AzTEC-3, and HDF\,850.1, respectively, i.e., a
factor of $\sim$2 variation. These values are consistent with
expectations for starburst galaxies at lower redshift (e.g., Stacey et
al.\ \citeyear{stacey10}, and references therein).\footnote{We caution
  that a direct comparison in terms of physical properties to lower
  redshifts (as typically done with photon dominated region models
  calculated at $z$=0) is not straight forward due to the reduction in
  the intensity of low-$J$ CO line emission at $z$$>$5 due to the
  warmer CMB.} The lowest ratio is found for AzTEC-3, i.e., the source
with the highest gas excitation (but the two lower-excitation sources
show comparable values), the highest dust temperature (see
Table~\ref{t5}), and the highest peak $\Sigma_{\rm SFR}$. This may
suggest a reduced [C{\sc ii}] line strength due to a stronger, more intense
interstellar radiation field, but it could also be due to dust
extinction of the [C{\sc ii}] line emission or a low brightness temperature
contrast between the [C{\sc ii}] and 158\,$\mu$m dust emission in the
nuclear starburst region.


\begin{figure}
\begin{deluxetable*}{ l c c c c c c c c c c }
\tabletypesize{\scriptsize}
\tablecaption{Properties of known $z$$>$5 DSFGs. \label{t5}}
\tablehead{
Name & redshift & lensed? & $\mu_{\rm L}$\tablenotemark{a} & selection\tablenotemark{b} & $S_{500}$ & $S_{870}$\tablenotemark{c} & $T_{\rm dust}$\tablenotemark{d} & $\mu_{\rm L}$$L_{\rm FIR}$\tablenotemark{e} & $L_{\rm FIR}$ & references \\
     &          &         &             &           & [mJy] & [mJy] & [K]          & [10$^{13}$\,\lsol ] & [10$^{12}$\,\lsol ] & }
\startdata
HXMM-30           & 5.094  & strongly & ...         & 250--500\,$\mu$m & 55$\pm$7      & 28$\pm$2      & ...                 & ...  & ... & 1, 2 \\
HELMS$\_$RED$\_$4 & 5.1612 & strongly & ...         & 250--500\,$\mu$m & 116.3$\pm$6.6 & 52.4$\pm$4.4  & 66.8$\pm$5.9        & 9.7$^{+2.3}_{-2.1}$ & ... & 3, 1 \\
{\bf HDF\,850.1}  & 5.1833 & weakly   & 1.6\tablenotemark{f} & 850\,$\mu$m      & $<$14         & 7.8$\pm$1.0   & 35$\pm$5            & 0.6$\pm$0.1 & 3.8$\pm$1.0 & 4, 5 \\
HLS0918           & 5.2430 & strongly & 8.9$\pm$1.9 & 250--500\,$\mu$m & 212$\pm$15    & 125$\pm$8     & 38$\pm$3            & 10.0$\pm$0.6 & 11.2$\pm$2.5 & 6 \\
HLock-102         & 5.2915 & strongly & 12.5$\pm$1.2 & 250--500\,$\mu$m & 140$\pm$7     & 55$\pm$4      & 54.7$\pm$5.0        & 9.9$\pm$1.2 & 7.9$\pm$1.2 & 7, 8, 1 \\
SPT\,2319--55     & 5.2929 & strongly & 6.9$\pm$0.6/13.9$\pm$1.8 & 1.4+2.0\,mm      & 49.0$\pm$6.6  & 38.1$\pm$2.9  & 42.1$\pm$2.1        & 2.9$^{+0.1}_{-0.2}$ & 3.7$\pm$0.8 & 9 \\
{\bf AzTEC-3}     & 5.2980 & no       & ---         & 1.1\,mm          & 14.4$\pm$8.0  & 8.7$\pm$1.5\tablenotemark{g}   & 92$^{+15}_{-16}$   & 1.1$\pm$0.2 & 11$\pm$2 & 10, 5 \\ 
{\bf GN10}        & 5.3031 & no       & ---         & 850\,$\mu$m      & 12.4$\pm$2.8  & 12.0$\pm$1.4  & 50.1$\pm$9.1        & 0.58$^{+0.11}_{-0.09}$ & 5.8$\pm$1.0 & 5 \\
SPT\,2353--50     & 5.576  & cluster  & ...         & 1.4+2.0\,mm      & 56.2$\pm$7.1  & 40.6$\pm$3.8  & 46.3$\pm$2.3        & 3.5$\pm$0.2 & ... & 9 \\
ADFS-27           & 5.6550 & weakly?  & ...         & 250--870\,$\mu$m & 24.0$\pm$2.7  & 25.4$\pm$1.8  & 55.3$^{+7.8}_{-7.6}$   & 1.6$\pm$0.3 & $\lesssim$16$\pm$3 & 11 \\
SPT\,0346--52     & 5.6559 & strongly & 5.6$\pm$0.1 & 1.4+2.0\,mm      & 203.7$\pm$8.3 & 130.8$\pm$7.6 & 50.5$\pm$1.9        & 13.1$^{+0.3}_{-0.6}$ & 23$\pm$1 & 9 \\
CRLE              & 5.6666 & weakly   & 1.09$\pm$0.02 & serendipitous\tablenotemark{h} & 31.1$\pm$1.4  & 16.7$\pm$2.0  & 41.2$^{+6.3}_{-2.2}$   & 1.6$\pm$0.1 & 15$\pm$1 & 12 \\
SPT\,0243--49     & 5.699  & strongly & 6.7$\pm$0.5/3.1$\pm$0.1 & 1.4+2.0\,mm      & 57.5$\pm$6.9  & 84.5$\pm$5.0  & 32.7$\pm$1.6        & 3.7$\pm$0.2 & 7.3$\pm$1.7 & 9 \\
SPT\,2351--57     & 5.811  & cluster  & ...         & 1.4+2.0\,mm      & 73.8$\pm$5.7  & 34.6$\pm$3.1  & 53.5$\pm$2.8        & 4.6$^{+0.2}_{-0.3}$ & ... & 9 \\
ID\,85001929      & 5.847  & no?      & ...         & SED template\tablenotemark{i} & 14.6$\pm$2.1   & 5.3$\pm$0.8  & 59.0$^{+7.7}_{-16.7}$\tablenotemark{j} & 0.58$^{+0.08}_{-0.07}$\tablenotemark{j} & 5.8$\pm$0.8 & 13 \\
HeLMS-54ab        & 5.880  & no?      & ...         & 250--500\,$\mu$m & 97$\pm$9\tablenotemark{k} & 36$\pm$4\tablenotemark{k} & ...                 & ...  & ... & 1, 2 \\
G09-83808         & 6.0269 & strongly & 8.2$\pm$0.3 & 250--500\,$\mu$m & 44.0$\pm$8.2  & 36.2$\pm$9.1  & 35.9$\pm$1.5        & 2.1$\pm$0.3 & 2.6$\pm$0.4 & 14, 15 \\
HFLS3             & 6.3369 & weakly   & 1.8$\pm$0.6\tablenotemark{l} & 250--500\,$\mu$m & 47.3$\pm$2.8   & 33.0$\pm$2.4  & 55.9$^{+9.3}_{-12.0}$ & 2.9$\pm$0.3 & 16$\pm$5 & 7, 1 \\
SPT\,0311--58     & 6.900  & weakly   & 2.0\tablenotemark{f} & 1.4+2.0\,mm      & 51.8$\pm$8.2  & 36.9$\pm$7.4  & 45.6$\pm$3.3        & 4.4$^{+0.4}_{-0.3}$ & 22$\pm$5 & 16
\enddata
\tablecomments{Uncertainties in $T_{\rm dust}$ are the statistical uncertainties from the fitting procedures adopted by different authors and do not account for systematic uncertainties due to differences between the procedures (see references provided for additional details). The method adopted here for GN10 is virtually identical to those used for fitting HELMS$\_$RED$\_$4, HLock-102, AzTEC-3, ADFS-27, CRLE, ID\,85001929, and HFLS3.}
\tablenotetext{\rm a}{Lensing magnification factor. Modeled with two source components when multiple values are given. The uncertainties of full SED-based quantities for strongly-lensed sources (i.e., those with $\mu_{\rm L}$$>$2 and/or evidence for multiple lensed images) may be limited by systematic effects, given that magnification factors are typically derived at a single wavelength only, and that the regions within the galaxies that are brightest at the selection wavelength could be preferentially magnified.}
\tablenotetext{\rm b}{Sources identified as ``red'' between the 250--500\,$\mu$m bands were typically preferentially followed up if they showed high 850\,$\mu$m--1.3\,mm fluxes.}
\tablenotetext{\rm c}{$S_{850}$ or $S_{890}$ are used where $S_{870}$ is not available.}
\tablenotetext{\rm d}{Dust temperatures with small uncertainties typically are due to keeping $\beta_{\rm IR}$, $\lambda_0$, or both fixed in the fitting process. HLS0918 was fitted with an optically-thin model only, which may underestimate $T_{\rm dust}$. HDF\,850.1 was fitted with a \textsc{magphys}-based model, and its dust peak is only constrained by upper limits shortward of 850\,$\mu$m.}
\tablenotetext{\rm e}{Apparent far-infrared luminosity, i.e., not corrected for lensing magnification where applicable.}
\tablenotetext{\rm f}{No uncertainties are quoted in the original works. Here we assume 20\% uncertainty for the calculation of derived quantities.}
\tablenotetext{\rm g}{Simpson et al.\ (\citeyear{simpson20}) report an updated value of 8.3$\pm$0.3\,mJy, which is consistent with the original measurement.}
\tablenotetext{\rm h}{Source was also independently identified as ``red'' at 250--500\,$\mu$m (D.\ Riechers et al., in prep.).}
\tablenotetext{\rm i}{Source was selected at 850\,$\mu$m (originally reported at 1.2\,mm; Bertoldi et al.\ \citeyear{bertoldi07}), but followed up based on a photometric redshift obtained by fitting a template SED based on HFLS3 (Riechers et al.\ \citeyear{riechers13b}).}
\tablenotetext{\rm j}{Values obtained by fitting the de-blended fluxes at 100\,$\mu$m--3\,mm using a method virtually identical to that adopted for GN10. The $T_{\rm dust}$ is compatible with the value of 61$\pm$8\,K reported by Jin et al.\ (\citeyear{jin19}).}
\tablenotetext{\rm k}{Blended with a lower-$z$ source that contributes $\sim$30\% of the flux at 870\,$\mu$m, and likely a higher fraction at 500\,$\mu$m.}
\tablenotetext{\rm l}{Updated value based on visibility-plane lens modeling of the dust and gas emission.}
\tablerefs{${}$[1] D.\ Riechers et al., in prep.; [2] Oteo et al.\ (\citeyear{oteo17}); [3] Asboth et al.\ (\citeyear{asboth16}); [4] Walter et al.\ (\citeyear{walter12b}); [5] this work; [6] Rawle et al.\ (\citeyear{rawle14}); [7] Riechers et al.\ (\citeyear{riechers13b}); [8] Dowell et al.\ (\citeyear{dowell14}); [9] Strandet et al.\ (\citeyear{strandet16}); [10] Riechers et al.\ (\citeyear{riechers14b}); [11] Riechers et al. (\citeyear{riechers17}); [12] Pavesi et al.\ (\citeyear{pavesi18a}); [13] Jin et al.\ (\citeyear{jin19}); [14] Fudamoto et al.\ (\citeyear{fudamoto17}); [15] Zavala et al.\ (\citeyear{zavala18}); [16] Strandet et al.\ (\citeyear{strandet17}).}
\end{deluxetable*}
\end{figure}


\begin{figure*}
  \epsscale{1.15}
  \plotone{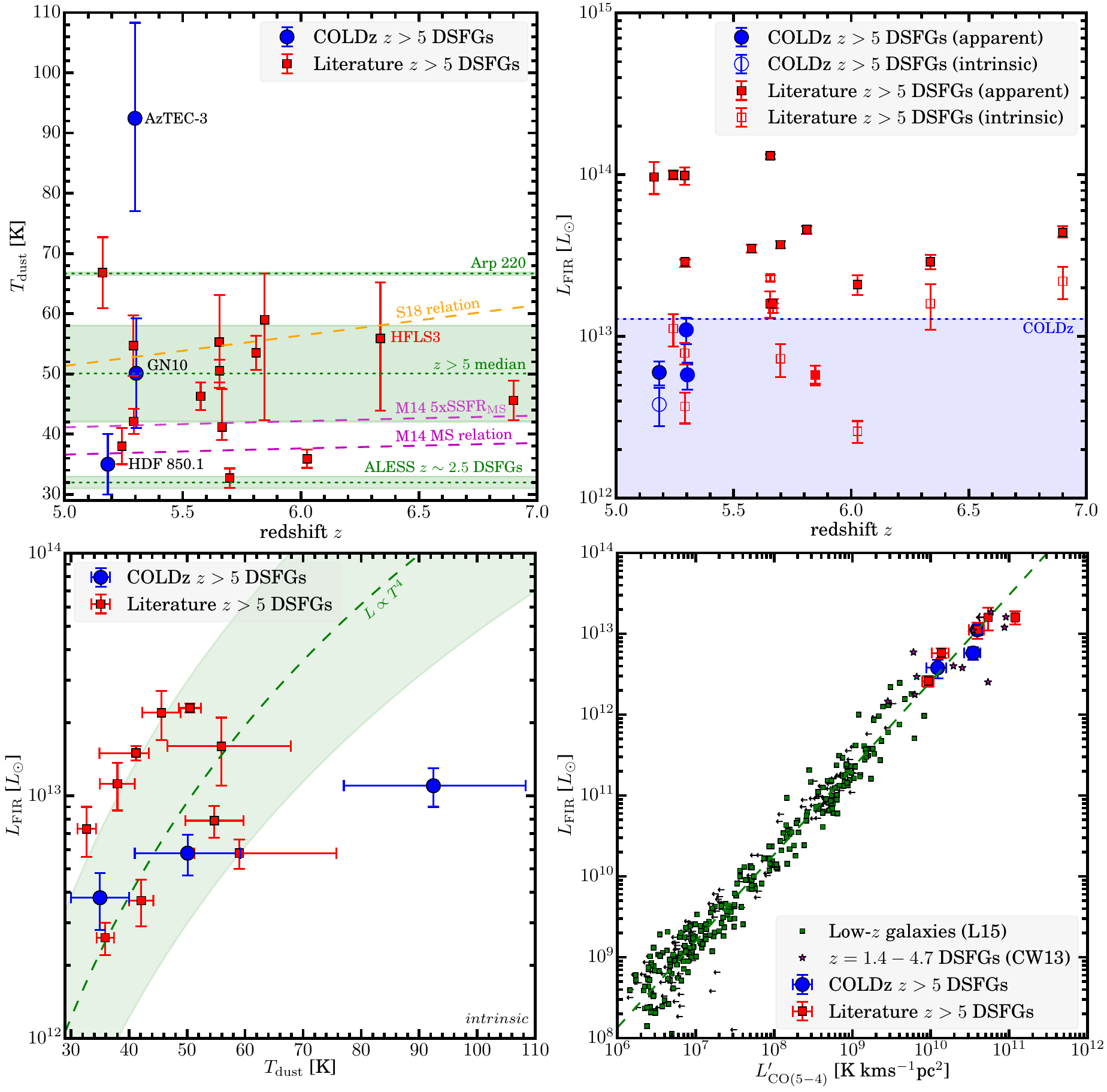}
\vspace{-2mm}
  
\caption{Dust temperature ({\em top left}) and far infrared luminosity
  ({\em top right}) of COLDz $z$$>$5 DSFGs, compared to literature
  DSFGs, showing that they cover a broad range in dust temperatures,
  and that they are less luminous than other DSFGs known at $z$$>$5
  and probe a luminosity range that is only accessible through
  gravitational lensing in other current samples with
    spectroscopic redshifts (dotted dividing line in {\em top right}
  panel; see Table~\ref{t5} for references), with the exception of one
  source discovered recently from de-blended single-dish catalogs (Jin
  et al.\ \citeyear{jin19}). There is a weak trend between dust
  temperature and far infrared luminosity within the current sample,
  which is largely consistent with a standard $L \propto T^4$ scaling
  (dashed line in {\em bottom left} panel, shown scaled to the median
  values and with $\pm$0.5\,dex scatter as the shaded region for
  reference).  The dotted lines and shaded regions in the {\em
    top left} panel show representative dust temperatures and
    uncertainty ranges for ALESS $z$$\sim$2.5 DSFGs (Swinbank et
  al.\ \citeyear{swinbank14}), the $z$$>$5 sample median, and the
  nearby dusty starburst Arp\,220. The dashed orange line shows the
  relation proposed by Schreiber et al.\ (\citeyear{schreiber18}). The
  lower and upper dashed magenta lines show the relation proposed by
  Magnelli et al.\ (\citeyear{magnelli14}) for ``main sequence'' (MS)
  galaxies, and for galaxies with specific star-formation rates
  (SSFRs) 5$\times$ higher than the MS, respectively. The COLDz
  $z$$>$5 DSFGs also closely follow the \eco\ -- FIR luminosity
  relation of nearby galaxies ({\em bottom right}; green symbols,
  upper limit arrows without symbols, and dashed line show the {\em
    Herschel}/SPIRE spectroscopy sample and relation by Liu et
  al.\ \citeyear{liu15}), suggesting that the properties of the dense,
  warm gas in their star-forming regions are as expected for starburst
  galaxies. Magenta stars show a lower redshift comparison DSFG
  sample, updated from the compilation of Carilli \& Walter
  (\citeyear{cw13}), featuring data from Andreani et
  al.\ (\citeyear{andreani00}), Wei\ss\ et
  al.\ (\citeyear{weiss05c,weiss09}), Carilli et
  al.\ (\citeyear{carilli10b}), Riechers et
  al.\ (\citeyear{riechers11c,riechers11d}), Cox et
  al.\ (\citeyear{cox11}), Danielson et al.\ (\citeyear{danielson11}),
  and Salome et al.\ (\citeyear{salome12}). Values are corrected for
  gravitational magnification unless mentioned otherwise. \label{f8}}
%
\end{figure*}

\subsection{Dust Temperatures}

We find that the COLDz $z$$>$5 DSFGs exhibit a broad range in dust
temperatures. HDF\,850.1 ($z$=5.18) has a comparatively low $T_{\rm
  dust}$=(35$\pm$5)\,K (Walter et al.\ \citeyear{walter12b}), while
GN10 ($z$=5.30) shows a moderate $T_{\rm dust}$=(50$\pm$9)\,K,
especially when compared to the relatively high $T_{\rm
  dust}$=(92$^{+15}_{-16}$)\,K displayed by AzTEC-3 ($z$=5.30;
Riechers et al.\ \citeyear{riechers14b}).\footnote{Note that an
  optically-thin fitting procedure suggests a more modest dust
  temperature of (53$\pm$5)\,K compared to the general fitting result
  adopted here, but it also yields a worse fit to the data.}

It is interesting to place these galaxies into the more general
context of all currently known $z$$>$5 DSFGs available in the
literature (see Table~\ref{t5} and Fig.~\ref{f8}, top
left).\footnote{Jin et al.\ (\citeyear{jin19}) report ID\,20010161, a
  source with a candidate redshift of $z$=5.051 based on a single
  emission line. While the identification is plausible, we do not
  include it in the current data compilation until the redshift is
  confirmed.}  The full $z$$>$5 literature sample contains other
  sources in the same category as HDF\,850.1 with relatively modest
$T_{\rm dust}$=33--46\,K such as CRLE ($z$=5.67; Pavesi et
al.\ \citeyear{pavesi18a}) and the gravitationally lensed HLS0918
($z$=5.24; Rawle et al.\ \citeyear{rawle14}), SPT\,2319--55, 2353--50,
0243--49, and 0311--58 ($z$=5.29, 5.58, 5.70, and 6.90; Strandet et
al.\ \citeyear{strandet16,strandet17}), and G09-83808 ($z$=6.03;
Fudamoto et al.\ \citeyear{fudamoto17}), sources with higher $T_{\rm
  dust}$=50--59\,K like GN10 such as ADFS-27 ($z$=5.66; Riechers
et al.\ \citeyear{riechers17}), ID\,85001929 ($z$=5.85; Jin et
al.\ \citeyear{jin19}), HFLS3 ($z$=6.34; Riechers et
al.\ \citeyear{riechers13b}), and the gravitationally lensed HLock-102
($z$=5.29; Riechers et al.\ \citeyear{riechers13b}; Dowell et
al.\ \citeyear{dowell14}) and SPT\,0346--52 and 2351--57 ($z$=5.66 and
5.81; Strandet et al.\ \citeyear{strandet16}), and sources with
  high $T_{\rm dust}$$>$60\,K closer to AzTEC-3 like the
strongly-lensed HELMS$\_$RED$\_$4 ($z$=5.16; $T_{\rm
  dust}$=67$\pm$6\,K; Asboth et al.\ \citeyear{asboth16}; D.~Riechers
et al., in prep.).

All currently known $z$$>$5 DSFGs have relatively high dust
temperatures (median value:\ 50.1$\pm$8.0\,K based on 17
galaxies,\footnote{Uncertainties are given as the median absolute
  deviation.}  with an average value of 50.3\,K) in comparison to the
bulk of the population at $z$$\sim$2--3 (32$\pm$1\,K, i.e.,
$\sim$1.57$\pm$0.25 times lower; e.g., Swinbank et
al.\ \citeyear{swinbank14}; see also Simpson et
al.\ \citeyear{simpson17a}; Dudzeviciute et
al.\ \citeyear{dudzeviciute20}).\footnote{Luminosity-averaged $T_{\rm
    dust}$ from energy balance modeling for ALESS galaxies are higher,
  with an average of (43$\pm$2)\,K (da Cunha et
  al.\ \citeyear{dacunha15}). We here adopt those from Swinbank et
  al., since their derivation is more consistent with the methods used
  for the $z$$>$5 DSFG sample.} Given the heterogenous selection and
some differences in the fitting methods used, we investigate in the
following to what degree they could be responsible for this
difference, but we find no trends that would be sufficient to explain
this difference. We then explore if the higher $T_{\rm dust}$ in the
$z$$>$5 sample is due to a trend with redshift or $L_{\rm FIR}$, and
find that the latter is likely sufficient to explain the observed
differences to lower-redshift samples.

\subsubsection{Potential Biases due to Gravitational Lensing, Selection Wavelength, or SED Fitting Method}

Removing all strongly-lensed and cluster-lensed galaxies from the full
$z$$>$5 sample to investigate potential differential lensing bias, we
find a median value of (52.7$\pm$6.7)\,K based on 8 galaxies. This
value is higher than the full sample median, and thus, does not
provide direct evidence for preferential magnification of higher
$T_{\rm dust}$ regions in the lensed subsample. Removing all galaxies
selected at wavelengths shorter than 850\,$\mu$m to investigate
potential SED peak selection bias, we find (46.0$\pm$4.7)\,K based on
10 galaxies. We thus do not find significant evidence that the
different selection methods strongly bias the median $T_{\rm dust}$
for this $z$$>$5 sample. Comparing sources that used the same SED
fitting method as GN10 and AzTEC-3 (8 galaxies; see Table~\ref{t5}
caption for details) to the SPT sample (6 galaxies) which was fitted
with a common but somewhat different method (using greybody-only fits
with fixed $\beta_{\rm IR}$=2 and $\lambda_0$=100\,$\mu$m; e.g.,
Wei\ss\ et al.\ \citeyear{weiss13}), we find median values of
(55.6$\pm$4.5) and (46.0$\pm$4.2)\,K, respectively. The former
subsample is selected at shorter wavelengths (except one serendipitous
discovery), while the latter contains a higher fraction of strongly-
and cluster-lensed systems (25\% vs.\ 83\%). Thus, the different
median $T_{\rm dust}$ cannot be directly attributed to differences in
the fitting methods, but our study would be consistent with it playing
a role. Overall, we thus find that the median value appears not to be
significantly biased towards high values based on the heterogenous
composition of the sample, while keeping in mind that even the
combination of all current selection methods could entirely miss
$z$$>$5 DSFGs in undersampled regions of the parameter space.

\subsubsection{Comparison to Proposed $T_{\rm dust}$--$z$ Relations}

Using the median redshift of the sample of 5.66, we find expected
median values of $T_{\rm dust}$ of 37.3 and 54.6\,K when applying the
$T_{\rm dust}$--$z$ relations by Magnelli et
al.\ (\citeyear{magnelli14}; their Equation 4) and Schreiber et
al.\ (\citeyear{schreiber18}; their Equation 15)\footnote{Value
  obtained after scaling by their Equation 6 to obtain modified
  black-body temperatures.} for galaxies on the star-forming ``main
sequence'' (Fig.~\ref{f8}, top left). At face value, the Magnelli et
al.\ relation appears consistent with the lowest $T_{\rm dust}$
sources, but the $z$$>$5 sample would need to exhibit median
star-formation rates that are by a factor of $\sim$90 higher
($\sim$240 when only considering unlensed and weakly-lensed systems)
than those of ``main sequence'' galaxies at the same redshift with the
same $M_\star$ for this relation to agree with the median $T_{\rm
  dust}$ value of our sample. While most of the current sample is
likely to be in excess of the ``main sequence'', there is no evidence
for a difference by more than a factor of several in SFR compared to
the ``main sequence''. As such, current observations of $z$$>$5 DSFGs
would appear to be in favor of a stronger evolution in $T_{\rm
  dust}$ with redshift than indicated by this relation when
  assuming that redshift is the main property responsible for
  explaining the observed differences, unless systematic effects in
the sample selection and SED fitting methods are more significant than
currently known. On the other hand, a significant fraction of the
$z$$>$5 sample appears to be in reasonable agreement with expectations
based on the Schreiber et al.\ relation, but the predicted median
value is higher than observed for the sample as a whole.\footnote{The
  relation also suggests 38.7\,K at $z$$\sim$2.5, which is $\sim$20\%
  higher than the ALESS sample median.} At least half the sample have
lower $T_{\rm dust}$ than expected from this relation.

Both of the proposed $T_{\rm dust}$--$z$ relations have to be
extrapolated significantly in redshift to match our sample, such that
the observed mismatches are likely consistent with the true
uncertainties of these relations. In particular, the $T_{\rm dust}$ of
``main sequence'' galaxies at $z$$>$5 is currently only poorly
constrained by observations (e.g., Pavesi et
al.\ \citeyear{pavesi16}), such that it remains unclear that an offset
in SFR from the ``main sequence'' can currently be meaningfully
translated to expected differences in $T_{\rm dust}$ for dusty
starbursts at these redshifts. Also, the increasing CMB temperature
toward $z$$>$5 provides a natural cutoff at low $T_{\rm dust}$, both
due to a reduced brightness temperature contrast and due to a higher
contribution to dust heating (e.g., da Cunha et
al.\ \citeyear{dacunha13b}). This effect leads to an overall change in
SED shape and increase in the measured $T_{\rm dust}$ with redshift,
but is not taken into account in current extrapolations of the $T_{\rm
  dust}$--$z$ relations. On the other hand, it has recently been
suggested that previously proposed $T_{\rm dust}$--$z$ relations could
be an observational artifact due to selection effects (Dudzeviciute et
al.\ \citeyear{dudzeviciute20}). If there were to be no trend in
$T_{\rm dust}$ with redshift, another explanation would be required to
explain the high median $T_{\rm dust}$ of the $z$$>$5 sample.

\subsubsection{Investigation of $T_{\rm dust}$--$L_{\rm FIR}$ Relations}

We find a trend between $T_{\rm dust}$ and $L_{\rm FIR}$ in our sample
(Fig.~\ref{f8}), which spans about an order of magnitude in intrinsic
$L_{\rm FIR}$, and a factor of 2.8 in $T_{\rm dust}$.  The data are
consistent with an overall increase of $L_{\rm FIR}$ with $T_{\rm
  dust}$. For $z$$>$5 DSFGs with $T_{\rm dust}$$<$50.1\,K (i.e., below
the sample median), we find a median $L_{\rm
  FIR}$=(7.3$\pm$3.9)$\times$10$^{12}$\,\lsol, but for those with
$>$50.1\,K, we find (11.0$\pm$5.0)$\times$10$^{12}$\,\lsol, as is
consistent with the general trends found for lower-redshift DSFG
samples (e.g., da Cunha et al.\ \citeyear{dacunha15}; Simpson et
al.\ \citeyear{simpson17a}; Dudzeviciute et
al.\ \citeyear{dudzeviciute20}).  Taken at face value, this trend may
at least partially, and perhaps entirely explain the high median
$T_{\rm dust}$ of the current $z$$>$5 DSFG sample as being due to
their relatively high median $L_{\rm FIR}$. Indeed, the three
intrinsically least luminous sources in the sample with an average
$L_{\rm FIR}$=(3.4$\pm$0.7)$\times$10$^{12}$\,\lsol\ have an average
$T_{\rm dust}$=(38$\pm$4)\,K, which is much closer to lower-redshift
samples with comparable $L_{\rm FIR}$ like ALESS and
AS2UDS.\footnote{For reference, the $z$$\sim$2--3 ALESS sample has a
  median $L_{\rm FIR}$=(3.0$\pm$0.3)$\times$10$^{12}$\,\lsol\ at
  $T_{\rm dust}$=(32$\pm$1)\,K (Swinbank et
  al.\ \citeyear{swinbank14}).} In the simplest scenario, the
  higher $T_{\rm dust}$ in the $z$$>$5 DSFG sample thus could be due
  to their high median SFRs. This subject thus warrants further study
based on the detailed dust properties of larger galaxy samples in the
future.

\subsection{Number Densities and Space Densities}

The overall population of DSFGs was found to be sufficient to explain
the early formation of most luminous local and intermediate redshift
elliptical galaxies, under the assumption that DSFGs are their
intensely star-forming progenitors (e.g., Simpson et
al.\ \citeyear{simpson14}). On the other hand, the highest-redshift
luminous DSFGs are the likely progenitors of some of the most massive
compact ``quiescent'' galaxies at $z$$>$2 (e.g., Toft et
al.\ \citeyear{toft14}). As such, a comparison of their space
densities to such galaxies can help us to understand the duty cycles,
formation mechanisms and buildup of massive galaxies through cosmic
history. Thus, it is important to better understand how reliable
current constraints on their space densities are. The COLDz survey
provides a unique selection method that significantly differs from
traditional DSFG surveys, which provides new insights compared to
previous estimates.

Searching the COLDz data, we have detected three luminous $z$$>$5
DSFGs in \bco\ emission in an area of $\sim$60\,arcmin$^2$, compared
to four independently-confirmed sources detected in \aco\ at
$z$=2.0--2.8 (one of which is a DSFG) in the same area. On the one
hand, the comoving volume covered by the survey is $\sim$6.4 times
larger for searches of \bco\ emission at $z$=4.9--6.7 compared to
those of \aco, at a comparable excitation-corrected $L^\prime_{\rm
  CO}$ limit (P18, R19). Also, one of the fields was chosen to include
AzTEC-3 at $z$=5.298, which thus needs to be removed from comparisons
to avoid potential bias.\footnote{The redshift of HDF\,850.1 was known
  at the time the COLDz survey was carried out, but the field was
  chosen based on the availability of the deep {\em HST}/WFC3-IR
  CANDELS data, not the presence of this source. Thus, it is not
  excluded from the space density discussion, since it should not
  introduce a bias.} As such, there are significantly fewer
serendipitously discovered CO-rich galaxies per unit volume in the
higher-redshift bin. On the other hand, once AzTEC-3 is removed from
further consideration, the two other sources alone still imply a
significant excess in the CO luminosity function at $z$$\sim$5--7
compared to (admittedly uncertain) model expectations (R19; their
Fig.~4). Also, given other recent studies such as the serendipitous
discovery of another luminous $z$$>$5 DSFG in the COSMOS field
(Pavesi et al.\ \citeyear{pavesi18a}), such sources may indeed be more
common than expected based on previous predictions. As such, a closer
look at these expectations is warranted.

\paragraph{COLDz Number Density and Space Density}

Conservatively restricting the analysis to the GOODS-North field
(i.e., excluding AzTEC-3 and its environment), the two detected
sources alone correspond to a number density of
$\sim$(150$\pm$100)\,deg$^{-2}$ for $z$$>$5 DSFGs, or a space density
of (1.0$\pm$0.7)$\times$10$^{-5}$\,Mpc$^{-3}$, within the volume
probed in \bco\ emission by COLDz ($\sim$200,000\,Mpc$^3$ in
GOODS-North alone; R19).\footnote{Quoted uncertainties here and in the
  following are the standard deviation of a Poisson distribution given
  the number of sources detected.}

\begin{figure}
\epsscale{1.15}
\plotone{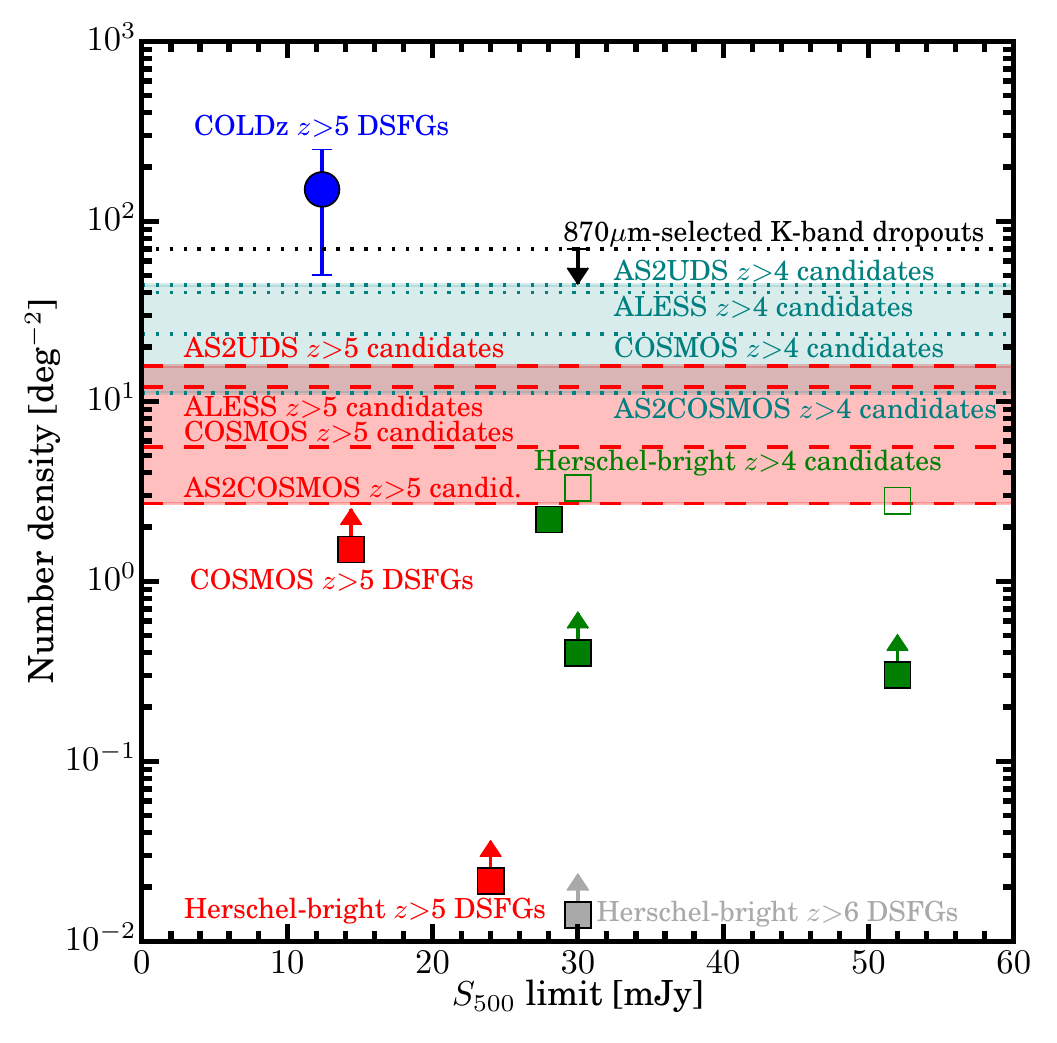}
\vspace{-2mm}

\caption{Number densities of high-redshift dusty galaxy samples down
  to a given 500\,$\mu$m flux limit, showing that the values found for
  COLDz $z$$>$5 DSFGs (blue points; GOODS-North field only) are high
  even when compared to lower-$z$ DSFG samples. Green and gray squares
  show red {\em Herschel} source samples identified by Riechers et
  al.\ (\citeyear{riechers13b}), Dowell et al.\ (\citeyear{dowell14}),
  Asboth et al.\ (\citeyear{asboth16}), and Ivison et
  al.\ (\citeyear{ivison16}). Open symbols show the full samples, and
  filled symbols show estimated fractions or lower limits of $z$$>$4
  ($z$$>$6) sources where applicable. Red symbols show lower limits at
  $z$$>$5 based on the spectroscopically confirmed sources in
  Table~\ref{t5}, using the full size of the COSMOS field (2\,deg$^2$)
  and the HerMES fields from which sources were selected (325\,deg$^2$
  total). The dotted black line shows ``K-band dropout'' galaxies
  identified at 870\,$\mu$m for reference (Dudzeviciute et
  al.\ \citeyear{dudzeviciute20}), many of which are likely at
  $z$$\sim$3--4 (line thus is an upper limit for $z$$>$4). Dotted teal
  and dashed red lines and shaded areas are candidate $z$$>$4 and
  $z$$>$5 DSFGs in the COSMOS, ALESS, AS2UDS, and
    $S_{870}$$>$6.2\,mJy AS2COSMOS samples, selected at 1.1\,mm,
  870\,$\mu$m, 850, and 850\,$\mu$m, respectively (Simpson
    et al.\ \citeyear{simpson14,simpson20}; Miettinen et
  al.\ \citeyear{miettinen17}; Dudzeviciute et
  al.\ \citeyear{dudzeviciute20}). The COLDz and COSMOS samples are
  not selected at 500\,$\mu$m; the symbols are shown at the lowest
  detected 500\,$\mu$m flux in the samples instead of a formal flux
  limit. \label{f10}}
%
\end{figure}

\paragraph{Comparison to Bright Samples from Large-Area Surveys}

At bright flux levels (i.e., down to a limit of $S_{500}$$>$52\,mJy at
500\,$\mu$m), Asboth et al.\ (\citeyear{asboth16}) find 477 candidate
$z$$>$4 DSFGs in an area of 274\,deg$^2$ based on the selection of red
{\em Herschel} sources (i.e., sources with
$S_{250}$$<$$S_{350}$$<$$S_{500}$). Based on a power law fit to their
sample, they find a number density of 2.8\,deg$^{-2}$. Follow-up
observations of a subsample of 188 sources at longer wavelengths
suggest that at least 21, or $\sim$11\%, are highly probable to be at
$z$$>$4 (Duivenvoorden et al.\ \citeyear{duivenvoorden18}). This
corresponds to a number density of $>$0.3\,deg$^{-2}$. Applying a
similar selection down to a limit of $S_{500}$$>$30\,mJy over an area
of 21\,deg$^2$, Dowell et al.\ (\citeyear{dowell14}) find a number
density of 3.3\,deg$^{-2}$ based on 38 such sources, of which at least
$\sim$11\% are spectroscopically confirmed to be at $z$$>$4. This
corresponds to a number density of $>$0.4\,deg$^{-2}$. Based on the
$z$=6.34 DSFG HFLS3 alone, Riechers et al.\ (\citeyear{riechers13b})
estimate that the number density of such sources may drop to values as
low as 0.014\,deg$^{-2}$ at $z$$>$6.  For a sample of 109 out of 708
sources found with similar selection criteria
($S_{500}$/$S_{250}$$\geq$1.5 and $S_{500}$/$S_{350}$$\geq$0.85) and
28$<$$S_{500}$$<$73\,mJy (with the exception of one source at
$S_{500}$=118\,mJy) selected over $\sim$600\,deg$^2$, Ivison et
al.\ (\citeyear{ivison16}) find that about one-third of their
candidates (or $\sim$2.2\,deg$^{-2}$ when correcting for completeness)
are likely at $z$$>$4. This leads them to infer a space density of
$\sim$6$\times$10$^{-7}$\,Mpc$^{-3}$ for $z$$>$4 DSFGs after applying
a duty cycle correction. All of these estimates likely imply number
densities of $<$1\,deg$^{-2}$ for $z$$>$5 DSFGs down to
$S_{500}$$>$28\,mJy (Fig.~\ref{f10}). Indeed, the number density of
spectroscopically confirmed $z$$>$5 red {\em Herschel} sources in the
HerMES fields provides a lower limit of only
$>$(0.021$\pm$0.008)\,deg$^{-2}$ down to $S_{500}$$>$24\,mJy. In
addition, some simulations appear to suggest that the number densities
of red {\em Herschel} sources could be biased towards high values due
to selection effects introduced by the noise level and limited
resolution of the {\em Herschel} data (e.g., Bethermin et
al.\ \citeyear{bethermin17}). If significant, this would render
current space density estimates upper limits.

With $S_{500}$=(12.4$\pm$2.8), (14.4$\pm$8.0), and $<$14\,mJy for
GN10, AzTEC-3, and HDF\,850.1 (Table~\ref{t5}; Walter et
al.\ \citeyear{walter12b}; Riechers et al.\ \citeyear{riechers14b};
Liu et al.\ \citeyear{liu18}), respectively, all of the COLDz $z$$>$5
DSFGs would have been missed by all of these surveys by factors of
$>$2--4 in $S_{500}$ (Fig.~\ref{f10}; see also Fig.~\ref{f8} for a
comparison to other $z$$>$5 samples in $L_{\rm FIR}$).\footnote{These
  surveys however may contain strongly-lensed analogs of COLDz sources
  (Fig.~\ref{f8}).}  However, the more than two orders of magnitude
difference in the implied space and number densities likely either
suggests that such sources are overrepresented in the COLDz survey
area due to large scale structure even after neglecting AzTEC-3, that
the counts fall very steeply with flux in the
10$\lesssim$$S_{500}$$\lesssim$30\,mJy regime at $z$$>$5, or that
current selection methods underlying these previous estimates are more
incomplete than currently thought. Substantially larger survey areas
than accessible with surveys like COLDz would be required to
distinguish between these possibilities. Until such surveys are
available, further insight may be gained through a comparison to
samples selected over smaller areas, but down to lower flux levels,
than the bright samples discussed so far.

\paragraph{Comparison to Faint Samples}

Targeted studies of DSFG populations down to $S_{870}$$\gtrsim$1\,mJy
(e.g., Simpson et al.\ \citeyear{simpson14}; i.e., typically a factor
of a few fainter than our $z$$>$5 sample with
$S_{870/890}$=(12.0$\pm$1.4), (8.7$\pm$1.5), and (7.8$\pm$1.0)\,mJy
for GN10, AzTEC-3, and HDF\,850.1, respectively; Wang et
al.\ \citeyear{wang07}; Younger et al.\ \citeyear{younger07}; Cowie et
al.\ \citeyear{cowie09}; see Table~\ref{t5}) appear to indicate an
order of magnitude or more higher space densities
($>$5$\times$10$^{-6}$\,Mpc$^{-3}$; Cooke et
al.\ \citeyear{cooke18})\footnote{Also see Aravena et
  al.\ (\citeyear{aravena10c}) for comparable estimates at $z$=3--5
  based on single-dish 1.2\,mm-selected sources.}  for 4$<$$z$$<$5
DSFGs than found for the typically an order of magnitude or more
brighter Ivison et al.\ (\citeyear{ivison16}) sample.\footnote{The
  Ivison et al.\ (\citeyear{ivison16}) sample has single-dish fluxes
  of $S_{850}$=8--71\,mJy, including upper limits consistent with this
  range. Thus, it reaches down to the long wavelength fluxes of the
  COLDz $z$$>$5 sample, despite having substantially higher fluxes at
  its prime selection wavelength of 500\,$\mu$m.} This lower limit
provides a closer match to the space density implied by the COLDz data
at face value, but it likely still falls short given the lower
redshift and lower flux limit of this comparison
sample.\footnote{Simpson et al.\ (\citeyear{simpson17a}) estimate a
  comoving space density of $\sim$10$^{-5}$\,Mpc$^{-3}$ for AS2UDS
  DSFGs with a median $S_{870}$=(8.0$\pm$0.4)\,mJy. While more
  comparable in $S_{870}$ to the COLDz $z$$>$5 DSFG sample, only one
  of their sources has an estimated photometric redshift of $z$$>$5.}
On the other hand, interferometric follow-up surveys of flux-limited
DSFG samples such as ALESS and AS2UDS find number densities of
``K-band dropout'' sources of $\sim$0.02\,arcmin$^{-2}$, or
$\sim$70\,deg$^{-2}$ (Fig.~\ref{f10}; e.g., Simpson et
al.\ \citeyear{simpson14,simpson17a}; Dudzeviciute et
al.\ \citeyear{dudzeviciute20}; also see discussion of similar
sources, e.g., by Dannerbauer et al.\ \citeyear{dannerbauer04}; Frayer
et al.\ \citeyear{frayer04}; Franco et
al.\ \citeyear{franco18}). These sources may be considered analogs of
HDF\,850.1 and GN10 in some respects, but they typically have lower
870\,$\mu$m fluxes and currently mostly lack spectroscopic
confirmation and molecular gas mass measurements. Also, many of them
are consistent with $z$$\sim$3--4 (see discussion in Dudzeviciute et
al.\ \citeyear{dudzeviciute20}), such that the true number density of
``COLDz analogs'' among them is likely significantly
lower.\footnote{The Simpson et al.\ (\citeyear{simpson17a}) AS2UDS
  sample contains two ``K-band dropout'' sources with
  $S_{870}$$>$8.0\,mJy but no redshift estimates, which could be the
  closest analogs to sources like GN10 and HDF\,850.1.} Based on
largely photometric redshifts, the ALMA-COSMOS, ALESS, and AS2UDS
samples selected at 1.1\,mm, 870\,$\mu$m, and 850\,$\mu$m provide
space density estimates of (24$\pm$6), (40$\pm$13), and
(42$\pm$7)\,deg$^{-2}$ for candidate $z$$>$4 DSFGs, and (5.6$\pm$2.8),
(12$\pm$7), and (16$\pm$4)\,deg$^{-2}$ for $z$$>$5 candidates,
respectively (Simpson et al.\ \citeyear{simpson14}; Miettinen et
al.\ \citeyear{miettinen17}; Dudzeviciute et
al.\ \citeyear{dudzeviciute20}). Taking the full photometric redshift
probability functions into account results in space density estimates
of (11$\pm$1) and (2.7$\pm$0.4)\,deg$^2$ down to $S_{870}$$>$6.2\,mJy
at $z$$>$4 and $z$$>$5, respectively, based on the AS2COSMOS survey
(Simpson et al.\ \citeyear{simpson20}). These samples suggest that the
space densities of DSFGs decline by factors of $\sim$3--4 from $z$=4
to 5, and that at most $\sim$1--3\% of 850\,$\mu$m--1.1\,mm selected
DSFGs at the currently achieved flux limits\footnote{There appears to
  be a trend that the brightest $S_{870}$-selected DSFGs tend to be
  found at higher redshifts (e.g., Younger et
  al.\ \citeyear{younger07}; Simpson et al.\ \citeyear{simpson20}).}
appear to be at $z$$>$5. Moreover, spectroscopic follow-up of two of
the three $z$$>$5 candidates in the ALESS survey finds them to be at
4$<$$z$$<$5, reducing the true ALESS $z$$>$5 space density estimate by
a factor of three (Danielson et al.\ \citeyear{danielson17}). Adopting
instead the spectroscopically confirmed $z$$>$5 DSFGs found across the
full COSMOS field provides a lower limit on their space density of
only $>$(1.5$\pm$0.9)\,deg$^{-2}$, i.e., $\lesssim$3--4 times below
the estimate based on the Miettinen et al.\ (\citeyear{miettinen17})
sample. As such, the space density of $z$$>$5 DSFGs in COLDz appears
to remain high by factors of $\sim$6--55 when compared to faint DSFG
samples, but the differences are significantly reduced relative to the
comparison to bright DSFG samples. A higher completeness in
spectroscopic confirmation of faint DSFGs will be critical to allow
for a more detailed comparison, especially in the
$S_{870}$$\gtrsim$5\,mJy regime probed by the current COLDz $z$$>$5
detections.\footnote{Values after accounting for gravitational
  magnification of HDF\,850.1.}

\paragraph{Potential Role of Selection Effects}

A possible concern for the comparisons between COLDz and continuum
surveys (beyond the increased impact of gravitational lensing on the
brighter populations) is that samples selected at a single
sub/millimeter wavelength may be incomplete at a given $L_{\rm FIR}$
due to selection effects, in particular those related to dust
temperature or optical depth, which could lead to an underestimate in
the volume density of luminous $z>5$ DSFGs. On the one hand, samples
selected at short wavelengths, e.g., at 500\,$\mu$m (i.e.,
$\lesssim$80\,$\mu$m rest-frame at $z$$>$5) and below, could miss
sources with low dust temperatures or high dust optical depths at the
highest redshifts, for which the dust SED peak shifts to significantly
longer wavelengths than 500\,$\mu$m. An example of such sources are
``870\,$\mu$m risers'', which are required to either be extremely
luminous or strongly gravitationally lensed to remain detectable at
$\leq$500\,$\mu$m (Riechers et al.\ \citeyear{riechers17}). Indeed,
despite its moderately high dust temperature, GN10 has a 870\,$\mu$m
flux that is just below the ``870\,$\mu$m riser'' selection
criterion. On the other hand, samples selected at long wavelengths,
e.g., 2\,mm (i.e., $\lesssim$330\,$\mu$m rest-frame at $z$$>$5) and
above, may be anticipated to be more complete at the highest redshifts
(see, e.g., Staguhn et al.\ \citeyear{staguhn14}, or similar
discussion by Casey et al.\ \citeyear{casey18}), but they could miss
sources at high dust temperatures. As an example, a recent sensitive
2\,mm survey in COSMOS finds several dusty sources which are claimed
to be at a relatively high median redshift compared to shorter
wavelength selected samples as expected, but it missed AzTEC-3 (i.e.,
the highest significance CO detection in the COLDz survey and the most
distant DSFG currently known in the area surveyed at 2\,mm) due to its
relatively high dust temperature (Magnelli et
al.\ \citeyear{magnelli19}).\footnote{GN10 and HDF\,850.1 however are
  solidly and tentatively detected in a 2\,mm survey in GOODS-North,
  consistent with their lower dust temperatures (Staguhn et
  al.\ \citeyear{staguhn14}).} Surveys at 870\,$\mu$m (i.e.,
$\lesssim$150\,$\mu$m in the rest frame) are less incomplete due to
variations in $T_{\rm dust}$ at a given $L_{\rm FIR}$ at $z$$>$5 than
at lower redshift (see, e.g., discussion by Simpson et
al.\ \citeyear{simpson17a}; Dudzeviciute et
al.\ \citeyear{dudzeviciute20}), but they are also affected. Such
selection effects can be largely addressed by a comprehensive
multi-wavelength selection of DSFGs, but they can contribute to the
uncertainties in the current estimates of the space density of $z$$>$5
DSFGs.

\paragraph{Potential Contribution of Cosmic Variance due to Clustering}

In principle, the COLDz survey may have led to the identification of
an unexpectedly high number of $z$$>$5 DSFGs due to cosmic variance in
the survey fields. On the one hand, the fields containing AzTEC-3 and
HDF\,850.1 are known to contain overdensities in star-forming galaxies
(e.g., Capak et al.\ \citeyear{capak11}; Walter et
al.\ \citeyear{walter12b}). In particular, the AzTEC-3 proto-cluster
environment (which was removed from all comparisons of space
densities) corresponds to one of the highest matter density peaks
currently known in the very early universe (e.g., Capak et
al.\ \citeyear{capak11}; Smol{\v c}i\'c et
al.\ \citeyear{smolcic17}). Also, the redshift difference between the
close-by HDF\,850.1 and GN10 is only d$z$$\simeq$0.12 (i.e.,
$\Delta$$v$$\simeq$5800\,\kms ), which corresponds to a comoving
distance of 61.4\,Mpc. As such, there already is evidence for large
scale structure, which, in principle, could include both DSFGs in
GOODS-North. On the other hand, the redshift difference between GN10
and AzTEC-3 is only d$z$$\simeq$0.005 (i.e.,
$\Delta$$v$$\simeq$240\,\kms, or 2.5\,Mpc comoving distance), and
thus, much smaller, but they are located in widely different parts of
the sky. As such, a physical connection between these sources is not
considered to be possible.  Thus, the modest redshift difference
between GN10 and HDF\,850.1 does not necessarily imply a physical
connection (and indeed, would require a rather large correlation
length). Moreover, it remains plausible to infer that modest
overdensities such as the one associated with HDF\,850.1 may simply be
highlighted by the presence of DSFGs, and thus, a common feature of
the environments of such sources at $z$$>$5 (see also, e.g., Pavesi et
al.\ \citeyear{pavesi18b}), rather than representing regions on the
sky that contain unusually high densities of DSFGs (which are also
known to exist at $z$$>$4, but perhaps are much less common; e.g.,
Oteo et al.\ \citeyear{oteo18}; Miller et al.\ \citeyear{miller18},
but also see Robson et al.\ \citeyear{robson14}). From these
considerations, it remains unclear that the field selection alone is
sufficient to explain the apparent excess in CO-bright galaxies at
$z$$>$5 in the COLDz volume.

\section{Summary and Conclusions} \label{sec:conclusions}

We have detected \bco\ emission toward three $z$$>$5 massive dusty
starburst galaxies by searching the $\sim$60\,arcmin$^2$ VLA COLDz
survey data (see P18, R19, for a complete description), including a
new secure redshift identification of the ``optically-dark'' source
GN10 at $z$=5.303. Despite star-formation rates of
$\sim$500--1000\,\msol\,yr$^{-1}$, two of the sources (which are
separated by only $\sim$5$'$ on the sky) remain undetected at
rest-frame ultraviolet to optical wavelengths, below observed-frame
3.6\,$\mu$m, due to dust obscuration. Molecular line scans such as
COLDz thus are an ideal method to determine the redshifts for such
sources, which will remain challenging to study in their stellar light
at least until the launch of the {\em James Webb Space Telescope}
({\em JWST}).

By carrying out a multi-wavelength analysis including new NOEMA and
VLA observations of GN10, AzTEC-3, and HDF\,850.1, we find a broad
range of physical properties among this CO-selected $z$$>$5 DSFG
sample, including a factor of $\sim$2.5 difference in dust
temperatures ($T_{\rm dust}$=35--92\,K), a range of a factor of
$\sim$3 in gas masses ($M_{\rm gas}$=2.2--7.1$\times$10$^{10}$\,\msol
) and gas-to-dust ratios (65--215), a factor of up to $\sim$30
difference in SFR surface densities ($\Sigma_{\rm
  SFR}$=60--1800\,\msol\,yr$^{-1}$\,kpc$^{-2}$), factors of $\sim$4
and $\sim$2 difference in the $L_{\rm CII}$/$L_{\rm FIR}$ and $L_{\rm
  CII}$/$L_{\rm CO(1-0)}$ ratios (0.6--2.5$\times$10$^{-3}$ and
2400--4500, respectively), and significant differences in CO line
excitation and the implied gas densities and kinetic temperatures. In
particular, we find a trend that appears to suggest a decrease in
$L_{\rm CII}$/$L_{\rm FIR}$ with increasing CO excitation. We also
find that the gas depletion times vary by a factor of $\sim$3 across
the sample ($\tau_{\rm dep}$=22--70\,Myr), consistent with short
starburst phases. Given the high inferred $\Sigma_{\rm SFR}$, we
cannot rule out a contribution of heavily obscured AGN to the dust
heating and/or gas excitation in these compact systems.

At the resolution of the current follow-up data, GN10 appears to
consist of a compact, $\sim$1.6\,kpc diameter, at least moderately
optically-thick ``maximum starburst'' nucleus embedded in a more
extended, $\sim$6.4\,kpc diameter rotating cold gas disk. This finding
is qualitatively consistent with those for lower-redshift DSFG samples
(e.g., Riechers et al.\ \citeyear{riechers11e}; Ivison et
al.\ \citeyear{ivison11}; Calistro Rivera et
al.\ \citeyear{calistro18}). AzTEC-3 appears to consist of a compact,
optically-thick $\sim$0.9\,kpc region exhibiting $\sim$75\% of the
dust luminosity, embedded in a more extended, $\sim$3.9\,kpc diameter
gas reservoir, which makes it an even more extreme nuclear starburst
than GN10. HDF\,850.1, on the other hand, appears to exhibit more
moderate properties for a massive starburst, with a $\sim$3$\times$
lower SFR than AzTEC-3 spread across a $\sim$6.7\,kpc diameter cold
gas reservoir, which contains two kinematic components (Neri et
al.\ \citeyear{neri14}).

By placing the COLDz $z$$>$5 DSFGs into context with all other $z$$>$5
DSFGs currently known, we find that their dust temperatures are
typically a factor of $\sim$1.5 higher than the bulk of the population
at $z$$\sim$2--3. On the one hand, such a trend is expected if $T_{\rm
  dust}$--$z$ relations proposed in the literature were to hold (e.g.,
Magnelli et al.\ \citeyear{magnelli14}; Schreiber et
al.\ \citeyear{schreiber18}), but the level of increase in $T_{\rm
  dust}$ with redshift is not captured well by these relations (which
require significant extrapolation in redshift). We investigate
potential biases due to preferential gravitational magnification,
sample selection wavelength, and SED fitting methods, but find no
clear evidence that the median $T_{\rm dust}$ is biased towards high
values due to the heterogenous composition of the parent sample. At
the same time, it cannot be ruled out that currently employed
selection methods miss $z$$>$5 DSFGs in undersampled regions of the
parameter space. On the other hand, the sample shows a trend
suggesting an increase in $L_{\rm FIR}$ towards higher $T_{\rm
  dust}$. Given their typically very high dust luminosities, this
trend is consistent with findings for lower-$z$ DSFG samples (e.g., da
Cunha et al.\ \citeyear{dacunha15}; Simpson et
al.\ \citeyear{simpson17a}) without requiring any redshift
evolution. This perhaps suggests that the observed trends in $T_{\rm
  dust}$ could be mostly, if not entirely due to the relatively high
median $L_{\rm FIR}$ (and thus, SFR) of the current $z$$>$5 DSFG
sample.

If the COLDz survey area is representative, the space density of
$z$$>$5 DSFGs could be significantly higher than previously thought
based on observations and simulations of the brightest DSFGs found in
large area sub/millimeter surveys. These sources appear to produce a
significant bright-end excess to the CO luminosity function compared
to models (R19; their Fig.~4). This appears consistent with recent
serendipitous discoveries of other $z$$>$5 DSFGs in targeted studies,
but statistics at the sub/millimeter flux levels of our sample are
currently too limited to allow for firm conclusions. Future large-area
molecular line scan surveys with the VLA, the \alma, and ultimately,
the Next Generation Very Large Array (ngVLA; e.g., Bolatto et
al.\ \citeyear{bolatto17}) will be required to put these findings on a
statistically more solid footing.

\acknowledgments

We thank the anonymous referee for a careful reading of the manuscript
and helpful comments that led to improvements in the structure and
content of this work. We also thank Christian Henkel for the original
version of the LVG code, Daizhong Liu for sharing results on the
de-blended photometry of GN10 in an early stage of the analysis and
the source data used in Fig.~\ref{f8}, Zhi-Yu Zhang for enlightening
discussions, and Ugne Dudzeviciute for help with the AS2COSMOS number
density calculations. D.R.\ and R.P.\ acknowledge support from the
National Science Foundation under grant numbers AST-1614213 and
AST-1910107 to Cornell University. D.R. also acknowledges support from
the Alexander von Humboldt Foundation through a Humboldt Research
Fellowship for Experienced Researchers. J.H. acknowledges support of
the VIDI research program with project number 639.042.611, which is
(partly) financed by the Netherlands Organization for Scientific
Research (NWO). I.R.S. acknowledges support from STFC
(ST/P000541/1). H.D.\ acknowledges financial support from the Spanish
Ministry of Science, Innovation and Universities (MICIU) under the
2014 Ram\'on y Cajal program RYC-2014-15686 and AYA2017-84061-P, the
latter one being co-financed by FEDER (European Regional Development
Funds). The National Radio Astronomy Observatory is a facility of the
National Science Foundation operated under cooperative agreement by
Associated Universities, Inc.

\appendix

\section{GN10 Line Parameters}

\begin{figure*}[b]
\epsscale{1.15}
\plotone{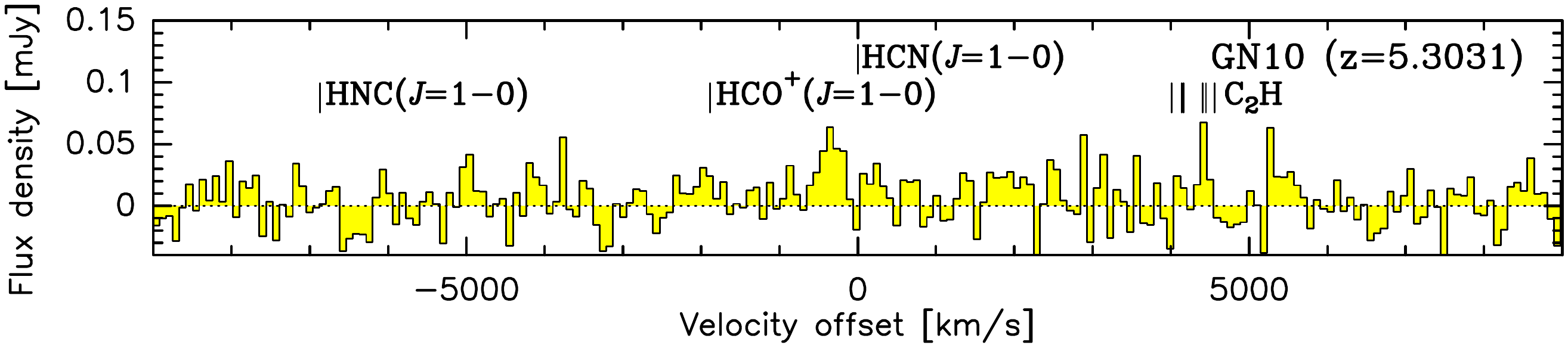}
\vspace{-2mm}

\caption{VLA line limit spectrum of GN10 ($z$=5.3031;
  histogram). Spectrum is shown at a resolution of 85\,\kms\ (4\,MHz),
  referenced to the expected redshift of the HCN($J$=1$\to$0)
  line. Velocities where the peaks of the HCN, HCO$^+$, and HNC
  $J$=1$\to$0 lines are expected to appear are indicated, as well as
  the hyperfine-structure transitions of the CCH($N$=1$\to$0)
  line. \label{fa2}}
%
\end{figure*}

Here we provide a table including the full line parameters from
Gaussian fitting to the line profiles for GN10 (Table~\ref{ta1}). We
also provide a figure showing the upper limit spectrum for the HCN,
HCO$^+$, and HNC $J$=1$\to$0 lines (Fig.~\ref{fa2}).

\begin{figure}
\begin{deluxetable}{ l c c c c c }
\tabletypesize{\scriptsize}
\tablecaption{GN10 line parameters. \label{ta1}}
\tablehead{
 Line & & $S_{\nu}$  & d$v_{\rm FWHM}$ & $v_0$\tablenotemark{a} & $I_{\rm line}$ \\
      & & [$\mu$Jy] & [\kms ]      & [\kms ] & [Jy\,\kms ] } 
\startdata
 \aco & & 74$\pm$13   & 687$\pm$144  & $-$23$\pm$60 & 0.054$\pm$0.017 \\
 \bco & & 544$\pm$63   & 512$\pm$72  & 0$\pm$30      & 0.295 $\pm$0.035 \\
 \eco & & 1046$\pm$205 & 772$\pm$220 & $-$27$\pm$71  & 0.86$\pm$0.20 \\
 \fco & & 719$\pm$144  & 681$\pm$173 & $-$78$\pm$70  & 0.52$\pm$0.11 \\
 \cii\ & (A) & (24.7$\pm$1.6)$\times$10$^3$ & 617$\pm$67  & $-$173$\pm$24 & 17.6$\pm$1.9\tablenotemark{b} \\
       & (B) & (6.0$\pm$4.2)$\times$10$^3$  & 227$\pm$243 & $-$875$\pm$128 & \\
 \ahcn & & & {\em (512)}\tablenotemark{c} & & $<$0.017 \\
 \ahco & & & {\em (512)}\tablenotemark{c} & & $<$0.017 \\
 \ahnc & & & {\em (512)}\tablenotemark{c} & & $<$0.017 \\
 \acch & & & {\em (512)}\tablenotemark{c} & & $<$0.025\tablenotemark{d} \\
\enddata 
\tablenotetext{\rm a}{Velocity offset relative to \bco\ redshift and uncertainty from Gaussian fitting.}
\tablenotetext{\rm b}{Summed over both components. Component (A) alone is (16.2$\pm$1.4)\,Jy\,\kms.}
\tablenotetext{\rm c}{Fixed to \bco\ line width.}
\tablenotetext{\rm d}{We conservatively assume equal strength of the hyperfine-strucutre transitions to obtain this limit. Assuming that the three strongest components dominate would yield a 3$\sigma$ limit of $<$0.021\,Jy\,\kms.}
\end{deluxetable}
\end{figure}


\section{GN10 \textsc{magphys} SED Fit}

\begin{figure*}
\epsscale{1.15}
\plotone{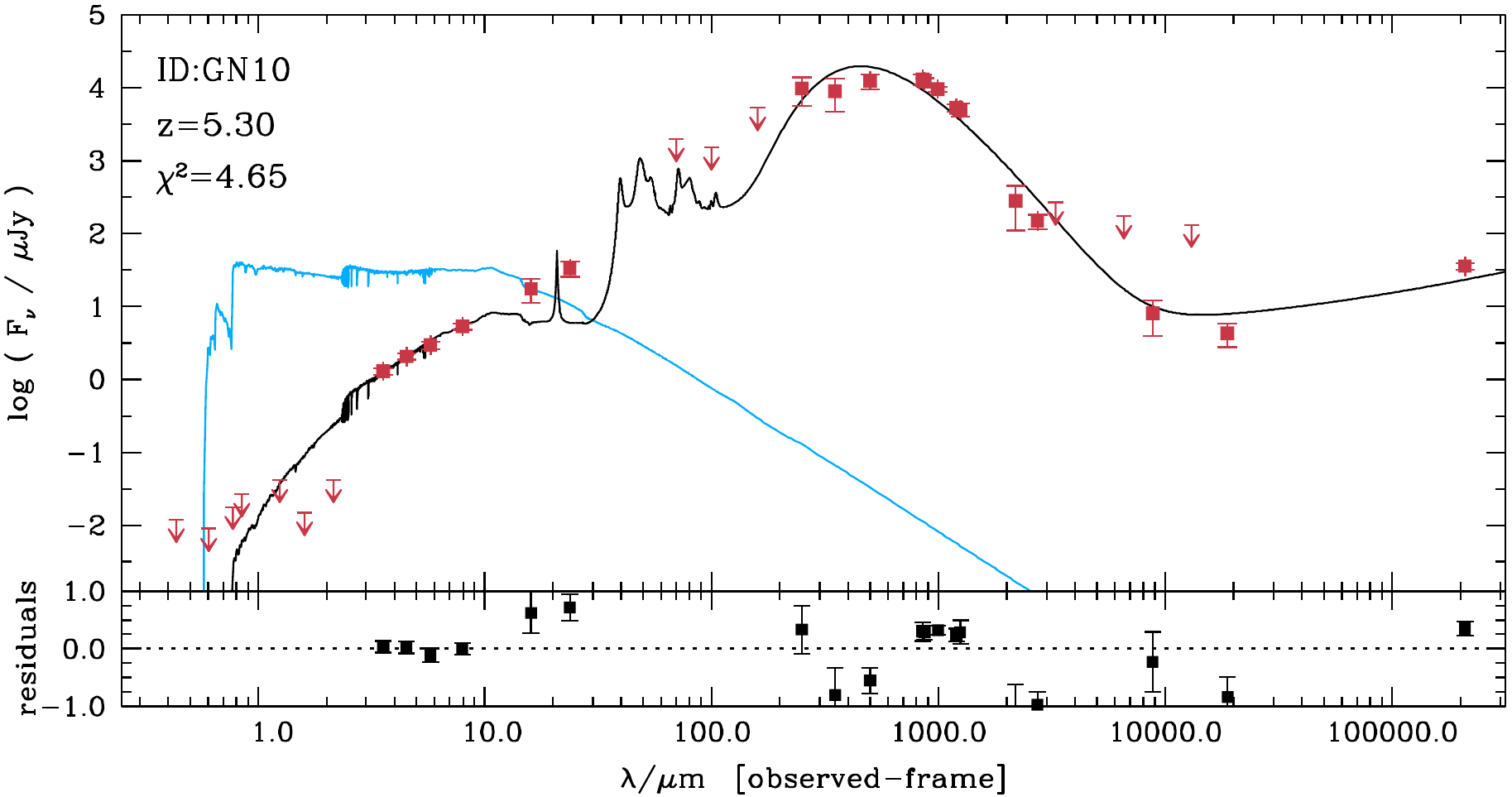}
\vspace{-2mm}

\caption{Spectral energy distribution of GN10 ({\em top;} red
  symbols), overlaid with best-fit \textsc{magphys} model (black line)
  and unattenuated stellar light emission spectrum before dust
  reprocessing (blue line), and residuals after subtracting the
  best-fit model ({\em bottom}). \label{fa3}}
%
\end{figure*}

In addition to \textsc{cigale,} we have also used the \textsc{magphys}
code (da Cunha et al.\ \citeyear{dacunha15}) to fit the full optical
to radio wavelength photometry of GN10. The fit to the spectral energy
distribution is shown in Fig.~\ref{fa3}, and the resulting physical
parameters are provided in Table~\ref{ta2}. \textsc{magphys} suggests
a dust luminosity $L_{\rm dust}$ that is $\sim$25\% higher than the
$L_{\rm IR}$ found from MBB fitting, but consistent within the
uncertainties. It also finds a total SFR that is comparable to the
SFR$_{\rm IR}$ found from the MBB fit, consistent with the expectation
that dust-obscured star formation dominates the SFR of GN10. $M_{\rm
  dust}$ is about half the value found from the MBB fit, and $T_{\rm
  dust}$ is $\sim$20\% lower (but consistent within the
uncertainties). These differences are likely due to the fact that
\textsc{magphys} uses multiple dust components in the fitting. In
particular, $T_{\rm dust}$ corresponds to a luminosity-averaged value,
calculated over multiple dust components. We adopt the values from the
MBB fit in the main text to enable a more straight forward comparison
to other $z$$>$5 sources, for which similar methods were used. We also
record these alternative values here to allow for comparison to other
samples modeled with \textsc{magphys} (e.g., da Cunha et
al.\ \citeyear{dacunha15}; Dudzeviciute et
al.\ \citeyear{dudzeviciute20}; Simpson et
al.\ \citeyear{simpson20}). \textsc{magphys} suggests approximately
two times the $M_\star$ found by \textsc{cigale}. We adopt the value
determined using \textsc{cigale} in the main text, as \textsc{cigale}
provides a better fit to the break between 2.2 and 3.6\,$\mu$m and to
the 16--24\,$\mu$m photometry.\footnote{We adopt the de-blended
  photometry throughout, but we caution that uncertainties due to
  deblending are significant for photometry in the latter wavelength
  range. {\em JWST} will be critical to overcome these uncertainties.}
We consider the two values to be consistent within the expected
uncertainties in determining $M_\star$ for highly-obscured $z$$>$5
galaxies like GN10.

\begin{figure}
\begin{deluxetable}{ l l l }
\tabletypesize{\scriptsize}
\tablecaption{GN10 \textsc{magphys} SED modeling parameters. \label{ta2}}
\tablehead{
Fit Parameter\hspace{1.5cm} & unit\hspace{1.5cm} & value\tablenotemark{a}\hspace{1.5cm} }
\startdata
$T_{\rm dust}$               & K     & 41.7$^{+5.0}_{-1.6}$ \\
$M_{\rm dust}$               & 10$^9$\,\msol\ & 0.58$^{+0.06}_{-0.04}$ \\
$L_{\rm dust}$               & 10$^{13}$\,\lsol\ & 1.26$^{+0.15}_{-0.16}$ \\
SFR$_{\rm total}$            & \msol\,yr$^{-1}$ & 1020$^{+150}_{-150}$ \\
$M_{\star}$\tablenotemark{e}     & 10$^{11}$\,\msol\ & 2.19$^{+1.28}_{-0.93}$ \\
\enddata 
\tablenotetext{\rm a}{Median values are given. Lower and upper error bars are stated as 16$^{\rm th}$ and 84$^{\rm th}$ percentiles, respectively.}
\end{deluxetable}
\end{figure}


\section{1.2\,mm continum imaging of GN20.2a and b}

\begin{figure*}
\epsscale{0.85}
\plotone{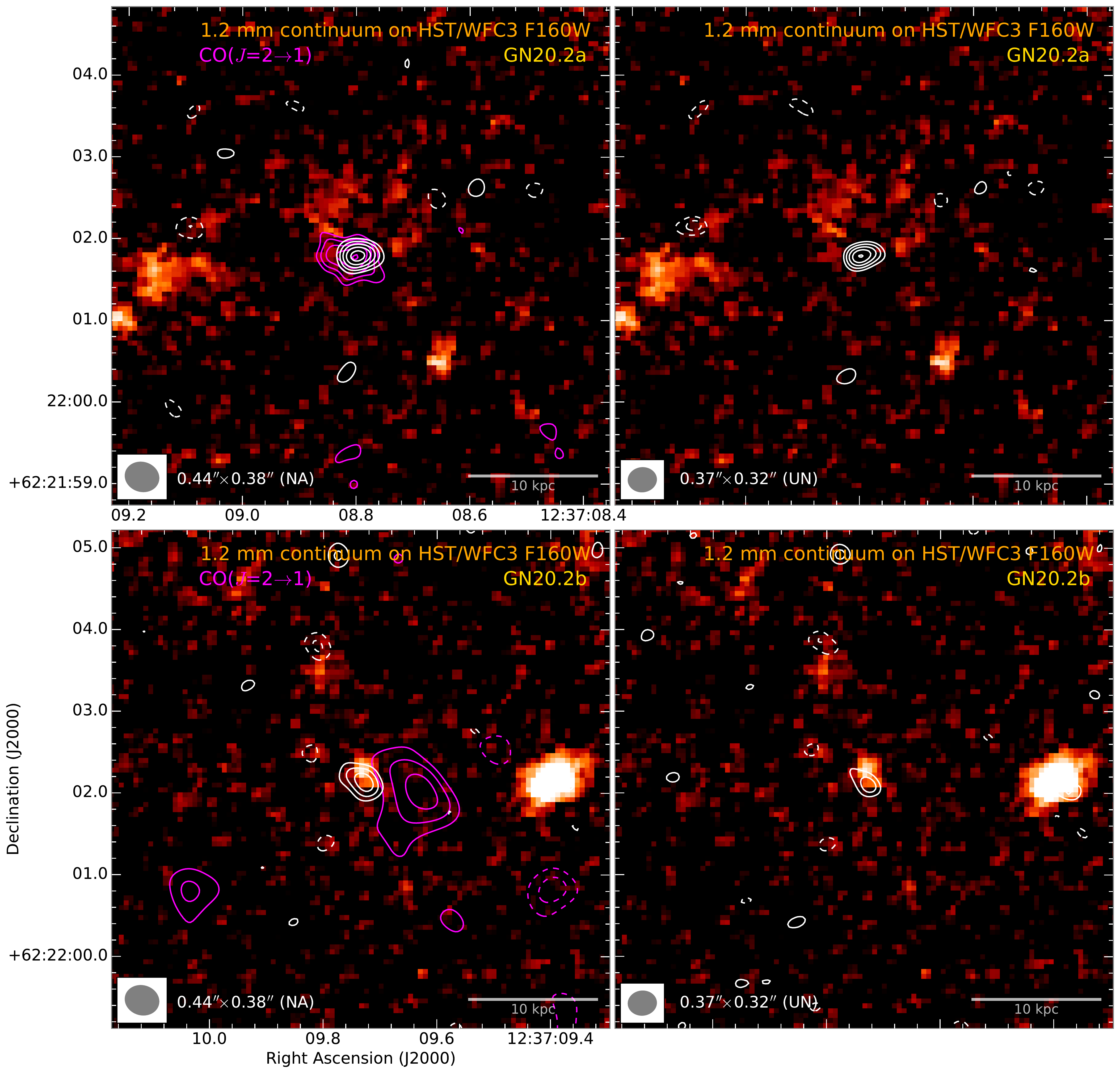}
\vspace{-2mm}

\caption{Rest-frame far-infrared continuum contour maps at
  observed-frame 1.2\,mm overlaid on a {\em HST}/WFC3 F160W continuum
  image toward GN20.2a (top) and GN20.2b (bottom). Contours start at
  $\pm$3$\sigma$, and are shown in steps of 1$\sigma$=0.35 (left;
  imaged using natural baseline weighting) and 0.41\,mJy\,beam$^{-1}$
  (right; imaged using uniform weighting), respectively. In the left
  panels, magenta \bco\ contours tapered to 0.38$''$ (GN20.2a) and
  0.77$''$ (GN20.2b; Hodge et al.\ \citeyear{hodge13b}) resolution are
  shown for comparison. Contour steps are the same, where 1$\sigma$=20
  and 28\,$\mu$Jy\,beam$^{-1}$ for GN20.2a and b, respectively. The
  synthesized beam size for the 1.2\,mm observations is indicated in
  the bottom left corner of each panel.\label{fa1}}
%
\end{figure*}

As part of the NOEMA 1.2\,mm observations of GN10 (project ID:\ T0B7;
PI:\ Riechers), we also observed GN20.2a ($z$=4.0508) and GN20.2b
($z$=4.0563), two member galaxies of the GN20 protocluster environment
(e.g., Daddi et al.\ \citeyear{daddi09a}; Hodge et
al.\ \citeyear{hodge13b}; Tan et al.\ \citeyear{tan14}) in track
sharing, leading to nearly the same amount of on source time and $u-v$
coverage (8936 visibilities).\footnote{We also observed GN20
  ($z$=4.0553) as part of this project in a second setup. Results from
  these data were reported by Hodge et al.\ (\citeyear{hodge15}).} The
pointing was centered between the two galaxies. From circular Gaussian
fitting to the visibility data, we find primary beam-corrected 1.2\,mm
fluxes of (3.85$\pm$0.71) and (4.24$\pm$0.95)\,mJy and source
diameters of 0.21$''$$\pm$0.08$''$ and 0.36$''$$\pm$0.09$''$ for
GN20.2a\footnote{Two-dimensional Gaussian fits suggest that GN20.2a is
  not resolved along its minor axis ($<$0.11$''$, or $<$0.8\,kpc),
  with a best-fit major axis diameter of 0.30$''$ (2.1\,kpc), but the
  fit does not converge well. As such, this estimate is considered to
  be a weak constraint at best.} and b,\footnote{For GN20.2b, a
  two-dimensional Gaussian fit suggests a size of
  (0.42$''$$\pm$0.11$''$)$\times$(0.28$''$$\pm$0.14$''$), or
  (3.0$\pm$0.8)$\times$(2.0$\pm$1.0)\,kpc$^2$.}  corresponding to
surface areas of (1.74$\pm$0.63) and (5.06$\pm$1.24)\,kpc$^2$,
respectively. The 1.2\,mm fluxes thus correspond to source-averaged
rest-frame brightness temperatures of $T_{\rm b}$=(8.5$\pm$1.9) and
(3.2$\pm$0.7)\,K, respectively. These modest values suggest that both
sources have significant substructure on scales below the resolution
of our observations. Both sources thus appear marginally resolved by
our observations (Fig.~\ref{fa1}).

Using the infrared luminosities measured by Tan et
al.\ (\citeyear{tan14}), we find infrared luminosity surface densities
of $\Sigma_{\rm IR}$=(2.6$\pm$1.0) and
(0.8$\pm$0.2)$\times$10$^{12}$\,\lsol\,kpc$^{-2}$ and SFR surface
densities of $\Sigma_{\rm SFR}$=(260$\pm$100) and
(80$\pm$20)\,\msol\,yr$^{-1}$\,kpc$^{-2}$ for GN20.2a and b,
respectively. This suggests that the star-formation activity in
GN20.2b is almost as intense as that in the central region of GN20
($\Sigma_{\rm SFR}$=(120$\pm$10)\,\msol\,yr$^{-1}$\,kpc$^{-2}$; Hodge
et al.\ \citeyear{hodge15}), while GN20.2a appears to be a more
intense starburst, approaching the activity level of ``maximum
starbursts''.

The dust continuum emission in GN20.2a appears to be more compact than
the \bco\ emission imaged by Hodge et al.\ (\citeyear{hodge13b}),
which has an extent of
(0.7$''$$\pm$0.1$''$)$\times$(0.4$''$$\pm$0.1$''$). The dust and cold
molecular gas emission appear to peak at the same position, such that
the rest-frame 237\,$\mu$m luminosity associated with the intense
starburst is likely dominantly emerging from the regions containing
the highest-density gas. The dust emission in GN20.2b also appears to
be more compact than the \bco\ emission, which has an extent of
(1.1$''$$\pm$0.4$''$)$\times$(0.7$''$$\pm$0.4$''$) as measured by
Hodge et al.\ (\citeyear{hodge13b}). Interestingly, the gas and dust
emission appear to be spatially offset by $<$1$''$, with the dust
emission peaking much closer to the likely near-infrared counterpart
of the dusty galaxy. This may indicate the presence of multiple galaxy
components, where the brightest CO-emitting component identified by
Hodge et al.\ (\citeyear{hodge13b}) is not the same as that dominating
the dust emission and stellar light. Another, perhaps less likely
possibility is that the dust-emitting component is not at the redshift
of the GN20 protocluster. On the other hand, the \fco\ emission
appears to peak at a position that is more consistent with the 1.2\,mm
dust continuum peak and the stellar light, albeit observed at about
two times lower linear spatial resolution (Tan et
al.\ \citeyear{tan14}). More sensitive CO observations are required to
further investigate the nature of this offset.

\facilities{VLA, NOEMA}

\bibliographystyle{yahapj}
\bibliography{ref.bib}

\end{document}